\documentclass[journal]{IEEEtran}
\usepackage[utf8]{inputenc}
\usepackage{latexsym}
\usepackage{amsfonts,amssymb,amsmath}
\usepackage{hyperref}
\def\BibTeX{{\rm B\kern-.05em{\sc i\kern-.025em b}\kern-.08em T\kern-.1667em\lower.7ex\hbox{E}\kern-.125emX}}
\usepackage{subfigure}
\usepackage{algorithm}
\usepackage{algpseudocode}
\usepackage{amsmath}
\usepackage{amsfonts}
\usepackage{graphicx}
\usepackage{epsfig}
\usepackage{cite}
\usepackage{txfonts}
\usepackage{relsize}
\usepackage{color}
\usepackage{color}
\usepackage{CJK}
\usepackage{times}
\usepackage{multicol}
\usepackage{stfloats}
\usepackage{float}
\usepackage{lipsum}
\usepackage{bm}

\begin{document}

\title{Performance Analysis of STAR-RIS-Assisted Cell-Free Massive MIMO Systems with Electromagnetic Interference and Phase Errors
\thanks{This work was supported by the Hong Kong Research Grants Council with Area of Excellence grant AoE/E-601/22-R.}}%Cell-Free Massive MIMO with Space-Constrained Access Points: The Analysis of Mutual Coupling and Spatial Correlation}
% The Effect of Mutual Coupling and Spatial Correlation on Cell-Free Massive MIMO with Space-Constrained Access Points
\author{Jun~Qian,~\IEEEmembership{Member,~IEEE,}    Ross~Murch,~\IEEEmembership{Fellow,~IEEE,}~and~Khaled~B.~Letaief,~\IEEEmembership{Fellow,~IEEE}
        
%\thanks{Jun Qian is with the Department of Electronic and Computer Engineering, The Hong Kong University of Science and Technology, Hong Kong (e-mail: eejunqian@ust.hk).}
\thanks{The authors are with the Department of Electronic and Computer
Engineering, The Hong Kong University of Science and Technology,
Hong Kong (e-mail: eejunqian@ust.hk, eermurch@ust.hk, eekhaled@ust.hk).}}

\maketitle
%\IEEEtitleabstractindextext
{\begin{abstract} 

Simultaneous Transmitting and Reflecting Reconfigurable Intelligent Surfaces (STAR-RISs) are being explored for sixth-generation (6G) wireless networks. A promising configuration for their deployment is within cell-free massive multiple-input multiple-output (MIMO) systems. However, despite the advantages that STAR-RISs could bring, challenges such as electromagnetic interference (EMI) and phase errors may lead to significant performance degradation.
In this paper, we investigate the impact of EMI and phase errors on STAR-RIS-assisted cell-free massive MIMO systems and propose techniques to mitigate these effects. We introduce a tailored projected gradient descent (GD) algorithm for STAR-RIS coefficient matrix design by minimizing the local channel estimation normalized mean square error (NMSE). We also derive the novel closed-form expressions of the uplink and downlink spectral efficiency (SE) to analyze system performance with EMI and phase errors, in which fractional power control methods are introduced for performance improvement. The results reveal that the projected GD algorithm can effectively tackle EMI and phase errors to improve estimation accuracy and compensate for performance degradation with nearly $30\%$ NMSE improvement and over $10\%$ SE improvement.
Moreover, increasing the number of access points (APs), antennas per AP, and STAR-RIS elements can also improve SE performance. However, the advantages of employing STAR-RIS are reduced when EMI and phase errors are severe. Notably, compared to conventional RISs, the incorporation of STAR-RIS in the proposed system yields better performance and presents less performance degradation in highly impaired environments.

\end{abstract}

}

\maketitle

\section{Introduction}
\IEEEPARstart{T}{here} is a growing interest in researching sixth-generation (6G) wireless communication to address the ever-increasing demand for ubiquitous connectivity and higher data rates\cite{9570143,7827017,9665300,8808168}. In past developments, massive multiple-input multiple-output (MIMO) has been one of the cornerstone technologies for enhancing data rates, reliability and coverage\cite{10297571,10225319,10167480}. However, in cellular networks with cell boundaries, massive MIMO experiences harsh inter-cell interference\cite{9665300}. 

To address the limit of massive MIMO in cellular networks, cell-free massive MIMO has been proposed to help diminish inter-cell interference, provide more extensive coverage and leverage favorable propagation configurations \cite{10058895,10422885,10535986}. Cell-free massive MIMO utilizes a distributed architecture with a central processing unit (CPU) that facilitates numerous geographically distributed access points (APs), which can synchronously provide service to users \cite{7917284,9416909,9665300}.
Cell-free massive MIMO integrates distributed networks and massive MIMO, so that the spectral efficiency (SE) performance can be greatly enhanced, even while using conjugate beamforming \cite{7827017,10297571,10621117}. In \cite{7917284,10167480}, a CPU-based large-scale fading decoding (LSFD) receiver has been proposed for uplink SE performance enhancement to utilize the benefits of cell-free massive MIMO. However, more APs introduce higher network overhead and power consumption, and the proposed system can experience poor quality of service (QoS) under severe propagation conditions \cite{9665300,10167480,10058895}. Therefore, advanced technologies must be developed to meet the required QoS and handle these challenges in cell-free massive MIMO.

Reconfigurable intelligent surfaces (RISs) can be used to shape electromagnetic waves smartly without increasing the number of APs or power consumption. They are thus regarded as a promising technology to assist cell-free massive MIMO systems \cite{9326394,10058895,10297571}. RISs are two-dimensional structures containing numerous passive reflecting elements that have the potential to reshape the propagation environment using the amplitude and phase shift adjustment at the RIS surface. RISs also do not need digital signal processing or active power amplifiers, so they can be implemented with lower complexity than a regular massive MIMO base station \cite{10621117,9905943,9875036}. RISs can enhance
wireless network performance by introducing additional controllable and reconfigurable cascaded links with low cost and power consumption \cite{9665300,10225319,9863732}. Integrating RIS with cell-free massive MIMO could allow the advantages of both to be obtained jointly, attracting research interests \cite{9322151,10001167,10556753}. Thus, the SE performance of spatially correlated RIS-assisted cell-free massive MIMO systems has been investigated \cite{10001167,10264149}. In \cite{10225319,9322151,10621117}, channel estimation improvement methods were introduced. \cite{9352948} introduced hybrid beamforming for energy efficiency (EE) optimization to enhance RIS-assisted cell-free massive MIMO system performance. Moreover, \cite{10167480,10621117} evaluated the system performance of electromagnetic interference (EMI)-aware RIS-assisted cell-free massive MIMO and \cite{10621117} introduced channel estimation improvement methods to mitigate the adverse effects of EMI. 

In their conventional form, RIS is configured to reflect incoming signals, and therefore, in wireless configurations, the receiver and transmitter are located on the same side of the RIS \cite{10316600,10297571}. However, in practice, users may be positioned on both RIS sides \cite{9570143,10297571}. To tackle this limitation, \cite{9437234,9690478} introduced the novel simultaneous transmitting and reflecting RIS (STAR-RIS) concept. 
STAR-RIS can provide full space coverage since the incoming signals incident on the STAR-RIS could be separated into reflected and transmitted signals \cite{9570143,10297571}. Moreover, the amplitudes and phases of the reflected and transmitted signals can be controlled by reflection and transmission coefficients, respectively, of the STAR-RIS \cite{9570143,9786058,10297571}. To realize the benefits of STAR-RISs, \cite{9462949} analyzed the fundamental coverage characterization. \cite{9740451} studied passive beamforming design and resource allocation to optimize the sum rate and coverage range of STAR-RIS-assisted orthogonal multiple access (OMA) and non-orthogonal multiple access (NOMA) networks. STAR-RIS was also introduced in \cite{10001603} to enhance full-duplex communication system performance. \cite{9786058,10373089} formulated a STAR-RIS-assisted massive MIMO system model to study the performance with phase errors caused by unavoidable phase-estimation imperfections \cite{8869792,9534477}.

Motivated by the interplay of RISs and cell-free massive MIMO systems, the joint benefits of STAR-RISs and cell-free massive MIMO systems may also be exploited \cite{10297571}. Research on STAR-RIS-assisted cell-free massive MIMO systems is at an early stage. It includes \cite{10297571}, which introduced spatial correlation, and \cite{10316600}, which studied the sum-rate optimization of STAR-RIS-assisted cell-free massive MIMO systems. \cite{10264149} studied the SE performance of active STAR-RIS-assisted cell-free massive MIMO systems considering spatial correlation. Similar to conventional RIS-assisted systems \cite{10167480,10621117,10133717,9598875}, when STAR-RIS elements are highly correlated, EMI impinging on the STAR-RIS will degrade system performance. Currently, no existing analyses focus on EMI-aware STAR-RIS-assisted cell-free massive MIMO systems. This includes the absence of the analysis of performance limits and the introduction of design guidelines for STAR-RIS-assisted cell-free massive MIMO systems suffering from EMI. Moreover, as far as we know, the study of STAR-RIS/RIS-assisted cell-free massive MIMO with phase errors has also been limited with only \cite{10841966} introducing the non-negligible phase error effect on system performance. 
Although EMI has been studied in RIS-assisted networks and phase errors have been introduced in STAR-RIS-assisted networks, STAR-RIS-assisted cell-free massive MIMO presents a different
landscape due to the specific nature of STAR-RIS, the unique interaction of two technologies, and unavoidable impairment factors. This novel landscape introduces specific system variables, poses system and beamforming challenges, and displays a different problem structure\cite{10556753}. Motivated by these observations, it is critical to analyze EMI and phase errors jointly to determine the feasibility of deploying STAR-RIS-assisted cell-free massive MIMO systems, introducing performance limits and design guidelines to promote realistic 6G implementations \cite{10167480,10373089}.

To the best of our knowledge, this paper is the first to focus on the performance analysis of spatially correlated STAR-RIS-assisted cell-free massive MIMO systems, where STAR-RISs are non-ideal, experiencing EMI and phase errors. 
In our work, based on the slowly varying statistical channel state information (CSI) in terms of large-scale statistics such as correlation matrices and
path-losses\cite{10297571,9875036,10326460}, we introduce a projected gradient descent (GD) algorithm to enhance the channel estimation normalized mean square error (NMSE) and reduce the performance degradation caused by EMI and phase errors. Subsequently, we derive the data transmission model and the closed-form expressions of the uplink and downlink SE adopting fractional power control. The uplink transmission utilizes the LSFD at the CPU, maximum ratio (MR) and local minimum mean square
error (MMSE) combining at APs, while the downlink transmission applies conjugate beamforming and regularized zero-forcing (RZF). 

The major contributions of the work are listed as follows:

\begin{itemize}
\item We establish a spatially correlated STAR-RIS-assisted cell-free massive MIMO model with EMI and phase errors, where STAR-RIS-assisted channels experience correlated Rician fading. As far as we know, this is the first work to introduce EMI and phase errors in STAR-RIS-assisted cell-free massive MIMO. These non-negligible factors impair system performance and require analysis to provide system limits and design guidelines.

\item We propose a tailored projected GD algorithm for the uplink channel estimation to optimize the STAR-RIS coefficient matrix based on slowly varying statistical CSI, which contains the optimization of the amplitudes and the phase shifts of the STAR-RIS. Our proposed GD algorithm, tailored to STAR-RIS-assisted networks, can minimize NMSE and introduce an additional channel estimation accuracy gain, addressing the unique challenges of STAR-RIS tied to EMI and phase errors.

\item We develop both uplink and downlink data transmission models experiencing EMI and phase errors.
MR and local
MMSE combiners are invoked for uplink transmission with a tailored LSFD in the CPU. Conjugate beamforming and RZF precoding are invoked for downlink transmission.
Based on the MR combiner and conjugate beamforming, closed-form expressions of uplink and downlink SE are derived with corresponding fractional power control, revealing insights into how EMI and phase errors impact system behaviour, facilitating efficient and targeted STAR-RIS-assisted system design.

\item  We demonstrate that the proposed projected GD algorithm can effectively tackle EMI and phase errors to improve estimation accuracy and compensate for performance degradation with nearly $30\%$ NMSE improvement and over $10\%$ SE improvement.
Meanwhile, increasing the number of APs, antennas per AP, and STAR-RIS elements can also improve SE performance. Moreover, STAR-RISs cannot bring significant gain when EMI and phase errors are severe, and the STAR-RIS is more susceptible to phase errors. Notably, applying STAR-RIS can achieve better performance in cell-free massive MIMO systems and experience less performance degradation in highly impaired environments compared to conventional RISs.

\end{itemize}

The remainder of this paper is organized as follows. Section II introduces the spatially correlated STAR-RIS-assisted channel model with EMI and phase errors. Section III introduces uplink channel estimation with pilot contamination and the projected GD algorithm. Section IV and Section V derive closed-form analytical expressions for the uplink and downlink SE with fractional power control, respectively. We provide numerical results and discussions in Section VI, and Section VII summarizes the paper and proposes subsequent works.

{\textit{Notation:} In this work, we utilize ${\textbf H}^T$, ${\textbf H}^H$, ${\textbf H}^*$ and ${\textbf H}^{-1}$ to represent the transpose, conjugate-transpose, conjugate and inverse of a matrix $\textbf H$, respectively. Moreover, ${\textbf I}_N$ denotes an $N \times N$ identity matrix. In addition, $|\cdot|$, $||\cdot||$ and $||\cdot||_2$ denote the respective  Absolute value, Standard norm and Euclidean
norm. $ \mathcal{CN}\left( {{\textbf{0}},{{\boldsymbol{\Sigma}}}} \right)$ is the circularly symmetric complex Gaussian distribution with zero mean and covariance $\boldsymbol{\Sigma}$, $\circ$ represents the element-wise product, and $\otimes$ denotes the Kronecker product. Finally, $\mathbb  E\{\cdot \}$ is the expectation operator, $\mathfrak{RE}\{\cdot \}$ is the real part and $\text {diag}(\textbf{h} )$ transform the vector $\textbf{h}$ into a diagonal matrix. }

\begin{figure}[!t]
\centering
\includegraphics[width=0.79\columnwidth]{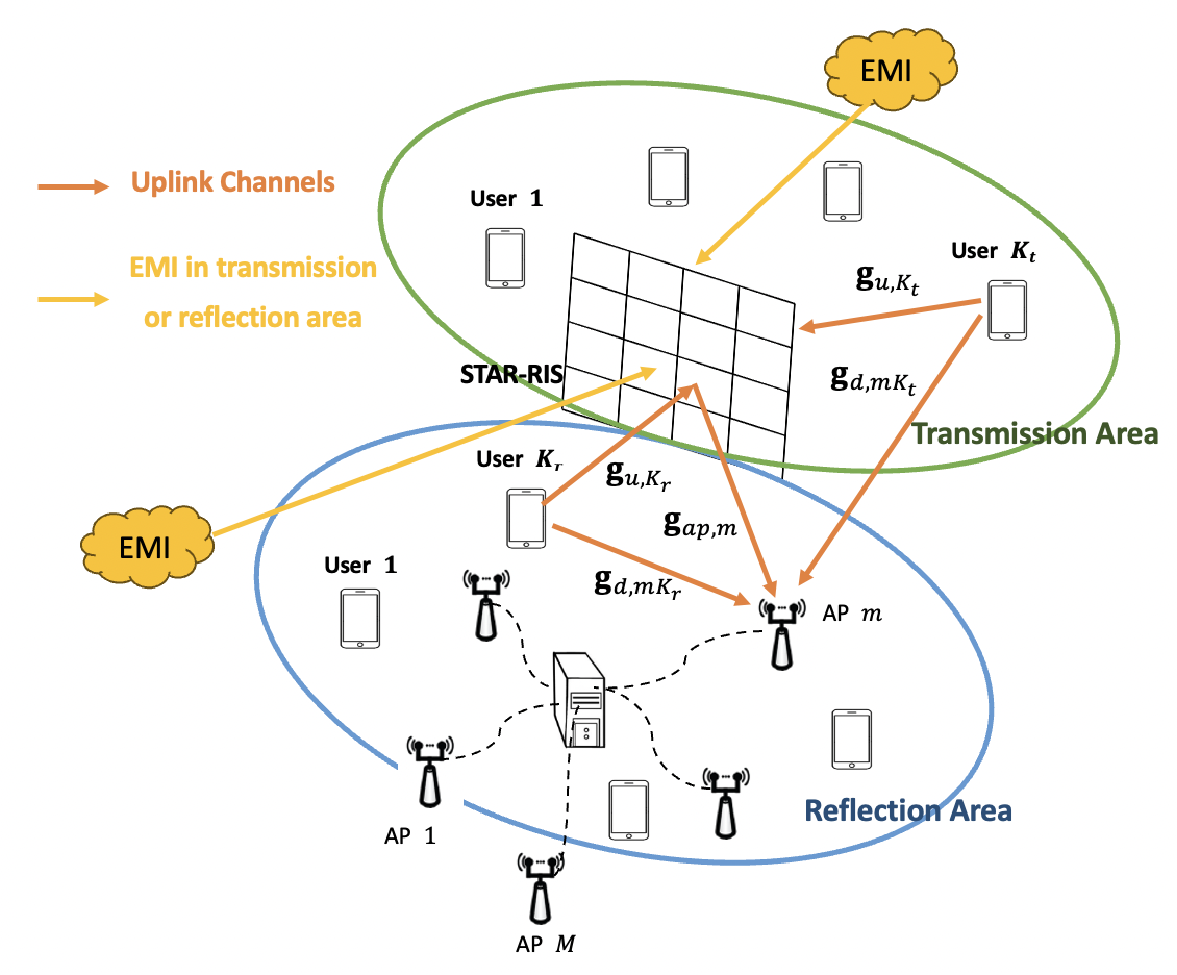}
\caption{Illustration of a STAR-RIS-assisted cell-free massive MIMO system.} 
\label{fig_1}
\vspace{-11 pt}
\end{figure}
\section{System Model}
Fig. \ref{fig_1} illustrates a STAR-RIS-assisted cell-free massive MIMO system\cite{10297571,10264149}. $M$ APs with $N$ antennas each, connected to the CPU through optimal backhaul links, can serve $K$ single-antenna users synchronously.
The $L$-element STAR-RIS assists in the communication between APs and users. First, the group of users indexed by set $\mathcal{K}_r$ with cardinality $|\mathcal{K}_r|=K_r$, are located on the same side of the APs and STAR-RIS in the reflection area. Similarly, users indexed by set $\mathcal{K}_t$ with $|\mathcal{K}_t|=K_t$, are located in the transmission area of the STAR-RIS. Note that $K_r+K_t=K$ and $\mathcal{K}_r\cap\mathcal{K}_t=\varnothing$. We also define the STAR-RIS operation mode of the $k$-th user as $\omega_k,~\forall k$. That is, if the $k$-th user is located in the reflection area experiencing the reflection mode of the STAR-RIS ($k \in \mathcal{K}_r$), $\omega_k=r$. Similarly, $\omega_k=t$ if the $k$-th user is located in the transmission area experiencing the transmission mode of the STAR-RIS ($k \in \mathcal{K}_t$) \cite{10297571}. Moreover, $\mathcal{W}_k$ is the set of users sharing the same STAR-RIS operation mode, including the $k$-th user. Note that all APs are randomly distributed in the reflection area in the light of \cite{10297571,10841966}. Adopting APs in both reflection and transmission areas will be left for our future work.
 
\subsection{STAR-RIS Protocols}

In general, energy splitting (ES), mode switching (MS), and time switching (TS) have been introduced as three feasible protocols for STAR-RIS operation \cite{9570143,10297571}. In this work, we mainly focus on the ES protocol, in which all STAR-RIS elements are operated jointly in reflection and transmission modes to serve all users regardless of their locations. As such, the STAR-RIS coefficient matrices follow
$\boldsymbol{\Theta}_{t}=\text{diag}(u_{1}^t{\theta}_{1}^t,u_{2}^t{\theta}_{2}^t,...,u_{L}^t{\theta}_{L}^t)\in\mathbb{C}^{L\times L}$ for the transmission mode and $\boldsymbol{\Theta}_{r}=\text{diag}(u_{1}^r{\theta}_{1}^r,u_{2}^r{\theta}_{2}^r,...,u_{L}^r{\theta}_{L}^r)\in\mathbb{C}^{L\times L}$ for the reflection mode, respectively, where ES amplitude coefficients are $u_{l}^t,~u_{l}^r \in[0,1]$, $(u_{l}^t)^2+(u_{l}^r)^2=1$, and the induced phase shifts are ${\theta}_{l}^t=e^{i\varphi_{l}^t},~{\theta}_{l}^r=e^{i\varphi_{l}^r}$ with $\varphi_{l}^t,~\varphi_{l}^r \in[0,2\pi),~\forall l$ \cite{9570143,10264149}. Note that the consideration that the reflecting coefficient phase and transmitting coefficient phase are coupled is left for future works \cite{10297571}. 
Moreover, it is possible to treat the MS protocol as a special case of the ES protocol, in which $u_{l}^t,~u_{l}^r$ are restricted to binary values, namely $u_{l}^t,~u_{l}^r \in\{0,1\}$
\cite{9570143,10297571}.

\subsection{STAR-RIS-assisted Channel Model with Phase Errors}
We assume that the system operates in time-division duplex (TDD) mode \cite{9905943,10621117}, experiencing channel reciprocity. In practice, RISs and STAR-RISs are deployed to enhance the AP-RIS and User-RIS communication by introducing Rician fading channels\cite{10167480}. Therefore, this work assumes that the direct channels between the $k$-th user and the $m$-th AP experience Rayleigh fading under the multi-scatterer distribution. The channels from the $k$-th user to the STAR-RIS  and from the STAR-RIS to the $m$-th AP experience Rician fading \cite{10167480,10901346}. The cascaded STAR-RIS-assisted uplink channel between the $k$-th user and the $m$-th AP, $\textbf{g}_{mk}\in \mathbb{C}^{N\times 1}$, can be provided by
\begin{equation}
     \displaystyle
\textbf{g}_{mk}=\textbf{g}_{d,mk}+\textbf{g}_{ap,m}\bar{\boldsymbol{\Theta}}_{\omega_k}\boldsymbol{\Theta}_{\omega_k}\textbf{g}_{u,k},
    \label{cascaded_uplink_channel_via_RIS}
   \end{equation}
   where $\textbf{g}_{d,mk}\in \mathbb{C}^{N\times 1}$ is the direct channel from the $k$-th user to the $m$-th AP, following $\mathcal{CN}(0,\beta_{d,mk}\textbf{R}_{d,mk})$. $\beta_{d,mk}$ is the large-scale fading coefficient between the $k$-th user and the $m$-th AP. $\textbf{R}_{d,mk}$ is the spatial correlation matrix at the $m$-th AP \cite{9099874,10167480}.
$\boldsymbol{\Theta}_{\omega_k}$ is the STAR-RIS coefficient matrix defined in Sec. II-A. The phase error matrix is defined as $\bar{\boldsymbol{\Theta}}_{\omega_k}=\text{diag}(\bar{{\theta}}_{1}^{\omega_k},\bar{{\theta}}_{2}^{\omega_k},...,\bar{{\theta}}_{L}^{\omega_k})\in\mathbb{C}^{L\times L}$, where $\bar{{\theta}}_{l}^{\omega_k}=e^{i\bar{\varphi}_{l}^{\omega_k}},~\forall l,~\forall k$. Note that the phase errors are typically modelled by independent and identically distributed (i.i.d.) random variables with a zero mean satisfying the uniform distribution or the von Mises distribution\cite{10025392,9534477,9786058}. In this paper, we consider $\bar{\varphi}_{l}^{\omega_k}\in[-a,a], ~\forall l$, following the uniform distribution with zero mean and a characteristic function as $\mathbb{E}\{e^{i\bar{\varphi}_{l}^{\omega_k}}\}=\frac{\text{sin}(a)}{a}=\phi,~\forall l$\cite{10373089,10025392}.
Furthermore,
the uplink channel from the STAR-RIS to the $m$-th AP, $\textbf{g}_{ap,m}\in \mathbb{C}^{N\times L}$, is obtained as
\begin{equation}
     \displaystyle \textbf{g}_{ap,m}=\sqrt{\bar{\beta}_{ap,m}}
     \bar{\textbf{g}}_{ap,m}
+\sqrt{\tilde{\beta}_{ap,m}}\underbrace{{\textbf{R}}_{ap,m}^{{1/2}}\textbf{v}_{ap,m}{\textbf{R}}_\text{SRIS}^{{1/2}}}_{\tilde{\textbf{g}}_{ap,m}},
     \label{channel_RIS_AP}
   \end{equation}
   where $\bar{\beta}_{ap,m}=\frac{\beta_{ap,m}\kappa_{ap,m}}{\kappa_{ap,m}+1}$, $\tilde{\beta}_{ap,m}=\frac{\beta_{ap,m}}{\kappa_{ap,m}+1}$. $\kappa_{ap,m}$ is the rice factor, $\beta_{ap,m}$ is the large-scale fading coefficient between the $m$-th AP and the STAR-RIS, $\textbf{v}_{ap,m}\in \mathbb{C}^{N\times L}$ consists of i.i.d. random variables following $\mathcal{CN}(0,1)$. $\bar{\textbf{g}}_{ap,m}$ is the line-of-sight (LoS) component. The respective spatial correlation matrices at the $m$-th AP and STAR-RIS are ${\textbf{R}}_{ap,m}$ and ${\textbf{R}}_\text{SRIS} =A{\textbf{R}} \in \mathbb{C}^{L\times L}$ with SRIS representing STAR-RIS\cite{10621117,10297571}. In particular, $d_H$ and $d_V$ are the respective horizontal width and the vertical height to introduce the STAR-RIS element area $A=d_Hd_V$ \cite{9300189}. The $(x,y)$-th element in $\textbf{R}$ is given by $\displaystyle [\textbf{R}]_{x,y}=\text{sinc}\Bigg{(}\frac{2||\textbf{e}_x-\textbf{e}_y||}{\lambda_c}\Bigg{)}$ \cite{9598875},
 where $\text{sinc}(a)=\text{sin}(\pi a)/(\pi a)$ is the sinc function, $\lambda_c$ is the carrier wavelength \cite{9598875,10167480,10001167}. The position vector of the $x$-th STAR-RIS element is $\textbf{e}_x=[0,\text{mod}(x-1,L_h)d_h,\lfloor(x-1)/L_h\rfloor d_v]^T$ \cite{10167480,10621117}. $L_h$ is the number of column elements and $L_v$ is the number of row elements at STAR-RIS, with $L=L_h L_v$.
Moreover, the uplink channel between the $k$-th user and the STAR-RIS, $\textbf{g}_{u,k}\in \mathbb{C}^{L\times 1}$ is formulated as
 \begin{equation}
     \begin{array}{c@{\quad}c}
\textbf{g}_{u,k}=\sqrt{\bar{\beta}_{u,k}}  e^{i\phi_{k}}   \bar{\textbf{g}}_{u,k}+\sqrt{\tilde{\beta}_{u,k}}\underbrace{{\textbf{R}}_{u,k}^{1/2}\textbf{v}_{u,k}}_{\tilde{\textbf{g}}_{u,k}},
     \end{array}
     \label{channel_user_RIS}
   \end{equation}
   similar to \eqref{channel_RIS_AP}, $\bar{\beta}_{u,k}=\frac{\beta_{u,k}\kappa_{u,k}}{\kappa_{u,k}+1}$, $\tilde{\beta}_{u,k}=\frac{\beta_{u,k}}{\kappa_{u,k}+1}$. $\kappa_{u,k}$ is the rice factor, $\beta_{u,k}$ is the large-scale fading coefficient between the $k$-th user and the STAR-RIS. $\bar{\textbf{g}}_{u,k}$ is the LoS component. ${\textbf{R}}_{u,k} ={\textbf{R}}_\text{SRIS} =A{\textbf{R}} \in \mathbb{C}^{L\times L}$ is the spatial correlation matrix of STAR-RIS. $\textbf{v}_{u,k}\in \mathbb{C}^{L\times 1}$ is the independent fast-fading channel containing i.i.d random variables obeying $\mathcal{CN}(0,1)$. 
Meanwhile, $\phi_{k}\sim\mathcal{U}[-\pi,~\pi],~\forall k$ is the
phase-shift of the LoS component between the $k$-th user and the STAR-RIS caused by the small changes in the location
of users due to user mobility, which is unknown to the system. Thus, in the absence of known $\phi_{k}$, $\mathbb{E}\{\textbf{g}_{mk}\}=0$ \cite{9258425}.
In this case, the channel vector $\textbf{g}_{mk}$ has the covariance matrix $\displaystyle \mathbf{\Delta}_{mk} = \mathbb{E}\{\textbf{g}_{mk}\textbf{g}_{mk}^H\}$ distributed as \eqref{Delta} at the top of this page 
\begin{figure*}[t!]
\begin{equation}
     \begin{array}{ll}
\displaystyle \mathbf{\Delta}_{mk} 
\displaystyle
\displaystyle=\underbrace{
{\beta}_{d,mk}\textbf{R}_{d,mk}}_{{\mathbf{\Delta}}_{mk}^d}
\displaystyle+\underbrace{
\tilde{\beta}_{ap,m}\textbf{R}_{ap,m}\text{tr}\left(\textbf{R}_\text{SRIS}\left(\bar{\beta}_{u,k}\bar{\textbf{T}}_{\omega_k}+\tilde{\beta}_{u,k}{\textbf{T}}_{\omega_k}\right)\right)+\bar{\beta}_{ap,m}\bar{\textbf{g}}_{ap,m}\left(\bar{\beta}_{u,k}\bar{\textbf{T}}_{\omega_k}+\tilde{\beta}_{u,k}{\textbf{T}}_{\omega_k}\right)\bar{\textbf{g}}_{ap,m}^H
}_{{\mathbf{\Delta}}_{mk}^c},
     \end{array}
\label{Delta}
    \vspace{-5 pt}
   \end{equation}
    \vspace{-10 pt}
   \hrulefill
   \end{figure*}
with
\begin{equation}
     \begin{array}{ll}
\bar{\textbf{T}}_{\omega_k}=
\boldsymbol{\Theta}_{\omega_k}\bar{\textbf{A}}_{u,k}\boldsymbol{\Theta}_{\omega_k}^H,
%\boldsymbol{\Theta}_{\omega_k}\left[\phi^2\boldsymbol{\Phi}_{k}\bar{\textbf{g}}_{u,k}\bar{\textbf{g}}_{u,k}^H\boldsymbol{\Phi}_{k}^H+(1-\phi^2)\boldsymbol{\Phi}_{k}\bar{\textbf{g}}_{u,k}\bar{\textbf{g}}_{u,k}^H\boldsymbol{\Phi}_{k}^H\circ\textbf{I}_L\right]\boldsymbol{\Theta}_{\omega_k}^H,
     \end{array}
\label{T_bar_matrix}
   \end{equation}
\begin{equation}
     \begin{array}{ll}
\displaystyle \textbf{T}_{\omega_k}=
\boldsymbol{\Theta}_{\omega_k}\bar{\textbf{R}}_\text{SRIS}\boldsymbol{\Theta}_{\omega_k}^H,
%A\boldsymbol{\Theta}_{\omega_k}\left[\phi^2\textbf{R}+(1-\phi^2)\textbf{R}\circ\textbf{I}_L\right]\boldsymbol{\Theta}_{\omega_k}^H,
%A^2{\textbf{R}}^{1/2}\bar{\boldsymbol{\Theta}}_{\omega_k}\boldsymbol{\Theta}_{\omega_k}\textbf{R}\boldsymbol{\Theta}_{\omega_k}^H\bar{\boldsymbol{\Theta}}_{\omega_k}^H{\textbf{R}}^{1/2}=A^2{\textbf{R}}^{1/2}\bar{\textbf{R}}_{\omega_k}{\textbf{R}}^{1/2},
     \end{array}
\label{T_matrix_phase_mean}
   \end{equation}
 where $\bar{\textbf{A}}_{u,k}=\mathbb{E}\{\bar{\boldsymbol{\Theta}}_{\omega_k}\bar{\textbf{g}}_{u,k}\bar{\textbf{g}}_{u,k}^H\bar{\boldsymbol{\Theta}}_{\omega_k}^H\}=\phi^2\bar{\textbf{g}}_{u,k}\bar{\textbf{g}}_{u,k}^H+(1-\phi^2)(\bar{\textbf{g}}_{u,k}\bar{\textbf{g}}_{u,k}^H)\circ\textbf{I}_L$,  $\bar{\textbf{R}}_\text{SRIS}=\mathbb{E}\{\bar{\boldsymbol{\Theta}}_{\omega_k}\textbf{R}_\text{SRIS}\bar{\boldsymbol{\Theta}}_{\omega_k}^H\}=\phi^2\textbf{R}_\text{SRIS}+(1-\phi^2)\textbf{R}_\text{SRIS}\circ\textbf{I}_L$. Notably, $\mathbf{\Delta}_{mk}$ does not depend on the phase-shifts if $\phi=0$, namely, $a=\pi$ \cite{9786058}. Phase errors that are von Mises-distributed can also be utilized \cite{10373089}. 

\subsection{Electromagnetic Interference Model} 
Based on \cite{9598875}, successive incoming plane waves introduced by external sources can superpose to generate EMI to degrade the system performance. The electromagnetic waves are typically incident from directions spanning large angular intervals\cite{10167480,9598875,10621117}. Then, the STAR-RIS spatial correlation model applies to uniform-distribution isotropic scattering \cite{9665300,9598875}. The EMI impinging on the STAR-RIS can be given by
\begin{equation}
     \begin{array}{c@{\quad}c}
\displaystyle \textbf{n}_\omega\sim\mathcal{CN}(0,A\sigma_\omega^2\textbf{R}),
     \end{array}
     \label{EMI}
   \end{equation}
where $\omega\in\{t,r\}$ represents the transmission or reflection area EMI comes from, $\sigma_\omega^2$ is the EMI power in the relevant area. We introduce a modified EMI power expression referring to \cite{10167480,9598875,10621117} to make the EMI scalable and feasible. First, when $\omega=r$, the EMI in the reflection area has the power
\begin{equation}
     \begin{array}{c@{\quad}c}
\displaystyle \sigma_r^2= \displaystyle\sqrt{\frac{\displaystyle p_up_d\sum\nolimits_{m=1}^M\beta_{ap,m}\sum\nolimits_{k\in\mathcal{K}_r}\beta_{u,k}}{MK_r\rho_r^2}} ,
     \end{array}
     \label{EMI_r}
   \end{equation}
where $p_d$ and $p_u$ are the respective downlink and user transmit power. $\rho_r$ is the ratio of the received signal power and the EMI power in the STAR-RIS reflection area \cite{10167480,9598875}. Similarly, when $\omega=t$, the EMI in the transmission area has the power
\begin{equation}
     \begin{array}{c@{\quad}c}
\displaystyle \sigma_t^2= \frac{\displaystyle p_u\sum\nolimits_{k\in\mathcal{K}_t}\beta_{u,k}}{K_t\rho_t},
     \end{array}
     \label{EMI_t}
   \end{equation}
where $\rho_t$ represents the ratio of the received signal power and the EMI power in the STAR-RIS transmission area. For simplicity, $\rho_t=\rho_r=\rho$ is assumed in this work.

\section{Uplink Channel Estimation Utilizing the Projected GD Algorithm}
We propose a projected GD algorithm that minimizes the channel estimation NMSE to optimize the STAR-RIS coefficient matrix. The optimization of the STAR-RIS coefficient matrix is crucial since improving the channel estimation accuracy results in notable performance enhancement \cite{9665300,10297571}. In our approach, we adopt the cascaded channel estimation with an uplink channel estimation phase using pilot symbols \cite{10297571}.

\subsection{Uplink Channel Estimation}
For uplink channel estimation, $\sqrt{\tau_p}\boldsymbol{\varphi}_k\in \mathbb{C}^{\tau_p\times 1}$ is assumed to be the pilot sequence allocated to $k$-th user, in which $\tau_p$ is the pilot signal length and $\boldsymbol{\varphi}_k^{H}\boldsymbol{\varphi}_k=1,\forall k$. Note that $\tau_c$, the coherence interval, is much larger than $\tau_p$ for higher transmission efficiency \cite{9665300,10167480}. In practice, $K>\tau_p$ introduces users sharing the same orthogonal pilot sequences and results in pilot contamination \cite{9416909}. We adopt $\mathcal{P}_k$ with $\boldsymbol{\varphi}_k^{H}\boldsymbol{\varphi}_{k'}=1, \forall k' \in \mathcal{P}_k$, to represent the user set using the same pilot sequence including $k$ itself.
 Distinct from the conventional assumption \cite{9665300,10225319}, we consider that EMIs from the reflection area and transmission area impinging on the STAR-RIS jointly are received by APs in the channel estimation phase\cite{9598875,10167480}. Thus, we can obtain the received signal at $m$-th AP, $\textbf{Y}_{m,p}\in \mathbb{C}^{N\times \tau_p}$, as
\begin{equation}
\begin{array}{ll}
     \displaystyle \textbf{Y}_{m,p}&\displaystyle=\sqrt{\tau_pp_p}\sum\nolimits_{k=1}^{K}\textbf{g}_{mk}\boldsymbol{\varphi}_k^{H}\\&\displaystyle+\textbf{g}_{ap,m}\bar{\boldsymbol{\Theta}}_{r}\boldsymbol{\Theta}_{r}\textbf{N}_{r}+\textbf{g}_{ap,m}\bar{\boldsymbol{\Theta}}_{t}\boldsymbol{\Theta}_{t}\textbf{N}_{t}+\textbf{N}_{m,p},
\end{array}\label{received_pilot_signal}
   \end{equation}
where $p_p$ is the pilot transmit power, $\textbf{N}_r\in \mathbb{C}^{L\times \tau_p}$ and $\textbf{N}_t\in \mathbb{C}^{L\times \tau_p}$ are the respective EMI from the reflection area and transmission area collected over the $\tau_p$ samples, with the $v$-th column following
$\Big{[}\textbf{N}_\omega\Big{]}_v\sim \mathcal{CN}(\textbf{0},A\sigma_\omega^2\textbf{R})$, $\omega\in\{t,r\}$. Similarly, $\textbf{N}_{m,p}\in \mathbb{C}^{N\times \tau_p}$ is the additive white Gaussian noise (AWGN) collected over the $\tau_p$ samples, where the $v$-th column follows $\Big{[}\textbf{N}_{m,p}\Big{]}_v\sim \mathcal{CN}(\textbf{0},\sigma^2\textbf{I}_{N})$, in which $\sigma^2$ is the noise power.
Then, the projection of $\textbf{Y}_{m,p}$ on $\boldsymbol{\varphi}_k$ is given by
\begin{equation}
\begin{array}{ll}
     \displaystyle \textbf{y}_{mk,p}=\textbf{Y}_{m,p}\boldsymbol{\varphi}_k\\~~~~~~\displaystyle=\sqrt{\tau_pp_p}\sum\nolimits_{k'=1}^{K}\textbf{g}_{mk'}\boldsymbol{\varphi}_{k'}^{H}\boldsymbol{\varphi}_k\\~~~~~~\displaystyle+\textbf{g}_{ap,m}\left(\bar{\boldsymbol{\Theta}}_{r}\boldsymbol{\Theta}_{r}\textbf{N}_{r}+\bar{\boldsymbol{\Theta}}_{t}\boldsymbol{\Theta}_{t}\textbf{N}_{t}\right)\boldsymbol{\varphi}_k+\textbf{N}_{m,p}\boldsymbol{\varphi}_k\\ ~~~~~~\displaystyle=\sqrt{\tau_pp_p}\sum\nolimits_{k'\in\mathcal{P}_k}\textbf{g}_{mk'}+\textbf{g}_{ap,m}\left(\bar{\boldsymbol{\Theta}}_{r}\boldsymbol{\Theta}_{r}\textbf{N}_{r}+\bar{\boldsymbol{\Theta}}_{t}\boldsymbol{\Theta}_{t}\textbf{N}_{t}\right)\boldsymbol{\varphi}_k\\ \vspace{1.4 pt}~~~~~~\displaystyle+\textbf{N}_{m,p}\boldsymbol{\varphi}_k.
\end{array}\label{projected_pilot_signal_j-th_phase}
   \end{equation}
According to the MMSE estimation method \cite{9322151,10621117}, $\textbf{g}_{mk}$ is estimated by 
\begin{equation}
\begin{array}{ll}
     \displaystyle \hat{\textbf{g}}_{mk}=\frac{\mathbb{E}\Big{\{}\textbf{y}_{mk,p}\textbf{g}_{mk}^H\Big{\}}}{\mathbb{E}\Big{\{}\textbf{y}_{mk,p}\textbf{y}_{mk,p}^H\Big{\}}}\textbf{y}_{mk,p}=\sqrt{\tau_pp_p}\mathbf{\Delta}_{mk}\mathbf{\Psi}_{mk}^{-1}\textbf{y}_{mk,p},
\end{array}\label{LMMSE_ES}
   \end{equation}
where 
\begin{equation}
\begin{array}{ll}
     \displaystyle \mathbf{\Psi}_{mk}\displaystyle& \displaystyle=\tau_pp_p\sum\nolimits_{k'\in\mathcal{P}_k}\mathbf{\Delta}_{mk'}+
     \bar{\beta}_{ap,m}\bar{\textbf{g}}_{ap,m}\left(\sigma_r^2\textbf{T}_r+\sigma_t^2\textbf{T}_t\right)\bar{\textbf{g}}_{ap,m}^H
     \\
     & \displaystyle
     +\tilde{\beta}_{ap,m}{\textbf{R}}_{ap,m}\text{tr}\left(
     {\textbf{R}}_\text{SRIS}\left(\sigma_r^2\textbf{T}_r+\sigma_t^2\textbf{T}_t\right)
     \right)
     +\sigma^2\textbf{I}_N.
\end{array}\label{Psi}
   \end{equation}
 Based on the above-mentioned observations, the channel estimation $\hat{\textbf{g}}_{mk}$ and the estimation error ${\textbf{e}}_{mk}={\textbf{g}}_{mk}-\hat{\textbf{g}}_{mk}$ can be distributed by $\mathcal{CN}\left(\textbf{0},\textbf{Q}_{mk}\right)$ and $\mathcal{CN}\left(\textbf{0},\hat{\textbf{Q}}_{mk}\right)$, respectively, with
   \begin{equation}
\begin{array}{ll}
     \displaystyle \textbf{Q}_{mk}=\tau_pp_p\mathbf{\Delta}_{mk}(\mathbf{\Delta}_{mk}\mathbf{\Psi}_{mk}^{-1})^H.
     \end{array}
     \label{Num_direct}
   \end{equation}
      \begin{equation}
\begin{array}{ll}
   \hat{\textbf{Q}}_{mk}=\mathbf{\Delta}_{mk}-{\textbf{Q}}_{mk}.
     \end{array}
   \end{equation}
Then, referring to \cite{10225319,9570816}, the channel estimation accuracy can be verified by 
the channel estimation NMSE, given by
\begin{equation}
\begin{array}{ll}
\displaystyle\text{NMSE}=\displaystyle\frac{\displaystyle\sum\nolimits_{m=1}^M\sum\nolimits_{k=1}^K\mathbb{E}\Big{\{}|| \tilde{\textbf{g}}_{mk}||^2\Big{\}}}{\displaystyle\sum\nolimits_{m=1}^M\sum\nolimits_{k=1}^K\mathbb{E}\Big{\{}|| \textbf{g}_{mk}||^2\Big{\}}}=\displaystyle \frac{\displaystyle\sum\nolimits_{m=1}^M\sum\nolimits_{k=1}^K\text{tr}(\mathbf{\Delta}_{mk}-\textbf{Q}_{mk})}{\displaystyle\sum\nolimits_{m=1}^M\sum\nolimits_{k=1}^K\text{tr}(\mathbf{\Delta}_{mk})}.
     \end{array}
     \label{aggregate_uplink_channel_estimation}
   \end{equation}

\subsection{STAR-RIS Coefficient Matrix Optimization}
To improve the channel estimation accuracy, we focus on minimizing the NMSE from all users and APs in \eqref{aggregate_uplink_channel_estimation} to optimize STAR-RIS coefficient matrices. First, we model the optimization problem for the NMSE minimization of STAR-RIS-assisted cell-free massive MIMO systems regarding phase shifts and amplitudes as \cite{9665300,10297571}
\begin{subequations}
    \begin{align}
      P_1:~&\mathop {\min }\limits_{{\boldsymbol{\theta}},~\textbf{u}}\text{NMSE}
      \\
      & \text{subject~to} \nonumber \\
      &(u_{l}^t)^2+(u_{l}^r)^2=1,~\forall l\\
     &u_{l}^t\geq 0,~u_{l}^r\geq 0,~\forall l\\
      &|{\theta}_{l}^t|=|{\theta}_{l}^r|=1,~\forall l
    \end{align}
    \label{GD_optimization}
  \end{subequations}
  where ${\boldsymbol{\theta}}=[{\boldsymbol{\theta}}_t,{\boldsymbol{\theta}}_r]$ and $\textbf{u}=[\textbf{u}_t,\textbf{u}_r]$. ${\boldsymbol{\theta}}_\omega\in\mathbb{C}^{L\times L}$ and $\textbf{u}_\omega\in\mathbb{C}^{L\times L}$ are diagonal matrices with ${\boldsymbol{\Theta}}_\omega=\textbf{u}_\omega{\boldsymbol{\theta}}_\omega={\boldsymbol{\theta}}_\omega\textbf{u}_\omega,~\omega\in\{t,r\}$.
The optimization problem is non-convex \cite{10130156,10297571}. The amplitudes and phase shifts for reflection and transmission are coupled \cite{10297571}. Thus, we focus on the projected GD algorithm based on statistical CSI to achieve the optimal solution to the minimization problem locally \cite{8968400,10130156,9875036,10326460}. 
The proposed projected GD algorithm, decreasing the objective from the current iteration $({\boldsymbol{\theta}}^n,\textbf{u}^n)$ to the gradient direction, contains the following iterations\cite{10297571,10093070,10164189}
 \begin{equation}
\begin{array}{ll}
{\boldsymbol{\theta}}^{n+1}=\text{P}_{{\boldsymbol{\theta}}}\left({\boldsymbol{\theta}}^n-\mu_{\boldsymbol{\theta}}\nabla _{{\boldsymbol{\theta}}^n}\text{NMSE}({\boldsymbol{\theta}}^n,\textbf{u}^n)\right),
     \end{array}
     \label{Optimization_1}
   \end{equation}
  \begin{equation}
\begin{array}{ll}
\textbf{u}^{n+1}=\text{P}_{\textbf{u}}\left(\textbf{u}^n-\mu_\textbf{u}\nabla _{\textbf{u}^n}\text{NMSE}({\boldsymbol{\theta}}^n,\textbf{u}^n)\right),
     \end{array}
     \label{Optimization_2}
   \end{equation}  
where $\mu_{\boldsymbol{\theta}}$ and $\mu_\textbf{u}$ are the respective step size for ${\boldsymbol{\theta}}$ and $\textbf{u}$. The superscript is the iteration index. To meet the constraints, we apply the projection functions $\text{P}_{{\boldsymbol{\theta}}}\left({\boldsymbol{\theta}}\right)$ and $\text{P}_{\textbf{u}}\left(\textbf{u}\right)$\cite{10297571,9473848}
\begin{flalign}
& [\text{P}_{{\boldsymbol{\theta}}}\left({\boldsymbol{\theta}}\right)]_{i,z}=\displaystyle\frac{{\boldsymbol{\theta}}_{i,z}}{|{\boldsymbol{\theta}}_{i,z}|}, ~z\in\{i,i+L\},~i=1,...,L,&\\
&[\text{P}_{\textbf{u}}\left(\textbf{u}\right)]_{i,z}=\displaystyle\frac{\textbf{u}_{i,z}}{\sqrt{\textbf{u}_{i,i}^2+\textbf{u}_{i,i+L}^2}}, ~z\in\{i,i+L\},~i=1,...,L.
\end{flalign}
Based on the observations mentioned above, Algorithm \ref{Algorithm} illustrates the iterative procedure of the projected GD algorithm to converge to a stationary point of $P_1$\cite{10130156,10297571}. We initialize the step size $\mu_{\boldsymbol{\theta}}$ and $\mu_\textbf{u}$ and reduce them by increasing the iterations with the penalty parameter $0\leq\varrho\leq 1$ \cite{10130156}.
  \begin{algorithm}[t!]
      \caption{Gradient Descent Based STAR-RIS Design} 
\begin{algorithmic}[1]
\renewcommand{\algorithmicrequire}{\textbf{Inputs:}}
\Require

$\epsilon$ (tolerance), IterMax, $\varrho$;
\renewcommand{\algorithmicensure}{\textbf{Output:}}
\Ensure
${\boldsymbol{\theta}}$, $\textbf{u}$, $\text{NMSE}({\boldsymbol{\theta}},\textbf{u})$
\State Initialize ${\boldsymbol{\theta}}^0$, $\textbf{u}^0$, $\textbf{f}^0=\text{NMSE}({\boldsymbol{\theta}}^0,\textbf{u}^0)$, $\mu_{\boldsymbol{\theta}},~\mu_\textbf{u}$; 
\For{$\text{t}= 1 :\text{IterMax}$}
\State ${\boldsymbol{\theta}}=[]$, $\textbf{u}=[]$;
\State  ${\boldsymbol{\theta}}=\text{P}_{{\boldsymbol{\theta}}}\left({\boldsymbol{\theta}}^0-\mu_{\boldsymbol{\theta}}\nabla _{{\boldsymbol{\theta}}^0}\text{NMSE}({\boldsymbol{\theta}}^0,\textbf{u}^0)\right)$
\State  $\textbf{u}=\text{P}_{\textbf{u}}\left(\textbf{u}^0-\mu_\textbf{u}\nabla _{\textbf{u}^0}\text{NMSE}({\boldsymbol{\theta}}^0,\textbf{u}^0)\right)$
\State Calculate $\textbf{f}=\text{NMSE}({\boldsymbol{\theta}},\textbf{u})$
\If{$\textbf{f}>\textbf{f}^0-\mu_{\boldsymbol{\theta}}||\nabla _{{\boldsymbol{\theta}}^0}\text{NMSE}({\boldsymbol{\theta}}^0,\textbf{u}^0)||_2^2$}
\State $\mu_{\boldsymbol{\theta}}=\varrho\mu_{\boldsymbol{\theta}}$;
\EndIf
\If{$\textbf{f}>\textbf{f}^0-\mu_{\textbf{u}}||\nabla _{\textbf{u}^0}\text{NMSE}({\boldsymbol{\theta}}^0,\textbf{u}^0)||_2^2$}
\State $\mu_{\textbf{u}}=\varrho\mu_{\textbf{u}}$;
\EndIf
\If {$\big{|}\textbf{f}-\textbf{f}^0\big{|}\leq \epsilon$}
\State $\textbf{break}$
       \Else  
       \State
       ${\boldsymbol{\theta}}^0={\boldsymbol{\theta}}$, $\textbf{u}^0=\textbf{u}$, $\textbf{f}^0=\textbf{f}$;
\EndIf
\EndFor
\end{algorithmic}
\label{Algorithm}
  %\vspace{-15 pt}
  \end{algorithm}
  \\
{\textit{Proposition 1:}} When the MS protocol is in operation, for simplicity, the projected GD algorithm for $\textbf{u}$ is updated to
  \begin{equation}
\begin{array}{ll}
\textbf{u}^{n+1}=\bar{\text{P}}_{\textbf{u}}\left(\nabla _{\textbf{u}^n}\text{NMSE}({\boldsymbol{\theta}}^n,\textbf{u}^n)\right),
     \end{array}
     \label{Optimization_21}
   \end{equation}  
with $z\in\{i,i+L\}$, $\bar{\text{P}}_{\textbf{u}}\left(\textbf{u}\right)$ is defined as
\begin{equation}
    \begin{array}{cc}
[\bar{\text{P}}_{\textbf{u}}\left(\textbf{u}\right)]_{i,z}=\displaystyle\left\{\begin{array}{cc} \displaystyle 1,~~\text{max}\left(|\textbf{u}_{i,i}|,|\textbf{u}_{i,i+L}|
\right)=|\textbf{u}_{i,z}|
\\\displaystyle0,~~\text{max}\left(|\textbf{u}_{i,i}|,|\textbf{u}_{i,i+L}|
\right)>|\textbf{u}_{i,z}|.

\end{array}
\right.
\end{array}
\end{equation}
\\
{\textit{Proposition 2:}} Note that the complex gradients of $\text{NMSE}({\boldsymbol{\theta}},\textbf{u})$ in terms of ${\boldsymbol{\theta}}$, $\textbf{u}$ can be computed as \cite{10297571,10093070,10164189}
 \begin{equation}
\begin{array}{ll}
\nabla _{\textbf{x}}\text{NMSE}({\boldsymbol{\theta}},\textbf{u})=[\nabla _{\textbf{x}_t}\text{NMSE}({\boldsymbol{\theta}},\textbf{u})^T,\nabla _{\textbf{x}_r}\text{NMSE}({\boldsymbol{\theta}},\textbf{u})^T]^T,
     \end{array}
     \label{Optimization_3}
   \end{equation}
   \begin{figure*}
  \begin{equation}
\begin{array}{ll}
\displaystyle \nabla _{\textbf{x}_\omega}\text{NMSE}({\boldsymbol{\theta}},\textbf{u}) & \vspace{5 pt}\displaystyle=\nabla _{\textbf{x}_\omega}\displaystyle \frac{\displaystyle\sum\nolimits_{m=1}^M\sum\nolimits_{k=1}^K\text{tr}(\mathbf{\Delta}_{mk}-\textbf{Q}_{mk})}{\displaystyle\sum\nolimits_{m=1}^M\sum\nolimits_{k=1}^K\text{tr}(\mathbf{\Delta}_{mk})}
\\ &\displaystyle=\frac{\displaystyle\sum\nolimits_{m=1}^M\sum\nolimits_{k=1}^K\text{tr}(\mathbf{Q}_{mk})\left(\sum\nolimits_{m=1}^M\sum\nolimits_{k\in\mathcal{K}_{\omega}}\nabla _{\textbf{x}_\omega}\text{tr}(\mathbf{\Delta}_{mk})\right)-\sum\nolimits_{m=1}^M\sum\nolimits_{k=1}^K\text{tr}(\mathbf{\Delta}_{mk})\left(\sum\nolimits_{m=1}^M\sum\nolimits_{k\in\mathcal{K}_{\omega}}\nabla _{\textbf{x}_\omega}\text{tr}(\mathbf{Q}_{mk})\right)}{\displaystyle\left(\sum\nolimits_{m=1}^M\sum\nolimits_{k=1}^K\text{tr}(\mathbf{\Delta}_{mk})\right)^2},
     \end{array}
     \label{Optimization_4}
      \vspace{-5 pt}
   \end{equation} 
    \vspace{-10 pt}
   \hrulefill
   \end{figure*}where $\textbf{x}\in\{{\boldsymbol{\theta}},\textbf{u}\}$. The decomposition of $\nabla _{\textbf{x}_\omega}\text{NMSE}({\boldsymbol{\theta}},\textbf{u})^T$ is shown in \eqref{Optimization_4} at the top of the next page with $\omega\in\{t,r\}$. 
Referring to \cite{Minka2000OldAN,4203075}, we can obtain $\nabla _{{\boldsymbol{\theta}}_{\omega_k}}\text{tr}(\boldsymbol{\Delta}_{mk})$ and $\nabla _{\textbf{u}_{\omega_k}}\text{tr}(\boldsymbol{\Delta}_{mk})$ which are shown as \eqref{Optimization_5} and \eqref{Optimization_9} at the top of the next page. Meanwhile, $\nabla _{{\boldsymbol{\theta}}_{\omega_k}}\text{tr}(\textbf{Q}_{mk})$ and $\nabla _{\textbf{u}_{\omega_k}}\text{tr}(\textbf{Q}_{mk})$ are shown as \eqref{Optimization_6} and \eqref{Optimization_10} at the top of the next page with the assistance of $\mathbf{\Pi}_{mk}=\mathbf{\Psi}_{mk}^{-1}\mathbf{\Delta}_{mk}+\mathbf{\Delta}_{mk}\mathbf{\Psi}_{mk}^{-1}$, $\bar{\mathbf{\Pi}}_{mk}=\mathbf{\Psi}_{mk}^{-1}\mathbf{\Delta}_{mk}\mathbf{\Delta}_{mk}\mathbf{\Psi}_{mk}^{-1}$.
 \begin{figure*}
\begin{equation}
\begin{array}{ll}
\displaystyle \nabla _{{\boldsymbol{\theta}}_{\omega_k}}\text{tr}(\mathbf{\Delta}_{mk}) &\displaystyle
=
\Big{(}\textbf{u}_{\omega_k}\left(\bar{\beta}_{u,k}\bar{\textbf{A}}_{u,k}+\tilde{\beta}_{u,k}\bar{\textbf{R}}_\text{SRIS}\right){\boldsymbol{\Theta}}_{\omega_k}^H\left(\bar{\beta}_{ap,m}\bar{\textbf{g}}_\text{ap,m}^H\bar{\textbf{g}}_\text{ap,m}+\tilde{\beta}_{ap,m}\text{tr}\left(\textbf{R}_{ap,m}\right){\textbf{R}}_\text{SRIS}\right)\Big{)}^T\circ\textbf{I}_L,
\end{array}
     \label{Optimization_5}
        \vspace{-5 pt}
   \end{equation}
   \vspace{-10 pt}
   \hrulefill
   \end{figure*}
   \begin{figure*}
  \begin{equation}
    \begin{array}{ll}
\displaystyle \nabla _{{\boldsymbol{\theta}}_{\omega_k}}\text{tr}(\textbf{Q}_{mk})&\displaystyle=\tau_pp_p%\left[\begin{array}{ll}
\Big{(}\textbf{u}_{\omega_k}\left(\bar{\beta}_{u,k}\bar{\textbf{A}}_{u,k}+\tilde{\beta}_{u,k}\bar{\textbf{R}}_\text{SRIS}\right){\boldsymbol{\Theta}}_{\omega_k}^H\left(\bar{\beta}_{ap,m}\bar{\textbf{g}}_\text{ap,m}^H\mathbf{\Pi}_{mk}\bar{\textbf{g}}_\text{ap,m}+\tilde{\beta}_{ap,m}\text{tr}\left(\mathbf{\Pi}_{mk}\textbf{R}_{ap,m}\right){\textbf{R}}_\text{SRIS}\right)\Big{)}^T\circ\textbf{I}_L
\\&\displaystyle-
     \tau_pp_p\left[\begin{array}{ll}
\displaystyle\tau_pp_p\sum\limits_{k'\in\mathcal{P}_k\cap\mathcal{W}_k}%\left(\begin{array}{ll}
\Big{(}\textbf{u}_{\omega_{k'}}\left(\bar{\beta}_{u,{k'}}\bar{\textbf{A}}_{u,{k'}}+\tilde{\beta}_{u,{k'}}\bar{\textbf{R}}_\text{SRIS}\right){\boldsymbol{\Theta}}_{\omega_{k'}}^H\left(\bar{\beta}_{ap,m}\bar{\textbf{g}}_\text{ap,m}^H\bar{\mathbf{\Pi}}_{mk}\bar{\textbf{g}}_\text{ap,m}+\tilde{\beta}_{ap,m}\text{tr}\left(\bar{\mathbf{\Pi}}_{mk}\textbf{R}_{ap,m}\right){\textbf{R}}_\text{SRIS}\right)\Big{)}^T\circ\textbf{I}_L\\
\displaystyle+
\sigma_{\omega_k}^2\Big{(}\textbf{u}_{\omega_k}\bar{\textbf{R}}_\text{SRIS}{\boldsymbol{\Theta}}_{\omega_k}^H\left(\bar{\beta}_{ap,m}\bar{\textbf{g}}_\text{ap,m}^H\bar{\mathbf{\Pi}}_{mk}\bar{\textbf{g}}_\text{ap,m}+\tilde{\beta}_{ap,m}\text{tr}\left(\bar{\mathbf{\Pi}}_{mk}\textbf{R}_{ap,m}\right){\textbf{R}}_\text{SRIS}\right)\Big{)}^T\circ\textbf{I}_L\displaystyle
\end{array}
\right]
    ,
  \end{array}
   \label{Optimization_6}
      \vspace{-5 pt}
   \end{equation} 
   \vspace{-10 pt}
   \hrulefill
\end{figure*}
       \begin{figure*}
  \begin{equation}
\begin{array}{ll}
\nabla _{\textbf{u}_{\omega_k}}\text{tr}(\mathbf{\Delta}_{mk}) &\displaystyle
=
2\mathfrak{RE}\left\{\Big{(}\boldsymbol{\theta}_{\omega_k}\left(\bar{\beta}_{u,k}\bar{\textbf{A}}_{u,k}+\tilde{\beta}_{u,k}\bar{\textbf{R}}_\text{SRIS}\right){\boldsymbol{\Theta}}_{\omega_k}^H\left(\bar{\beta}_{ap,m}\bar{\textbf{g}}_\text{ap,m}^H\bar{\textbf{g}}_\text{ap,m}+\tilde{\beta}_{ap,m}\text{tr}\left(\textbf{R}_{ap,m}\right){\textbf{R}}_\text{SRIS}\right)\Big{)}^T\circ\textbf{I}_L\right\},
\end{array}
     \label{Optimization_9}
      \vspace{-5 pt}
   \end{equation}
   \vspace{-10 pt}
    \hrulefill
\end{figure*}
         \begin{figure*}[t!]
 \begin{equation}
    \begin{array}{ll}
\displaystyle \nabla _{\textbf{u}_{\omega_k}}\text{tr}(\textbf{Q}_{mk})&\displaystyle=
2\tau_pp_p%\left[\begin{array}{ll}
\mathfrak{RE}\left\{\Big{(}\boldsymbol{\theta}_{\omega_k}\left(\bar{\beta}_{u,k}\bar{\textbf{A}}_{u,k}+\tilde{\beta}_{u,k}\bar{\textbf{R}}_\text{SRIS}\right){\boldsymbol{\Theta}}_{\omega_k}^H\left(\bar{\beta}_{ap,m}\bar{\textbf{g}}_\text{ap,m}^H\mathbf{\Pi}_{mk}\bar{\textbf{g}}_\text{ap,m}+\tilde{\beta}_{ap,m}\text{tr}\left(\mathbf{\Pi}_{mk}\textbf{R}_{ap,m}\right){\textbf{R}}_\text{SRIS}\right)\Big{)}^T\circ\textbf{I}_L\right\}
\\&\displaystyle-
    2 \tau_pp_p\mathfrak{RE}\left\{\begin{array}{ll}
\displaystyle\tau_pp_p\sum\limits_{k'\in\mathcal{P}_k\cap\mathcal{W}_k}%\left(\begin{array}{ll}
\Big{(}\boldsymbol{\theta}_{\omega_{k'}}\left(\bar{\beta}_{u,{k'}}\bar{\textbf{A}}_{u,{k'}}+\tilde{\beta}_{u,{k'}}\bar{\textbf{R}}_\text{SRIS}\right){\boldsymbol{\Theta}}_{\omega_{k'}}^H\left(\bar{\beta}_{ap,m}\bar{\textbf{g}}_\text{ap,m}^H\bar{\mathbf{\Pi}}_{mk}\bar{\textbf{g}}_\text{ap,m}+\tilde{\beta}_{ap,m}\text{tr}\left(\bar{\mathbf{\Pi}}_{mk}\textbf{R}_{ap,m}\right){\textbf{R}}_\text{SRIS}\right)\Big{)}^T\circ\textbf{I}_L\\
\displaystyle+
\sigma_{\omega_k}^2\Big{(}\boldsymbol{\theta}_{\omega_k}\bar{\textbf{R}}_\text{SRIS}{\boldsymbol{\Theta}}_{\omega_k}^H\left(\bar{\beta}_{ap,m}\bar{\textbf{g}}_\text{ap,m}^H\bar{\mathbf{\Pi}}_{mk}\bar{\textbf{g}}_\text{ap,m}+\tilde{\beta}_{ap,m}\text{tr}\left(\bar{\mathbf{\Pi}}_{mk}\textbf{R}_{ap,m}\right){\textbf{R}}_\text{SRIS}\right)\Big{)}^T\circ\textbf{I}_L\displaystyle
\end{array}
\right\}
,  \end{array}
   \label{Optimization_10}
    \vspace{-5 pt}
   \end{equation} 
   \vspace{-10 pt}
   \hrulefill
\end{figure*}

     \textit{Proof:} Please see Appendix A.
     
\textit{Remark 1:} To evaluate the algorithm feasibility, we deliver the computational complexity analysis of Algorithm \ref{Algorithm} based on the big-O notation \cite{9399102,10297571}. First, to compute $\boldsymbol{\Delta}_{mk}$ in \eqref{Delta}, we find that the complexity to compute $\text{tr}\left(\textbf{R}_\text{SRIS}{\textbf{T}}_{\omega_k}\right)$ and $\text{tr}\left(\textbf{R}_\text{SRIS}\bar{\textbf{T}}_{\omega_k}\right)$ is $O\left(L^3\right)$ since $\boldsymbol{\Theta}_{\omega_k}$ is diagonal matrix \cite{10297571,8932556}. Meanwhile,  the complexity to compute $\bar{\textbf{g}}_{ap,m}{\textbf{T}}_{\omega_k}\bar{\textbf{g}}_{ap,m}^H$ and $\bar{\textbf{g}}_{ap,m}\bar{\textbf{T}}_{\omega_k}\bar{\textbf{g}}_{ap,m}^H$ is $O\left(N^2L\right)$. Thus, the complexity to compute $\boldsymbol{\Delta}_{mk}$ is $O\left(L^3+N^2+N^2L\right)$, where $O\left(N^2\right)$ are required for $\text{tr}\left(\textbf{R}_\text{SRIS}{\textbf{T}}_{\omega_k}\right)\textbf{R}_{ap,m}$. Moreover, since $\textbf{Q}_{mk}$ contains the matrix inverse of $\mathbf{\Psi}_{mk}$, it takes $O\left(N^3\right)$\cite{10297571}. Consequently, the complexity for each iteration of $\text{NMSE}({\boldsymbol{\theta}},\textbf{u})$ is $O\left(MK\left(N^3+L^3+N^2+N^2L\right)\right)$.\footnote{It is noted that the projected GD algorithm has increasing complexity as the number of users grows, which might become unfeasible in practice. Therefore, future efforts should focus on efficient resource allocation schemes and feasible optimization algorithm design to sustain the feasibility of the relevant techniques in real-world scenarios.}

As an overall summary, we can use \eqref{Num_direct} and \eqref{aggregate_uplink_channel_estimation} to determine the channel estimation NMSE by applying the projected GD algorithm with \eqref{Optimization_1}-\eqref{Optimization_2} to optimize \eqref{GD_optimization}.

\section{Uplink Transmission and Closed-form SE Expressions}
We derive novel closed-form uplink SE expressions with fractional power control to investigate the uplink performance of the proposed STAR-RIS-assisted cell-free massive MIMO systems. EMI and phase errors are included in the analysis. Local MMSE and MR combining at the APs and LSFD processing at the CPU during the uplink transmission are introduced \cite{10167480,10621117}.

\subsection{Uplink Data Transmission}
First, APs use the local channel estimation to estimate the corresponding uplink data. Next, APs forward the data estimates to the CPU for data detection \cite{9416909,8845768}. Thus, the received signal at the $m$-th AP, $\textbf{y}_m \in \mathbb{C}^{N\times 1}$ is obtained as
\begin{equation}
\begin{array}{ll}
     \displaystyle \textbf{y}_m  \displaystyle=\sqrt{p_u}\sum\nolimits_{k=1}^K\textbf{g}_{mk}\sqrt{\eta_k}s_k+\textbf{g}_{ap,m}\bar{\boldsymbol{\Theta}}_{r}\boldsymbol{\Theta}_{r}\textbf{n}_r+\textbf{g}_{ap,m}\bar{\boldsymbol{\Theta}}_{t}\boldsymbol{\Theta}_{t}\textbf{n}_t+\textbf{w}_{m},
\end{array}\label{uplink_received_signal}
   \end{equation}
   where $p_u$ is the user transmit power. $s_k\sim \mathcal{CN}(0,1)$ is the $k$-th user's transmit signal, $\eta_k $ is the uplink power control coefficient with $\eta_k \leq 1,~\forall k $. $\textbf{n}_\omega\sim\mathcal{CN}(0, A\sigma_\omega^2\textbf{R})$ is the EMI at the reflection area or transmission area, with respect to $\omega\in\{t,r\}$. $\textbf{w}_{m}\sim\mathcal{CN}(0,\sigma^2\textbf{I}_N)$ is the noise at the $m$-th AP.
Then, the $m$-th AP multiplies the uplink beamforming $\textbf{v}_{mk}\in\mathbb{C}^{1\times N}$ with $\textbf{y}_m$ to detect the symbol transmitted by the $k$-th user. Subsequently, all APs pass $\check{s}_{mk}\triangleq \textbf{v}_{mk}\textbf{y}_m$ to the CPU via the fronthaul link \cite{10621117,9665300,8845768,10001167}. The CPU uses weights $a_{mk}$ with $\textbf{a}_{k}=[a_{1k},...,a_{Mk}]^T\in\mathbb{C}^{M\times1}$ to obtain $\hat{s}_{k}=\sum\nolimits_{m=1}^Ma_{mk}^*\check{s}_{mk}$ as \eqref{Quantity_ES} at the top of the next page, where $\text{DS}_{\text{k}}$ is the desired signal, $\text{BU}_{\text{k}}$ is the beamforming gain uncertainty, $\text{UI}_{\text{kk}'}$ is the inter-user interference, $\text{EMI}_{\text{k}}$ and $\text{NS}_{\text{k}}$ are the respective EMI and noise.
\begin{figure*}[!t]
\begin{equation}
\begin{array}{ll}
\displaystyle \hat{s}_{k}&   \displaystyle =\sum\nolimits_{m=1}^Ma_{mk}^*\textbf{v}_{mk}\Bigg{(}\sqrt{p_u}\sum\nolimits_{k'=1}^K\textbf{g}_{mk'}\sqrt{\eta_{k'}}s_{k'}+\textbf{g}_{ap,m}\bar{\boldsymbol{\Theta}}_{r}\boldsymbol{\Theta}_{r}\textbf{n}_r+\textbf{g}_{ap,m}\bar{\boldsymbol{\Theta}}_{t}\boldsymbol{\Theta}_{t}\textbf{n}_t+\textbf{w}_{m}\Bigg{)}\\&=\underbrace {\sqrt{p_u\eta_k}\sum\nolimits_{m=1}^M a_{mk}^*\mathbb{E}\Big{\{}\textbf{v}_{mk}{\textbf{g}}_{mk}\Big{\}}s_k}_\text {$\text{DS}_{\text{k}}$}+\underbrace {\sqrt{p_u\eta_k}\sum\nolimits_{m=1}^M a_{mk}^*\Bigg{(}\textbf{v}_{mk}{\textbf{g}}_{mk}-\mathbb{E}\Big{\{}\textbf{v}_{mk}{\textbf{g}}_{mk}\Big{\}}\Bigg{)}s_k}_\text {$\text{BU}_{\text{k}}$}\\&+\displaystyle\sum\nolimits_{k' \neq k}^K \underbrace {\sqrt{p_u}\sum\nolimits_{m=1}^M \sqrt{\eta_{k'}}a_{mk}^*\textbf{v}_{mk}{\textbf{g}}_{mk'}s_{k'}}_\text {$\text{UI}_{\text{kk}'}$}+\underbrace {\sum\nolimits_{m=1}^M a_{mk}^*\textbf{v}_{mk}{\textbf{g}}_{ap,m}\bar{\boldsymbol{\Theta}}_{r}\boldsymbol{\Theta}_{r}\textbf{n}_r+\sum\nolimits_{m=1}^M a_{mk}^*\textbf{v}_{mk}{\textbf{g}}_{ap,m}\bar{\boldsymbol{\Theta}}_{t}\boldsymbol{\Theta}_{t}\textbf{n}_t}_\text {$\text{EMI}_{\text{k}}$}+\underbrace {\sum\nolimits_{m=1}^M a_{mk}^*\textbf{v}_{mk}{\textbf{w}}_{m}}_\text {$\text{NS}_{\text{k}}$}.
\end{array}\label{Quantity_ES}
   \vspace{-5 pt}
   \end{equation}
    \vspace{-10 pt}
\hrulefill
\end{figure*} 
\subsection{Performance Analysis and Closed-form SE Derivations}
The use-and-then-forget (UatF) bound \cite{8845768,9416909} is utilized to determine the uplink SE lower bound. Based on \eqref{Quantity_ES} at the top of the next page, the uplink SE of the $k$-th user is expressed as
\begin{equation}
\begin{array}{ll}
     \displaystyle \text{SE}_{\text{u},k}=\frac{\tau_c-\tau_p}{\tau_c}\text{log}_2\Big{(} 1+\text{SINR}_{\text{u},k}\Big{)},
\end{array}\label{uplink_SE}
   \end{equation}
where the effective signal-to-interference-plus-noise ratio (SINR) is denoted by \eqref{uplink_SINR_description_ES} at the top of this page with 
\begin{figure*}[t!]
\begin{equation}
\begin{array}{ll}
     \displaystyle \text{SINR}_{\text{u},k}
     \displaystyle\displaystyle=\frac{\displaystyle p_u\eta_k\Big{|}\textbf{a}_{k}^H\bar{\textbf{b}}_k\Big{|}^2}{\displaystyle\sum\nolimits_{k'=1}^Kp_u\eta_{k'}\textbf{a}_k^H\mathbb{E}\left\{\bar{\boldsymbol{\Omega}}_{kk'}\bar{\boldsymbol{\Omega}}_{kk'}^H\right\}\textbf{a}_k-p_u\eta_k\Big{|}\textbf{a}_{k}^H\bar{\textbf{b}}_k\Big{|}^2+\textbf{a}_{k}^H\mathbb{E}\left\{\bar{\mathbf{\Gamma}}_{t,k}\bar{\mathbf{\Gamma}}_{t,k}^H\right\}\textbf{a}_{k}+\textbf{a}_{k}^H\mathbb{E}\left\{\bar{\mathbf{\Gamma}}_{r,k}\bar{\mathbf{\Gamma}}_{r,k}^H\right\}\textbf{a}_{k}+\sigma^2\textbf{a}_{k}^H\bar{\mathbf{\Lambda}}_{k}\textbf{a}_{k}},
\end{array}\label{uplink_SINR_description_ES}
   \vspace{-5 pt}
   \end{equation}
    \vspace{-10 pt}
\hrulefill
\end{figure*} 
\begin{equation}
\begin{array}{ll}
    \bar{\textbf{b}}_k=\left[\mathbb{E}\left\{(\textbf{v}_{1k}{\textbf{g}}_{1k})^T\right\},...,\mathbb{E}\left\{(\textbf{v}_{Mk}{\textbf{g}}_{Mk})^T\right\}\right]^T\in\mathbb{C}^{M\times 1}.
    \end{array}
\end{equation}
   \begin{equation}
\begin{array}{ll}
\bar{\boldsymbol{\Omega}}_{kk'}=\left[(\textbf{v}_{1k}{\textbf{g}}_{1k'})^T,...,(\textbf{v}_{Mk}{\textbf{g}}_{Mk'})^T\right]^T\in\mathbb{C}^{M\times 1},
    \end{array}
\end{equation}
\begin{equation}
\begin{array}{ll}
\bar{\mathbf{\Gamma}}_{\omega,k}=\left[(\textbf{v}_{1k}{\textbf{g}}_{ap,1}\bar{\boldsymbol{\Theta}}_{\omega}\boldsymbol{\Theta}_{\omega}\textbf{n}_\omega)^T,...,(\textbf{v}_{Mk}{\textbf{g}}_{ap,M}\bar{\boldsymbol{\Theta}}_{\omega}\boldsymbol{\Theta}_{\omega}\textbf{n}_\omega)^T\right]^T\in\mathbb{C}^{M\times 1},%~\omega=t,r,
    \end{array}
\end{equation}
\begin{equation}
\begin{array}{ll}
  \bar{\mathbf{\Lambda}}_{k,mm}=\mathbb{E}\left\{\textbf{v}_{mk}\textbf{v}_{mk}^H\right\},%~\omega=t,r,
    \end{array}
\end{equation}
where $\omega\in\{t,r\}$, $\bar{\mathbf{\Lambda}}_{k}\in\mathbb{C}^{M\times M}$ is a diagnal matrix. The above formulas hold for any uplink beamforming schemes. In this work, we consider local MMSE and MR combining\cite{8845768,10167480}.
\subsubsection{Local MMSE Combining}
The design of local MMSE combining aims to minimize the mean square error and maximize the instantaneous SINR. Following\cite{8845768}, the local MMSE combining can be given by \eqref{MMSE} at the top of this page.
\begin{figure*}
\begin{equation}
\begin{array}{ll}
     \displaystyle \textbf{v}_{mk}=\sqrt{p_u\eta_{k}}\hat{\textbf{g}}_{mk}^H\left[\begin{array}{ll}
\displaystyle\sum\nolimits_{k'=1}^Kp_u\eta_{k'}\left(\hat{\textbf{g}}_{mk'}\hat{\textbf{g}}_{mk'}^H+ \hat{\textbf{Q}}_{mk'}\right)\displaystyle+\sum\nolimits_{\omega=t,r}\sigma^2_{\omega}\left(
\bar{\beta}_{ap,m}\bar{\textbf{g}}_{ap,m}\textbf{T}_{\omega}\bar{\textbf{g}}_{ap,m}^H+\tilde{\beta}_{ap,m}\text{tr}(\textbf{R}_\text{SRIS}\textbf{T}_{\omega})\textbf{R}_{ap,m}

\right)\displaystyle+\sigma^2\textbf{I}_{N}\end{array}\right]^{-1},
\end{array}\label{MMSE}
   \vspace{-5 pt}
   \end{equation}
   \vspace{-16 pt}
   \hrulefill
   \end{figure*}
The Monte Carlo method allows for the computation of achievable SE expressions.
\subsubsection{Maximum Ratio Combining} Since deriving closed-form expressions for the achievable SE with local MMSE combining is difficult due to the analytical intractability of inverting a random matrix, low-complexity MR combining $\textbf{v}_{mk}=\hat{\textbf{g}}_{mk}^H$ \cite{9737367,10201892} can be utilized to derive closed-form analytical results to facilitate system design and performance analysis.

After uplink beamforming, the CPU can optimize the weight vector for the uplink SE maximization. Compared to the matched filter (MF) receiver in \cite{9416909,7827017}, the LSFD receiver introduced in \cite{8845768,10167480} can maximize the $k$-th user's effective SINR for performance improvement. In this case, the corresponding LSFD weight vector $\textbf{a}_{k}\in\mathbb{C}^{M\times 1}$ can be given by
   \begin{equation}
\begin{array}{ll}
\displaystyle\textbf{a}_{k}=\left[\begin{array}{ll}
\displaystyle\sum\nolimits_{k'=1}^Kp_u\eta_{k'}\mathbb{E}\left\{\bar{\boldsymbol{\Omega}}_{kk'}\bar{\boldsymbol{\Omega}}_{kk'}^H\right\}\\\displaystyle+\mathbb{E}\left\{\bar{\mathbf{\Gamma}}_{t,k}\bar{\mathbf{\Gamma}}_{t,k}^H\right\}+\mathbb{E}\left\{\bar{\mathbf{\Gamma}}_{r,k}\bar{\mathbf{\Gamma}}_{r,k}^H\right\}+\sigma^2\bar{\mathbf{\Lambda}}_{k}\end{array}\right]^{-1}\bar{\textbf{b}}_{k}.
\end{array}
\label{LSFD_ES}
   \end{equation}
By adopting MR combining $\textbf{v}_{mk}=\hat{\textbf{g}}_{mk}^H$, we can derive the closed-form expression of SINR as \eqref{SE_CF} at the top of the next page with the assistance of $\textbf{b}_k$ in \eqref{b_k}, $\boldsymbol{\Omega}_{kk'}$ in \eqref{part_1}-\eqref{part_2}, $\boldsymbol{\Upsilon}_{kk'} $ in \eqref{part_3}-\eqref{part_4}, $\mathbf{\Gamma}_{\omega,k}$ in \eqref{EMI_CF} and $\boldsymbol{\Lambda}_k$ in \eqref{lambda}.
\begin{figure*}
\begin{equation}
\begin{array}{ll}
     \displaystyle \text{SINR}_{\text{u},k}\displaystyle=\frac{\displaystyle p_u\eta_k\Big{|}\textbf{a}_{k}^H\textbf{b}_k\Big{|}^2}{\displaystyle\sum\nolimits_{k'\in\mathcal{P}_k}p_u\eta_{k'}\textbf{a}_k^H\boldsymbol{\Omega}_{kk'}\textbf{a}_k+\sum\nolimits_{k'=1}^Kp_u\eta_{k'}\textbf{a}_k^H\boldsymbol{\Upsilon}_{kk'}\textbf{a}_k-p_u\eta_k\Big{|}\textbf{a}_{k}^H\textbf{b}_k\Big{|}^2+\sigma_{t}^2\textbf{a}_{k}^H\mathbf{\Gamma}_{t,k}\textbf{a}_{k}+\sigma_{r}^2\textbf{a}_{k}^H\mathbf{\Gamma}_{r,k}\textbf{a}_{k}+\sigma^2\textbf{a}_{k}^H\mathbf{\Lambda}_{k}\textbf{a}_{k}}.
\end{array}\label{SE_CF}
\vspace{-2 pt}
   \end{equation}
    \vspace{-6 pt}
\hrulefill
\end{figure*} 
The closed-form LSFD weight vector can be given by
 \begin{equation}
\begin{array}{ll}
\displaystyle\textbf{a}_{k}=\Biggl[\begin{array}{ll}\displaystyle\sum\nolimits_{k'\in\mathcal{P}_k}p_u\eta_{k'}\boldsymbol{\Omega}_{kk'}+\sum\nolimits_{k'=1}^Kp_u\eta_{k'}\boldsymbol{\Upsilon}_{kk'}\\\displaystyle+\sigma_{t}^2\mathbf{\Gamma}_{t,k}+\sigma_{r}^2\mathbf{\Gamma}_{r,k}+\sigma^2\mathbf{\Lambda}_{k}\end{array}\Biggr]^{-1}\textbf{b}_{k}.
\end{array}
\label{LSFD_ES}
   \end{equation}
   
\textit{Proof:} Please refer to Appendix B.

\vspace{-10 pt}
\subsection{Uplink Power Control}
Uplink fractional power control is introduced to reduce the near-far effects in the EMI-aware and phase error-aware environment\cite{8968623,10621117}. The $k$-th user's uplink power control coefficient, depending on its corresponding large-scale fading coefficients, is formulated as
\begin{equation}
\begin{array}{ll}
\displaystyle\eta_k=\left(\frac{\text{min}_{k'}\left(\displaystyle\sum\nolimits_{m=1}^M\text{tr}(\mathbf{\Delta}_{mk'})\right)}{\displaystyle\sum\nolimits_{m=1}^M\text{tr}(\mathbf{\Delta}_{mk})}\right)^{\alpha_u}, ~\forall k,
 \end{array}
 \label{uplink_power_control}
   \end{equation}
   where $0\leq {\alpha_u}\leq 1$ is the fractional power control parameter\cite{10167480}.
As a summary of the derivations in this section, note that we can use \eqref{uplink_SE}- \eqref{uplink_power_control} to determine the uplink SE performance.

\section{Downlink Data Transmission and Closed-form SE Expressions}
We derive novel closed-form downlink SE expressions and utilize fractional power control to investigate the downlink performance of the proposed STAR-RIS-assisted cell-free massive MIMO system suffering from EMI and phase errors.

\subsection{Downlink Data Transmission}

The precoding vector $\textbf{f}_{mk}\in\mathbb{C}^{N\times 1},~\forall m,~\forall k$ is utilized to assist the broadcast channel of the downlink data transmission\cite{9875036}.
Since the system is operated under TDD mode, the channel reciprocity characteristic can regard the uplink channel transpose as the downlink channel \cite{10621117,9570816}. Therefore, the transmitted signal by the $m$-th AP can be obtained as
\begin{equation}
\begin{array}{ll}
     \displaystyle \textbf{x}_m=\sqrt{p_d}\sum\nolimits_{k=1}^K \textbf{f}_{mk}\sqrt{\eta_{mk}}q_k,
\end{array}\label{DL_transmitted_signal_ES}
   \end{equation}
where $p_d$ is the downlink transmit power, $q_k\sim \mathcal{CN}(0,1)$ is the signal transmitted to the $k$-th user, $\eta_{mk}$ denotes downlink power control coefficients satisfing $\mathbb{E}\big{\{}|\textbf{x}_m|^2\big{\}}\leq p_d$. $\textbf{f}_{mk}\in\mathcal{C}^{N\times 1}$ is the downlink beamforming vector.
\eqref{downlink_received_signal_ES} at the top of this page expresses the received signal at the $k$-th user where $\textbf{n}_r\sim\mathcal{CN}(0,A\sigma_r^2\textbf{R})$ denotes the EMI from the reflection area, $z_k\sim \mathcal{CN}(0,\sigma^2)$ represents the receiver noise at the $k$-th user.
\begin{figure*}[!t]
\begin{equation}
\begin{array}{ll}
     \displaystyle r_k&   \displaystyle \vspace{2 pt}=\sum\nolimits_{m=1}^{M}\textbf{g}_{mk}^{T}\textbf{x}_m+(\bar{\boldsymbol{\Theta}}_{\omega_k}\boldsymbol{\Theta}_{\omega_k}\textbf{g}_{u,k})^T\textbf{n}_r+z_k=\sum\nolimits_{m=1}^{M}\sum\nolimits_{k'=1}^K\sqrt{p_d} \textbf{g}_{mk}^{T}\textbf{f}_{mk'}\sqrt{\eta_{mk'}}q_{k'}+(\bar{\boldsymbol{\Theta}}_{\omega_k}\boldsymbol{\Theta}_{\omega_k}\textbf{g}_{u,k})^T\textbf{n}_r+z_{k}\\&\displaystyle=\underbrace {\sqrt{p_d}\sum\nolimits_{m=1}^{M}\mathbb{E}\Big{\{}\textbf{g}_{mk}^{T}\textbf{f}_{mk}\Big{\}}\sqrt{\eta_{mk}}q_k}_{\text{DS}_k}+\underbrace {\sqrt{p_d}\sum\nolimits_{m=1}^{M}\left(\textbf{g}_{mk}^{T}\textbf{f}_{mk}-\mathbb{E}\Big{\{}\textbf{g}_{mk}^{T}\textbf{f}_{mk}\Big{\}}\right)\sqrt{\eta_{mk}}q_k}_{\text{BU}_k}\\&\displaystyle+\displaystyle\sum\nolimits_{k'\neq k}^K\underbrace {\sqrt{\rho_d} \sum\nolimits_{m=1}^{M}\textbf{g}_{mk}^{T}\textbf{f}_{mk'}\sqrt{\eta_{mk'}}q_{k'}}_{\text{UI}_{kk'}}+\underbrace{(\bar{\boldsymbol{\Theta}}_{\omega_k}\boldsymbol{\Theta}_{\omega_k}\textbf{g}_{u,k})^T\textbf{n}_r}_{\text{EMI}_k}+\underbrace {z_{k}}_{\text{NS}_k},
\end{array}\label{downlink_received_signal_ES}
\vspace{-5 pt}
   \end{equation}
    \vspace{-8 pt}
\hrulefill
\end{figure*}

\subsection{Performance Analysis and Closed-form SE Derivations}
Based on \eqref{downlink_received_signal_ES}, we can lower bound the downlink SE of the $k$-th user and obtain it by \cite{10225319,9416909}
\begin{equation}
\begin{array}{ll}
     \displaystyle \text{SE}_{\text{d},k}=\frac{\tau_c-\tau_p}{\tau_c}\text{log}_2\Big{(} 1+\text{SINR}_{\text{d},k}\Big{)},
\end{array}\label{downlink_SE_description}
   \end{equation}
   where $\text{SINR}_{\text{d},k}$ is given by \eqref{downlink_SINR} at the top of this page. 
   \begin{figure*}[t!]
\begin{equation}
\begin{array}{ll}
\displaystyle \text{SINR}_{\text{d},k}&\displaystyle=\frac{\displaystyle{p_d}\Big{|}\sum\nolimits_{m=1}^{M}\sqrt{\eta_{mk}}\mathbb{E}\bigg{\{}\textbf{g}_{mk}^{T}\textbf{f}_{mk}\bigg{\}}\Big{|}^2}{\displaystyle{p_d}\sum\nolimits_{k'=1}^K\mathbb{E}\bigg{\{}\Big{|}\sum\nolimits_{m=1}^{M}\sqrt{\eta_{mk'}}\textbf{g}_{mk}^{T}\textbf{f}_{mk'}\Big{|}^2\bigg{\}}-{p_d}\Big{|}\sum\nolimits_{m=1}^{M}\sqrt{\eta_{mk}}\mathbb{E}\bigg{\{}\textbf{g}_{mk}^{T}\textbf{f}_{mk}\bigg{\}}\Big{|}^2+\mathbb{E}\Big{\{}\Big{|}(\bar{\boldsymbol{\Theta}}_{\omega_k}\boldsymbol{\Theta}_{\omega_k}\textbf{g}_{u,k})^T\textbf{n}_r\Big{|}^2\Big{\}}+\sigma^2}.
\end{array}\label{downlink_SINR}
\vspace{-2 pt}
   \end{equation}
    \vspace{-8 pt}
   \hrulefill
   \end{figure*}
Then, we utilize the conjugate beamforming $\textbf{f}_{mk}=\hat{\textbf{g}}_{mk}^*$ \cite{9737367,10201892} to derive closed-form analytical results, thereby preserving the integrity of our conclusions. Meanwhile, RZF $\textbf{f}_{mk}=\left(\sum\nolimits_{k'=1}^K\hat{\textbf{g}}_{mk'}\hat{\textbf{g}}_{mk'}^H+\epsilon_{mk}\textbf{I}_{N}\right)\hat{\textbf{g}}_{mk}^*$ is adopted to indicate the benefits of efficient downlink beamforming, utilizing the Monte Carlo method to compute achievable SE expressions \cite{9841466}. Due to the limited length, the design of efficient downlink beamforming is left for our future work. Subsequently, with the assistance of $\textbf{c}_k=[\sqrt{\eta_{1k}},\sqrt{\eta_{2k}},...,\sqrt{\eta_{Mk}}]^T\in\mathbb{C}^{M\times 1},~\forall k$, \eqref{downlink_SINR_ES} at the top of this page shows the closed-form expression of $\text{SINR}_{\text{d},k}$ with the assistance of $\textbf{b}_k$ in \eqref{b_k}, $\boldsymbol{\Omega}_{k'k}$ in \eqref{part_1}-\eqref{part_2} and $\boldsymbol{\Upsilon}_{k'k} $ in \eqref{part_3}-\eqref{part_4}.
\begin{figure*}[t!]
\begin{equation}
\begin{array}{ll}
\displaystyle \text{SINR}_{\text{d},k}^{}\displaystyle=%\frac{\displaystyle{p_d}\Big{|}\sum\nolimits_{m=1}^{M}\sqrt{\eta_{mk}}\mathbb{E}\bigg{\{}\textbf{g}_{mk}^{T}\textbf{f}_{mk}\bigg{\}}\Big{|}^2}{\displaystyle{p_d}\sum\nolimits_{k'=1}^K\mathbb{E}\bigg{\{}\Big{|}\sum\nolimits_{m=1}^{M}\sqrt{\eta_{mk'}}\textbf{g}_{mk}^{T}\textbf{f}_{mk'}\Big{|}^2\bigg{\}}-{p_d}\Big{|}\sum\nolimits_{m=1}^{M}\sqrt{\eta_{mk}}\mathbb{E}\bigg{\{}\textbf{g}_{mk}^{T}\textbf{f}_{mk}\bigg{\}}\Big{|}^2+\mathbb{E}\Big{\{}\Big{|}(\bar{\boldsymbol{\Theta}}_{\omega_k}\boldsymbol{\Theta}_{\omega_k}\textbf{g}_{k})^T\textbf{n}_r\Big{|}^2\Big{\}}+\sigma^2}
\displaystyle\frac{\displaystyle p_d\Big{|}\textbf{c}_{k}^H\textbf{b}_k\Big{|}^2}{\displaystyle\sum\nolimits_{k'\in\mathcal{P}_k}p_d\textbf{c}_{k'}^H\boldsymbol{\Omega}_{k'k}\textbf{c}_{k'}+\sum\nolimits_{k'=1}^Kp_d\textbf{c}_{k'}^H\boldsymbol{\Upsilon}_{k'k}\textbf{c}_{k'}-p_d\Big{|}\textbf{c}_{k}^H\textbf{b}_k\Big{|}^2+\sigma_r^2\left(\bar{\beta}_{u,k}\bar{\textbf{g}}_{u,k}^H\textbf{T}_{\omega_k}\bar{\textbf{g}}_{u,k}+\tilde{\beta}_{u,k}\text{tr}(\textbf{R}_\text{SRIS}\textbf{T}_{\omega_k})\right)+\sigma^2}.
\end{array}\label{downlink_SINR_ES}
\vspace{-2 pt}
   \end{equation}
    \vspace{-8 pt}
   \hrulefill
   \end{figure*}
  Therefore, the closed-form EMI term is expressed as
\begin{equation}
\begin{array}{ll}
\displaystyle\mathbb{E}\Big{\{}\Big{|}(\bar{\boldsymbol{\Theta}}_{\omega_k}\boldsymbol{\Theta}_{\omega_k}\textbf{g}_{u,k})^T\textbf{n}_r\Big{|}^2\Big{\}}\displaystyle=\mathbb{E}\Big{\{}(\bar{\boldsymbol{\Theta}}_{\omega_k}\boldsymbol{\Theta}_{\omega_k}\textbf{g}_{u,k})^T\textbf{n}_r\textbf{n}_r^H(\bar{\boldsymbol{\Theta}}_{\omega_k}\boldsymbol{\Theta}_{\omega_k}\textbf{g}_{u,k})^*\Big{\}}\\ ~~~~~~~~~~~~~~~~~~~~~~~~\displaystyle=\sigma_r^2\left(\bar{\beta}_{u,k}\bar{\textbf{g}}_{u,k}^H\textbf{T}_{\omega_k}\bar{\textbf{g}}_{u,k}+\tilde{\beta}_{u,k}\text{tr}(\textbf{R}_\text{SRIS}\textbf{T}_{\omega_k})\right).
 \end{array}
 \label{EMI_downlink}
   \end{equation}  
   We omit the proof procedure since the proof is similar to Appendix B.  
\subsection{Downlink Power Control}
According to \cite{8968623,9875036,10621117}, we introduce the fractional power control to meet the downlink power constraint. This work assumes that $\mathbb{E}\big{\{}|\textbf{x}_m|^2\big{\}}\leq p_d$, and the downlink power control coefficients can be obtained as
\begin{equation}
\begin{array}{ll}
     \displaystyle \eta_{mk}=\left({\left(\sum\nolimits_{m'=1}^M\text{tr}(\mathbf{\Delta}_{m'k})\right)^{\alpha_d}\sum\nolimits_{k'=1}^K\frac{\text{tr}(\textbf{Q}_{mk'})}{\left(\sum\nolimits_{m'=1}^M\text{tr}(\mathbf{\Delta}_{m'k'})\right)^{\alpha_d}}}\right)^{-1},~\forall m,~k,
\end{array}\label{downlink_power_control}
   \end{equation}
where $0 \leq \alpha_d\leq 1$ is the fractional power control parameter. %\alpha=0 for our case
\\
As a summary of the derivations in this section, we can utilize \eqref{downlink_SE_description}-\eqref{downlink_power_control} to determine the downlink SE performance.

\section{Numerical Results and Discussions}
This section presents numerical results that demonstrate the impact of EMI and phase errors on the STAR-RIS-assisted cell-free massive MIMO system. These results provide performance limits and system design guidelines. We also introduce the innovative use of the projected GD algorithm, which optimizes the STAR-RIS coefficient matrix to improve system performance. Particularly, \eqref{Num_direct} and \eqref{aggregate_uplink_channel_estimation} allow us to determine the channel estimation NMSE by applying the projected GD algorithm with \eqref{Optimization_1}-\eqref{Optimization_2} to optimize \eqref{GD_optimization}. \eqref{uplink_SE}- \eqref{uplink_power_control} determine the uplink SE performance and \eqref{downlink_SE_description}-\eqref{downlink_power_control} determine the downlink SE performance.
The numerical results include Monte Carlo (MC) simulations and closed-form analytical results. The analytical results quantify a mathematical approximation of the average performance over the channel realizations to indicate how design variables affect performance visually.

\subsection{Parameter Setup}

In the results, a two-dimensional coordinate system similar to \cite{10841966,10297571} is utilized, and all coordinates are in meters. STAR-RIS is located at $(x^{\text{STAR-RIS}},y^{\text{STAR-RIS}})=(500,100)$, APs are randomly distributed within the region with $x^{\text{AP}}\in\left[-100,800\right]$, $y^{\text{AP}}\in\left[-100,100\right]$. Users in the reflection area are located with $x^{\text{user}}\in\left[300,600\right]$
and $y^{\text{user}}\in\left[0,100\right)$, while users in the transmission area are distributed with $x^{\text{user}}\in\left[300,600\right]$
and $y^{\text{user}}\in(100,200]$. The AP, user and RIS heights are 15m, 1.65m, and 30m \cite{10225319}. We assume that all APs employ a uniform linear array, where the inter-antenna distance is $d_{ap}=\lambda/2$. The spatial correlation
matrix at the $m$-th AP is constructed as the exponential correlation model adopted in \cite{951380,10621117}. Meanwhile,
adopting the normalized wave vector pointing to the direction of interest $[\text{cos}\theta\text{sin}\phi, \text{cos}\phi, \text{sin}\theta\text{sin}\phi]$, the
steering vectors, $\textbf{a}_{ap}\left(
   \theta,  \phi
   \right)\in\mathbb{C}^{N\times 1}$ related to APs, and $\textbf{a}_\text{SRIS}\left(
   \theta,  \phi
   \right)\in\mathbb{C}^{L\times 1}$ related to the STAR-RIS, are given by \cite{9570816,9099874}
 \begin{equation}
\begin{array}{ll}
   \displaystyle \textbf{a}_{ap}\left(
   \theta,  \phi\right)&\displaystyle=\left[1,~e^{j2\pi \frac{d_{ap}}{\lambda}\text{cos}\theta\text{sin}\phi}~...~e^{j2\pi(N-1) \frac{d_{ap}}{\lambda}\text{cos}\theta\text{sin}\phi}\right]^T,
   \end{array}
   \label{Rx_steering}
  \end{equation}
 \begin{equation}
\begin{array}{ll}
   \displaystyle \textbf{a}_\text{SRIS}\left(
   \theta,  \phi\right)&\displaystyle=\left[1,~e^{j2\pi \frac{d_h}{\lambda}\text{cos}\theta\text{sin}\phi}~...~e^{j2\pi\frac{(L_h-1) d_h}{\lambda}\text{cos}\theta\text{sin}\phi}\right]^T\\&\displaystyle \otimes \left[1,~e^{j 2\pi \frac{d_v}{ \lambda}\text{cos}\phi}~...~e^{j 2\pi\frac{(L_v-1)d_v}{\lambda}\text{cos}\phi}\right]^T.
   \end{array}
   \label{Tx_steering}
  \end{equation}
  Then, we can model LOS components as
   \begin{equation}
\begin{array}{ll}
\bar{\textbf{g}}_{ap,m}=\textbf{a}_{ap}\left(
   \theta_\text{AoA}^m,  \phi_\text{AoA}^m\right)\textbf{a}_\text{SRIS}\left(
   \theta_\text{AoD}^m,  \phi_\text{AoD}^m\right)^H,
   \end{array}
  \end{equation}
     \begin{equation}
\begin{array}{ll}
\bar{\textbf{g}}_{u,k}=\textbf{a}_\text{SRIS}\left(
   \theta_\text{AoA}^k,  \phi_\text{AoA}^k\right),
   \end{array}
  \end{equation}
where $\theta_\text{AoA}^m,~\phi_\text{AoA}^m,~\theta_\text{AoD}^m,~\phi_\text{AoD}^m$ are the relevant azimuth and elevation angle-of-arrival (AoA) and angle-of-departure (AoD) between the $m$-th AP and the STAR-RIS, respectively. Similarly, $\theta_\text{AoA}^k, ~ \phi_\text{AoA}^k$ are the relevant azimuth and elevation AoA between the STAR-RIS and the $k$-th user\cite{9570816,9099874}. For ease of simplicity, we assume that $\kappa_{u,k}=\kappa_{ap,m}=10,~\forall m,k$, making the LOS components dominant\cite{10167480}.
We apply the path loss model and the relevant settings in \cite{7827017,10621117} to express the large-scale fading coefficients as $\beta_x=\text{PL}_x\cdot z_x$, $x=(d,mk),~(ap,m),~(u,k)$. $\text{PL}_x$ is the three-slope path loss, $z_x$ is the log-normal shadowing. 
Unless mentioned, $p_{p}=p_u=20~\text{dBm}$, $p_{d}=23~\text{dBm}$, $\sigma^2=-91$ dBm, and $\alpha_u=\alpha_d=0$. $d_h=d_v=\lambda/2$ for the STAR-RIS elements. Moreover, 
$f_c=1.9$ GHz is the carrier frequency, and the coherence interval length is $\tau_c=200$, $\tau_p=3$ is applied for channel estimation\cite{10167480}. For the projected GD algorithm, we set the maximum number of iterations IterMax$=200$, the initial step size $\mu_{\boldsymbol{\theta}},~\mu_\textbf{u}=30$, the penalty parameter $\varrho=0.2$ \cite{10130156} and tolerance $\epsilon=10^{-4}$ until stated. 

\begin{figure}[!t]
\centering
\includegraphics[width=0.76\columnwidth]{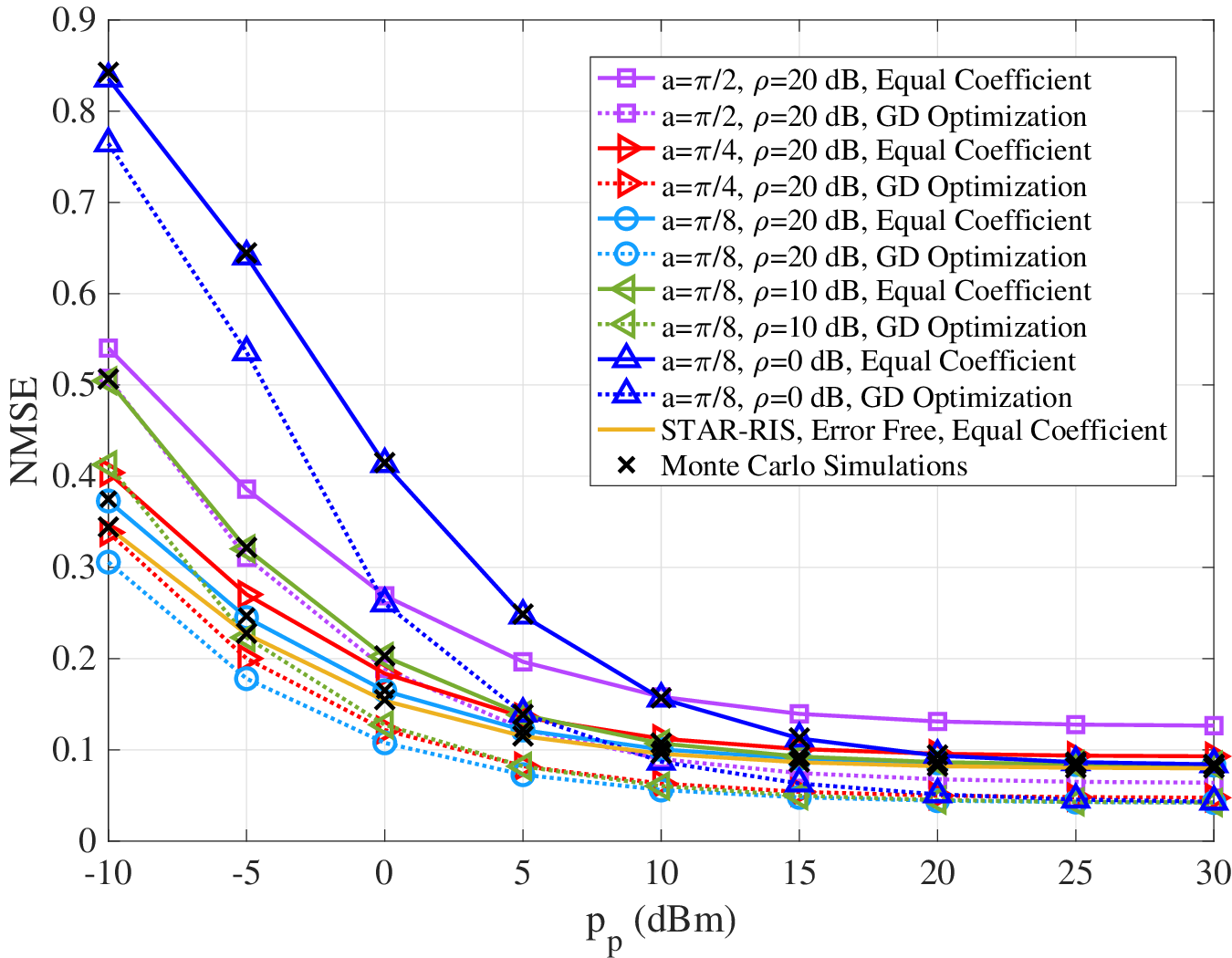}
\caption{ NMSE versus the pilot transmit power $p_p$ operating at different EMI powers and ranges of phase errors with $M=10$, $N=4$, $K=6$, $K_t=K_r=3$, $L=16$ (MC Simulations and Analytical Results).}
\label{fig_2}
\vspace{-11 pt}
\end{figure}

\subsection{Channel Estimation Accuracy}

Channel estimation NMSE versus $p_p$ is displayed in Fig. \ref{fig_2}. For comparison, the equal STAR-RIS coefficient matrix follows $u_l^\omega=1/\sqrt{2},~\varphi_l^r=\pi/4,~\varphi_l^t=\varphi_l^r+\pi/2,~\forall l$. The conventional MMSE scheme with an equal STAR-RIS coefficient matrix without EMI or phase errors, i.e., error-free, serves as the baseline scenario\cite{10167480,9875036}. The closed-form analytical results computed by \eqref{Delta}-\eqref{T_matrix_phase_mean} and \eqref{Num_direct}- \eqref{aggregate_uplink_channel_estimation} closely match the performance obtained from MC simulations. The joint effect of EMI and phase errors can significantly reduce channel estimation accuracy. To be more specific, increasing the range of phase errors from $a=\pi/8$ to $a=\pi/2$ with $\bar{\varphi}_{l}^{\omega_k}\in[-a,~a], ~\forall l$, can greatly impair the NMSE performance by over $50\%$-likely NMSE increase. Meanwhile, decreasing the ratio of the received signal and the EMI power from $\rho=20$ dB to $\rho=0$ dB can inflate EMI power levels and impair the NMSE performance by over $100\%$-likely NMSE increase when $p_p<5$ dBm. It reveals that these impairments can be compensated for by a larger $p_p$, but EMI dominates at low $p_p$. Moreover, a non-zero error floor remains due to pilot contamination occurring at $\tau_p=3<K$, even as $p_p$ is increased. 
Our projected GD algorithm indicates that a larger $p_p$ can lead to a greater decrease in NMSE. For example, -5 dBm introduces a $30\%$-likely NMSE decrease, and this trend increases to $100\%$-likely when $p_p$ increases to 30 dBm, resulting in significant gains in estimation accuracy. Our simulations also reveal that the convergence speed of the proposed projected GD algorithm increases with larger initial step sizes and penalty parameters. When $p_p=20$ dBm, with penalty parameter $\varrho=1$ and IterMax=1000, the algorithm with $\mu=20$ requires approximately 200 iterations, and with $\mu=30$ takes approximately 100 iterations to converge, that is, $\mu=20$ requires two times of running time of $\mu=30$, while the steady-state NMSE is unaffected. Other random initializations can also affect the convergence speed, with the above results agreeing with \cite{10130156}. The insights suggest that EMI and phase errors markedly impair the NMSE performance. Therefore, analyzing EMI and phase errors is necessary to delineate system constraints and provide design guidelines for real-world implementations. Advancing low-complexity optimization algorithms will be pursued as a future topic to mitigate these negative effects and maintain the feasibility of a fast-growing number of users in real-world scenarios.

\begin{figure}[!t]
\centering
\includegraphics[width=0.76\columnwidth]{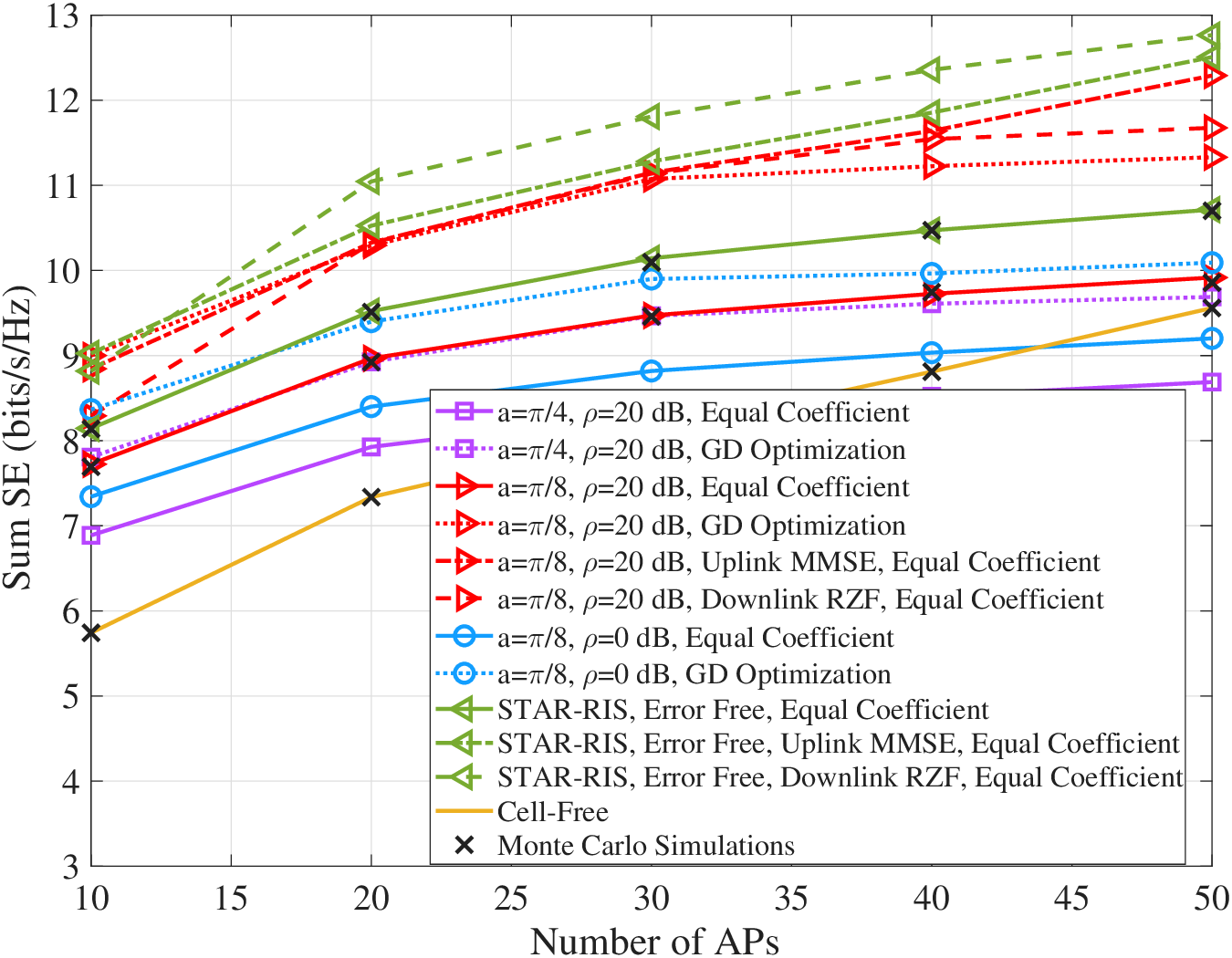}
\caption{Sum SE versus the number of APs $M$ operating at different uplink beamforming schemes with LSFD and $N=4$, $K=6$, $K_t=K_r=3$, $L=16$ (MC Simulations and Analytical Results).}
\label{fig_3}
\vspace{-11 pt}
\end{figure}
\begin{figure}[!t]
\centering
\includegraphics[width=0.76\columnwidth]{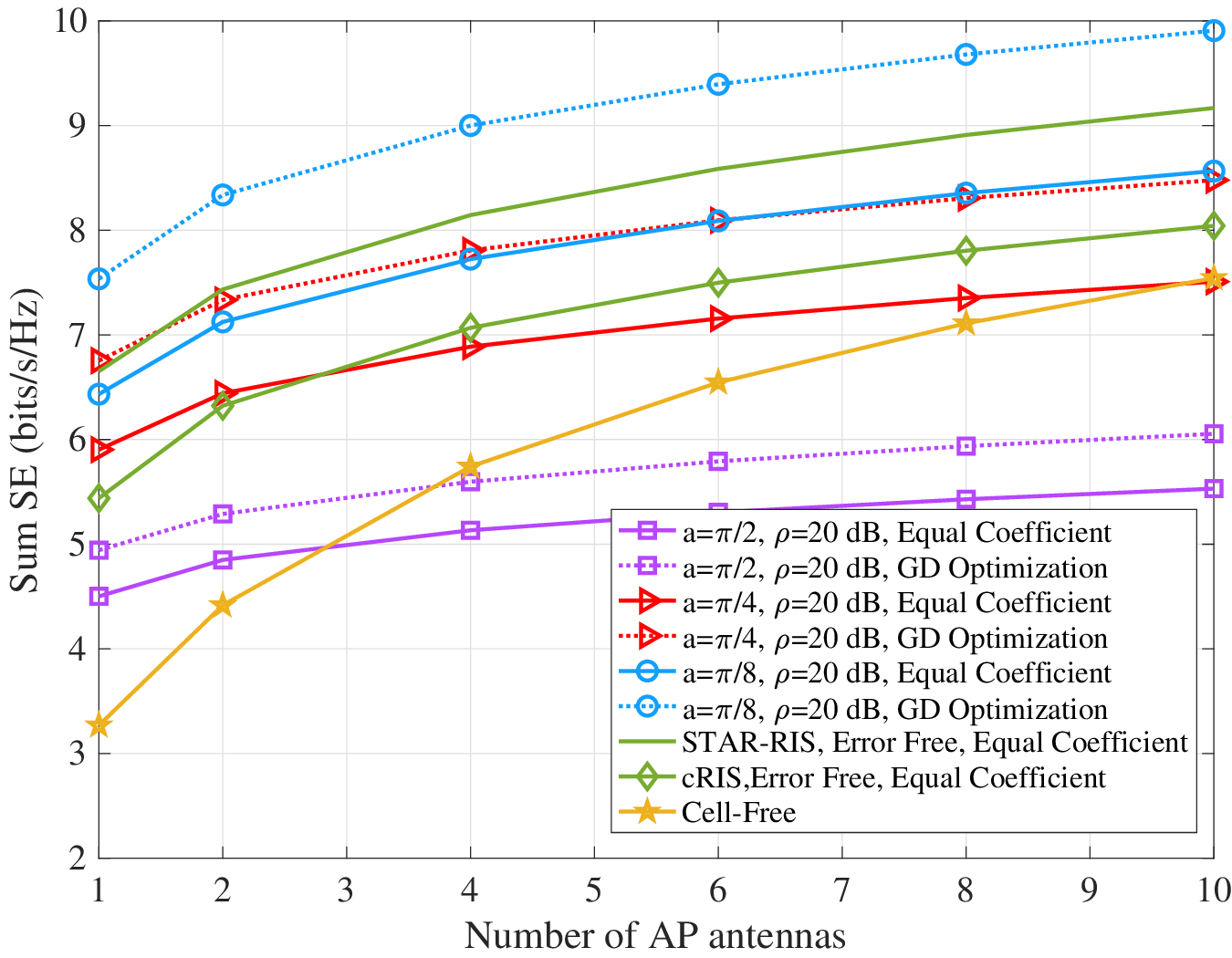}
\caption{Sum SE versus the number of antennas per AP $N$ operating at different ranges of phase errors with LSFD and $M=10$, $K=6$, $K_t=K_r=3$, $L=16$ (Analytical Results).}
\label{fig_4}
\vspace{-11 pt}
\end{figure}

\subsection{Effects of the Number of APs and Antennas Per AP}

Fig. \ref{fig_3} depicts the sum SE as a function of the number of APs, with the sum SE formulated as $\text{SE}_{\text{sum}}=\displaystyle\frac{1}{2}\sum\nolimits_{k=1}^K\left(\text{SE}_{\text{u},k}+\text{SE}_{\text{d},k}\right)$.
The closed-form analytical results utilizing MR combining derived from \eqref{uplink_SE}- \eqref{uplink_power_control} and \eqref{downlink_SE_description}-\eqref{downlink_power_control} can match the numerical results obtained from MC simulations closely. For clarity, only closed-form analytical results are shown in the subsequent figures. Note that MC simulations generate results with local MMSE combining and RZF precoding, respectively. Meanwhile, for the RZF precoding, we take $\epsilon_{mk}=\sigma_r^2\left(\bar{\beta}_{u,k}\bar{\textbf{g}}_{u,k}^H\textbf{T}_{\omega_k}\bar{\textbf{g}}_{u,k}+\tilde{\beta}_{u,k}\text{tr}(\textbf{R}_\text{SRIS}\textbf{T}_{\omega_k})\right)+\sigma^2,~\forall m,k$. It shows that increasing APs can enhance the sum SE. Consistent with \cite{10001167,9841466}, the sum SE with RZF preocding outperforms that with conjugate beamforming by around $16\%$-likey SE improvement. Meanwhile, the sum SE with local MMSE combining surpasses that with MR combining by around $11\%$-likey SE improvement with error-free STAR-RIS and $16\%$-likey SE improvement with error-impaired STAR-RIS. This emphasizes the need for efficient, low-complexity beamforming designs, to enhance performance under harsh propagation conditions while ensuring feasibility. Note that the harsh propagation conditions bring up the limited performance gain, e.g., $a=\pi/4$ and $\rho=0$ dB introduce $12\%$-likely and $6\%$-likely SE decrease than $a=\pi/8$ and $\rho=20$ dB with equal coefficient matrix when $M=10$, respectively. This indicates that the STAR-RIS-assisted performance is more susceptible to phase errors. Moreover, the error-free STAR-RIS-assisted cell-free massive MIMO systems can provide at least $6\%$-likely SE improvement compared to STAR-RIS-assisted cell-free massive MIMO with EMI and phase errors, applying MR combining. This limitation can be compensated by our proposed projected GD algorithm, which can introduce an around $12\%$ to $15\%$-likely SE improvement than that with the equal coefficient matrix. A $5\%$ to $10\%$-likely SE improvement is introduced compared to error-free STAR-RIS when $\rho=20$ dB and $a=\pi/8$, with growth trends decreasing as the number of APs rises. Meanwhile, error-free STAR-RIS-assisted cell-free massive MIMO can provide a nearly $40\%$-likely SE improvement compared to RIS-free cell-free when $M=10$, highlighting the advantages of STAR-RIS deployment. However, this growth trend decreases as $M$ increases.
Therefore, the number of APs should be carefully scaled to meet the required system performance. Future investigations should prioritize effective beamforming design and feasible resource allocation design to fully release the benefits of introducing APs under harsh propagation conditions.

Fig. \ref{fig_4} displays the sum SE as a function of the number of antennas per AP. The results reveal that increasing antennas per AP can enhance the sum SE. However, the growth of performance gains wanes as the number of AP antennas grows since the introduction of increased spatial freedom cannot make up for the increasing inter-user interference, EMI and network overhead. To further illustrate the benefits of STAR-RIS deployment, we present a benchmark scenario with a conventional transmitting-only RIS and a reflecting-only RIS positioned next to each other at the exact STAR-RIS location. For fairness, each conventional RIS (cRIS) is assumed to have $L/2$
elements \cite{9570143}. Given $N=4$, the error-free STAR-RIS-assisted cell-free massive MIMO systems can yield over $15\%$ and $40\%$ sum SE improvement compared to cRIS-assisted and RIS-free cell-free massive MIMO systems, respectively. This underscores the necessity of incorporating STAR-RISs into cell-free massive MIMO systems. Analogous to Fig. \ref{fig_2}, the range of phase errors increases from $a=\pi/8$ to $a=\pi/2$ can introduce around $50\%$-likely SE decrease. When phase errors are moderate, i.e., $a=\pi/8$, the proposed system with the projected GD algorithm can outperform the error-free STAR-RIS-assisted scenarios by around $10\%$-likely SE improvement, emphasizing the importance of efficient STAR-RIS coefficient design for performance enhancement. Moreover, our proposed projected GD algorithm achieves a larger $15\%$-likely SE improvement when $a=\pi/8$ and a larger $9\%$-likely SE improvement when $a=\pi/2$ compared to the equal coefficient matrix, indicating that the advantages of the projected GD algorithm cannot be fully released under harsh propagation conditions. Thus, the number of antennas per AP should be appropriately raised, accompanied by an effective optimization design to further boost system performance.

\begin{figure}[!t]
\centering
\includegraphics[width=0.76\columnwidth]{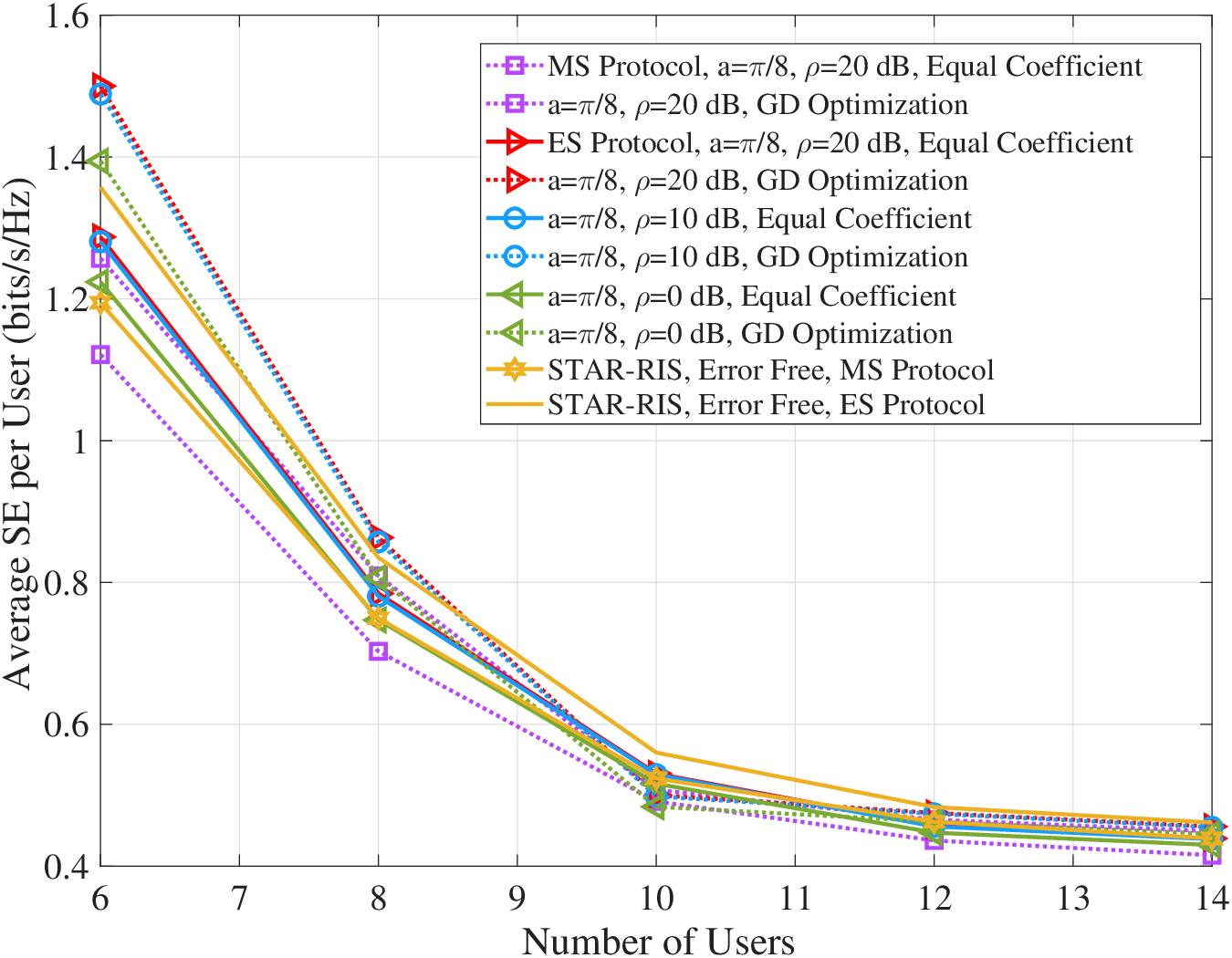}
\caption{Average SE per user versus the number of users $K$ operating at different EMI powers with LSFD and $M=10$, $N=4$, $K_t=K_r$, $L=16$ (Analytical Results).}
\label{fig_5}
\vspace{-11 pt}
\end{figure}

\subsection{Effects of the Number of users}
The average SE per user, defined by ${\text{SE}_\text{ave}}=\displaystyle{\text{SE}_\text{sum}}/K$, is displayed in Fig. \ref{fig_5} as a function of the number of users. The results reveal that an increase in users adversely affects average performance since more users elevate inter-user interference and introduce more severe pilot contamination. In parallel with \cite{10297571}, the ES protocol can exceed the MS protocol by around $15\%$-likely average SE when $K=6$, while this trend decreases to $5\%$-likely average SE when $K=16$, which shows the potential applications of the MS protocol with a larger number of users.
Similar to Fig. \ref{fig_3}, decreasing $\rho$ from $20$ dB to $0$ dB can yield an over $5\%$-likely average SE decrease when $K=8$. Moreover, the proposed projected GD algorithm can help to improve the average SE per user by at least $10\%$-likely average SE improvement compared to the equal coefficient matrix. The proposed system performs similarly when $\rho=20$ dB and $\rho=10$ dB since EMI does not introduce dominant negative effects with given configurations to introduce a significant performance gap. Moreover, when EMI powers are moderate, i.e., $\rho=20$ dB or $10$ dB, the proposed system
utilizing the projected GD algorithm can introduce more than $10\%$-likely SE improvement compared to the error-free scenario when $K=6$. However, this benefit is somewhat constrained, as the error-free STAR-RIS performs better with increasing $K$. This is because a larger $K$ heightens EMI powers to impair the propagation conditions and limit the effectiveness of the proposed GD algorithm compared to the error-free STAR-RIS scenario. Therefore, introducing feasible and efficient STAR-RIS coefficient design and resource allocation schemes to boost performance under harsh propagation conditions is critical. This leads to scalability concerns, as numerous users can introduce high computational complexity and processing overhead to make the implementation infeasible in practice\cite{10556753,10422885}. Therefore, it is essential to adopt advanced signal-processing methods to eliminate pilot contamination and inter-user interference. Moreover, a future topic might be introducing STAR-RISs in user-centric cell-free massive MIMO where fewer APs serve each user to achieve low-complexity computational feasibility to address scalability challenges.

\begin{figure}[!t]
\centering
\includegraphics[width=0.76\columnwidth]{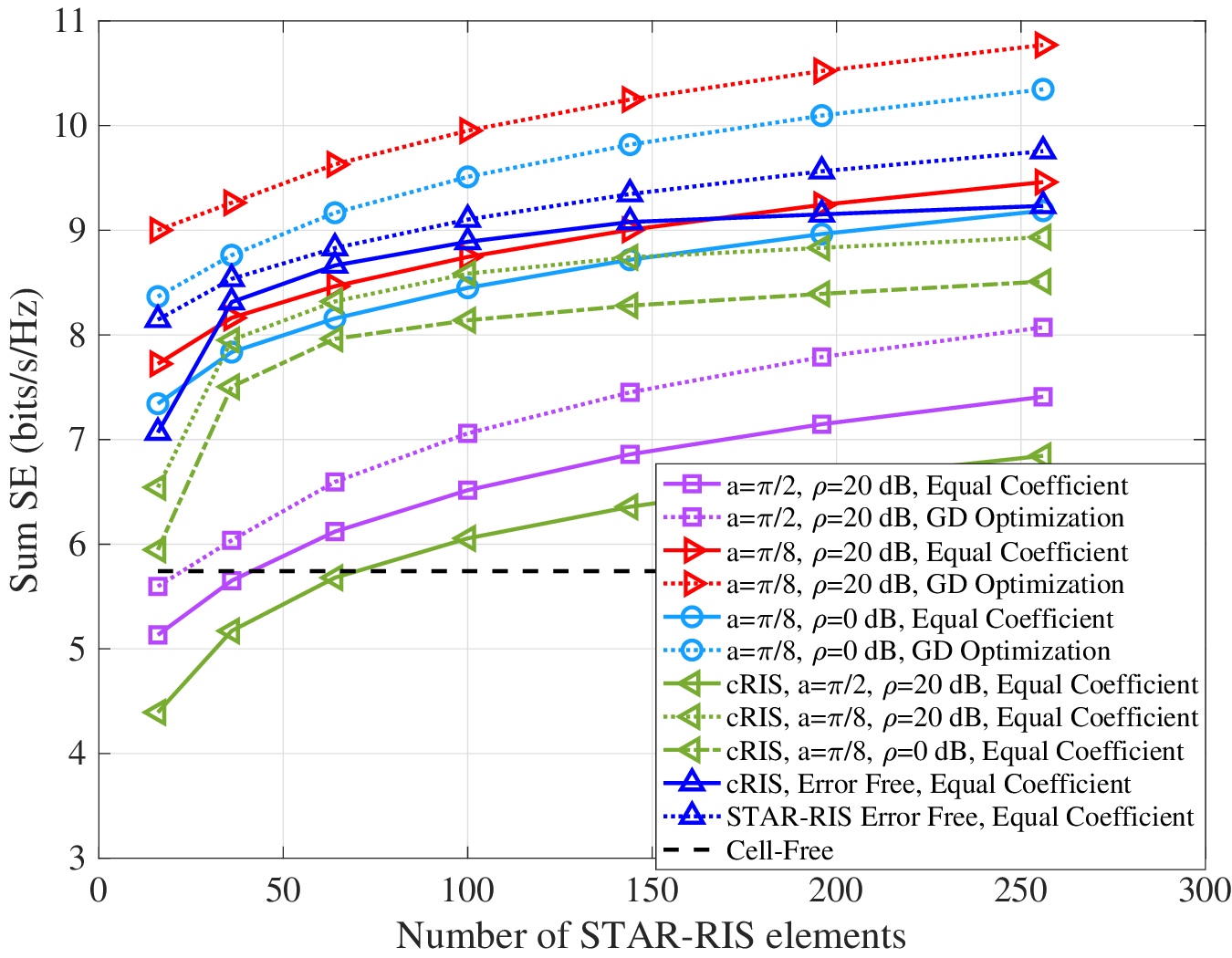}
\caption{Sum SE versus the number of STAR-RIS elements $L$ operating at different EMI powers and ranges of phase errors with LSFD and $M=10$, $N=4$, $K=6$, $K_t=K_r=3$ (Analytical Results).}
\label{fig_6}
\vspace{-11 pt}
\end{figure}

\subsection{Effects of the Number of STAR-RIS Elements}
Fig. \ref{fig_6} indicates the sum SE as a function of the number of STAR-RIS elements. It is evident that increasing the number of STAR-RIS elements could markedly enhance system performance.
Notably, with the same system configurations, the STAR-RIS can outperform cRIS by around $15\%$-likely SE improvement when $L=16$. However, when $50<L<100$, the performance of STAR-RIS and cRIS is comparable. When $L>100$, STAR-RIS increasingly outperforms cRIS, achieving a nearly $6\%$-likely SE improvement when $L=256$. Meanwhile, when $a=\pi/8$, $\rho=0$ dB introduces an extra $2\%$-likely SE decrease for cRIS than STAR-RIS, compared to $\rho=20$ dB. Similarly, when $\rho=20$ dB, $a=\pi/2$ introduces an extra $7\%$-likely SE decrease for cRIS than STAR-RIS, compared to 
$a=\pi/8$.
Therefore, cRIS is more sensitive to severe EMI and phase errors than STAR-RIS under the proposed scenarios, experiencing higher performance degradation accordingly. These highlight the benefits of introducing STAR-RIS in the severe EMI-aware and phase error-aware environment. 
Additionally, the STAR-RIS-assisted cell-free massive MIMO offers at least a $20\%$-likely SE improvement over the RIS-free cell-free massive MIMO, motivating the deployment of STAR-RISs in real-world applications. Meanwhile, when EMI and phase errors are severe ($a=\pi/2$, $\rho=0$ dB), the application of STAR-RISs with the equal coefficient matrix when $L<50$ will underperform conventional cell-free massive MIMO systems.
Our projected GD algorithm can help to mitigate this performance degradation and introduce a more than $13\%$-likely SE improvement compared to the equal coefficient matrix. Although severe EMI and phase errors can mitigate the benefits of introducing an increasing number of STAR-RIS elements, the projected GD algorithm consistently enhances performance under various EMI powers ($a=\pi/8$), surpassing the error-free performance with the equal coefficient matrix by at least $6\%$-likely SE improvement when $L=256$.
Thus, to fully exploit the benefits of STAR-RIS deployment in real-world applications, efficient coefficient matrix design and other advanced techniques are crucial to address practical impairments, including EMI elimination and phase error compensation methods.

\section{Conclusion}

STAR-RIS-assisted cell-free massive MIMO represents a sustainable and attractive architecture that leverages their combined advantages. To our knowledge, this work is the first to analyze the performance of spatially correlated STAR-RIS-assisted cell-free massive MIMO systems, considering the impacts of EMI and phase errors while conducting SE performance analysis. We introduced the projected GD algorithm for designing the STAR-RIS coefficient matrix to minimize the NMSE. We derived the closed-form expressions of the uplink and downlink SE with respective uplink LSFD processing and downlink conjugate beamforming for performance evaluation. The results reveal that EMI and phase errors cause non-negligible channel estimation accuracy and sum SE degradation. Our proposed projected GD algorithm can introduce nearly $ 30\%$-likely estimation accuracy improvement and over $ 10\%$-likely SE improvement. Increasing the number of APs, antennas per AP and STAR-RIS elements can reduce performance degradation. More users might introduce severe inter-user interference and pilot contamination to reduce the average SE per user. Severe EMI and phase errors could mitigate the advantages of increasing STAR-RIS elements, which the projected GD algorithm can compensate for. 
Moreover, compared to conventional RISs, STAR-RIS achieves a larger $15\%$-likely SE improvement and exhibits a smaller performance degradation in highly impaired environments, underscoring the necessity of deploying STAR-RISs in cell-free massive MIMO with proper system configurations. Inspired by these results, our future work will focus on scalable implementation and realistic scenarios, such as time-varying channels and user mobility. Furthermore, advanced channel estimation schemes, inter-user interference and EMI mitigation methods will be developed to realize the advantages of deploying STAR-RISs in the cell-free massive MIMO to provide feasible service to the ever-increasing number of users.

\ifCLASSOPTIONcaptionsoff
  \newpage
\fi

\begin{appendices}
\section{Proof of Complex Gradients}
This appendix provides the derivations of the complex gradients in \eqref{Optimization_3} regarding ${{\boldsymbol{\theta}}}$ and $\textbf{u}$ with the assistance of the fundamental formula\cite{Minka2000OldAN,4203075}
\begin{equation}
    \begin{array}{ll}
\displaystyle
d\text{tr}(\textbf{B}\textbf{X}\textbf{C}\textbf{X}^H\textbf{B}^H)\displaystyle=\text{tr}\bigg{(}\textbf{C}\textbf{X}^H\textbf{B}^H\textbf{B}d{\textbf{X}}+\left(\textbf{B}^H\textbf{B}\textbf{X}\textbf{C}\right)^Td{\textbf{X}}^*\bigg{)}.
   \end{array}
   \label{Lemma}
\end{equation}
First, we can obtain the derivation of $\nabla _{{\boldsymbol{\theta}}_{\omega_k}}\text{NMSE}({\boldsymbol{\theta}},\textbf{u})$. Based on \eqref{Delta} and \eqref{Lemma} with $\bar{\textbf{R}}=\mathbb{E}\{\bar{\boldsymbol{\Theta}}_{\omega_k}\textbf{R}\bar{\boldsymbol{\Theta}}_{\omega_k}^H\}=\phi^2\textbf{R}+(1-\phi^2)\textbf{R}\circ\textbf{I}_L$, where $\bar{\textbf{R}}_\text{SRIS}=A\bar{\textbf{R}}$. 
We can obtain the following differentials with respect to $\boldsymbol{\theta}_{\omega_k}$,
\begin{equation}
    \begin{array}{ll}
\displaystyle
d\text{tr}(\textbf{T}_{\omega_k})& \displaystyle=\text{tr}\Big{(}d\big{(}A\boldsymbol{\Theta}_{\omega_k}\bar{\textbf{R}}\boldsymbol{\Theta}_{\omega_k}^H\big{)}\Big{)}\\ &\displaystyle=A\text{tr}\left(
\textbf{u}_{\omega_k}\bar{\textbf{R}}_\text{SRIS}{\boldsymbol{\Theta}}_{\omega_k}^Hd{\boldsymbol{\theta}}_{\omega_k}+\left({\boldsymbol{\Theta}}_{\omega_k}\bar{\textbf{R}}\textbf{u}_{\omega_k}^H\right)^Td{\boldsymbol{\theta}}^*_{\omega_k}\right),
   \end{array}
   \label{derivative_pre1}
\end{equation}
\begin{equation}
    \begin{array}{ll}
\displaystyle
d\text{tr}(\textbf{R}_\text{SRIS}\textbf{T}_{\omega_k})& \displaystyle=\text{tr}\Big{(}d\big{(}A^2{\textbf{R}}\boldsymbol{\Theta}_{\omega_k}\bar{\textbf{R}}\boldsymbol{\Theta}_{\omega_k}^H\big{)}\Big{)}\\ &\displaystyle=A^2\text{tr}\bigg{(}\textbf{u}_{\omega_k}\bar{\textbf{R}}{\boldsymbol{\Theta}}_{\omega_k}^H{\textbf{R}}d{\boldsymbol{\theta}}_{\omega_k}+\left({\textbf{R}}{\boldsymbol{\Theta}}_{\omega_k}\bar{\textbf{R}}\textbf{u}_{\omega_k}^H\right)^Td{\boldsymbol{\theta}}^*_{\omega_k}\bigg{)}.
   \end{array}
   \label{derivative_pre}
\end{equation}
Similarly, $d\text{tr}(\bar{\textbf{T}}_{\omega_k})$ and $d\text{tr}(\textbf{R}_\text{SRIS}\bar{\textbf{T}}_{\omega_k})$ can obtained based on \eqref{T_bar_matrix}, \eqref{derivative_pre1}-\eqref{derivative_pre}.
Since ${{\boldsymbol{\theta}}_\omega}$ is a diagonal matrix, then so must $d{{\boldsymbol{\theta}}_\omega}$. With the help of \cite{Minka2000OldAN}, we can obtain
\begin{equation}
    \begin{array}{ll}
\displaystyle
\nabla _{{\boldsymbol{\theta}}_{\omega_k}}\text{tr}(\bar{\textbf{g}}_\text{ap,m}\textbf{T}_{\omega_k}\bar{\textbf{g}}_\text{ap,m}^H)\displaystyle=A\Big{(}\textbf{u}_{\omega_k}\bar{\textbf{R}}{\boldsymbol{\Theta}}_{\omega_k}^H\bar{\textbf{g}}_\text{ap,m}^H\bar{\textbf{g}}_\text{ap,m}\Big{)}^T\circ\textbf{I}_L,
   \end{array}
   \label{derivative_pre_11}
\end{equation}
\begin{equation}
    \begin{array}{ll}
\displaystyle
\nabla _{{\boldsymbol{\theta}}_{\omega_k}}\text{tr}(\textbf{R}_\text{SRIS}\textbf{T}_{\omega_k})\displaystyle=A^2\Big{(}\textbf{u}_{\omega_k}\bar{\textbf{R}}{\boldsymbol{\Theta}}_{\omega_k}^H{\textbf{R}}\Big{)}^T\circ\textbf{I}_L,
   \end{array}
   \label{derivative_pre_1}
\end{equation}
\begin{equation}
    \begin{array}{ll}
\displaystyle
\nabla _{{\boldsymbol{\theta}}_{\omega_k}}\text{tr}(\bar{\textbf{g}}_\text{ap,m}\bar{\textbf{T}}_{\omega_k}\bar{\textbf{g}}_\text{ap,m}^H)\displaystyle=\Big{(}\textbf{u}_{\omega_k}\bar{\textbf{A}}_{u,k}{\boldsymbol{\Theta}}_{\omega_k}^H\bar{\textbf{g}}_\text{ap,m}^H\bar{\textbf{g}}_\text{ap,m}\Big{)}^T\circ\textbf{I}_L,
   \end{array}
   \label{derivative_pre21}
\end{equation}
\begin{equation}
    \begin{array}{ll}
\displaystyle
\nabla _{{\boldsymbol{\theta}}_{\omega_k}}\text{tr}(\textbf{R}_\text{SRIS}\bar{\textbf{T}}_{\omega_k})\displaystyle=A\Big{(}\textbf{u}_{\omega_k}\bar{\textbf{A}}_{u,k}{\boldsymbol{\Theta}}_{\omega_k}^H{\textbf{R}}\Big{)}^T\circ\textbf{I}_L.
   \end{array}
   \label{derivative_pre211}
\end{equation}
Based on $\mathbf{\Delta}_{mk} $ in \eqref{Delta} and applying \eqref{derivative_pre_11}-\eqref{derivative_pre211}, we can obtain $ \nabla _{{\boldsymbol{\theta}}_{\omega_k}}\text{tr}(\mathbf{\Delta}_{mk})$ in \eqref{Optimization_5}\cite{Minka2000OldAN,4203075}.
Referring to \cite{Hjørungnes_2011,10297571},  we can derive $d(\mathbf{\Psi}_{mk}^{-1})$ as
\begin{equation}
    \begin{array}{ll}
\displaystyle
d(\mathbf{\Psi}_{mk}^{-1})=-\mathbf{\Psi}_{mk}^{-1}d(\mathbf{\Psi}_{mk})\mathbf{\Psi}_{mk}^{-1}.
   \end{array}
   \label{derivative_4}
\end{equation}

   Then, we can obtain the differential of $\text{tr}(\textbf{Q}_{mk})$ as \eqref{derivative_11} at the top of the next page \cite{Hjørungnes_2011,10297571}. Accordingly, we can obtain the derivative $\nabla _{{\boldsymbol{\theta}}_{\omega_k}}\text{tr}(\textbf{Q}_{mk})$ in \eqref{Optimization_6}.
\begin{figure*}
\begin{equation}
    \begin{array}{ll}
\displaystyle d\big{(}\text{tr}(\textbf{Q}_{mk})\big{)}&\displaystyle=\text{tr}\big{(}d(\tau_pp_p\mathbf{\Delta}_{mk}\mathbf{\Psi}_{mk}^{-1}\mathbf{\Delta}_{mk})\big{)}=\tau_pp_p\text{tr}\left(d(\mathbf{\Delta}_{mk})\mathbf{\Psi}_{mk}^{-1}\mathbf{\Delta}_{mk}-\mathbf{\Delta}_{mk}\mathbf{\Psi}_{mk}^{-1}d(\mathbf{\Psi}_{mk})\mathbf{\Psi}_{mk}^{-1}\mathbf{\Delta}_{mk}+\mathbf{\Delta}_{mk}\mathbf{\Psi}_{mk}^{-1}d(\mathbf{\Delta}_{mk})\right)\\\vspace{1.5 pt}& \displaystyle=\tau_pp_p\left[\begin{array}{ll}

\bar{\beta}_{ap,m}\bar{\beta}_{u,k}\text{tr}\left(\bar{\textbf{g}}_{ap,m}^H\mathbf{\Pi}_{mk}\bar{\textbf{g}}_{ap,m}d(\bar{\textbf{T}}_{\omega_k})\right)
\displaystyle+
\bar{\beta}_{ap,m}\tilde{\beta}_{u,k}\text{tr}\left(\bar{\textbf{g}}_{ap,m}^H\mathbf{\Pi}_{mk}\bar{\textbf{g}}_{ap,m}d({\textbf{T}}_{\omega_k})\right)
\\\vspace{1.5 pt}\displaystyle+
\tilde{\beta}_{ap,m}\bar{\beta}_{u,k}\text{tr}(\mathbf{\Pi}_{mk}\textbf{R}_{ap,m})d\text{tr}\left(\textbf{R}_\text{SRIS}\bar{\textbf{T}}_{\omega_k}\right)
\displaystyle+
\tilde{\beta}_{ap,m}\tilde{\beta}_{u,k}\text{tr}(\mathbf{\Pi}_{mk}\textbf{R}_{ap,m})d\text{tr}\left(\textbf{R}_\text{SRIS}{\textbf{T}}_{\omega_k}\right)
\end{array}
\right]\\\vspace{1.5 pt}&\displaystyle-
     \tau_pp_p\left[\begin{array}{ll}\displaystyle
\tau_pp_p\sum\nolimits_{k'\in\mathcal{P}_k}\left(\begin{array}{ll}
\bar{\beta}_{ap,m}\bar{\beta}_{u,k'}\text{tr}\left(\bar{\textbf{g}}_{ap,m}^H\bar{\mathbf{\Pi}}_{mk}\bar{\textbf{g}}_{ap,m}d(\bar{\textbf{T}}_{\omega_{k'}})\right)
\displaystyle+
\bar{\beta}_{ap,m}\tilde{\beta}_{u,k'}\text{tr}\left(\bar{\textbf{g}}_{ap,m}^H\bar{\mathbf{\Pi}}_{mk}\bar{\textbf{g}}_{ap,m}d({\textbf{T}}_{\omega_{k'}})\right)
\\\vspace{1.5 pt}\displaystyle+
\tilde{\beta}_{ap,m}\bar{\beta}_{u,k'}\text{tr}(\bar{\mathbf{\Pi}}_{mk}\textbf{R}_{ap,m})d\text{tr}\left(\textbf{R}_\text{SRIS}\bar{\textbf{T}}_{\omega_{k'}}\right)
\displaystyle+
\tilde{\beta}_{ap,m}\tilde{\beta}_{u,k'}\text{tr}(\bar{\mathbf{\Pi}}_{mk}\textbf{R}_{ap,m})d\text{tr}\left(\textbf{R}_\text{SRIS}{\textbf{T}}_{\omega_{k'}}\right)
\end{array}
\right)\\\vspace{1.5 pt}\displaystyle+
\sigma_r^2\bar{\beta}_{ap,m}\text{tr}\left(\bar{\textbf{g}}_{ap,m}^H\bar{\mathbf{\Pi}}_{mk}\bar{\textbf{g}}_{ap,m}d({\textbf{T}}_{r})\right)+\sigma_t^2\bar{\beta}_{ap,m}\text{tr}\left(\bar{\textbf{g}}_{ap,m}^H\bar{\mathbf{\Pi}}_{mk}\bar{\textbf{g}}_{ap,m}d({\textbf{T}}_{t})\right)
\\\vspace{1.5 pt}\displaystyle+
\sigma_r^2\tilde{\beta}_{ap,m}\text{tr}(\bar{\mathbf{\Pi}}_{mk}\textbf{R}_{ap,m})d\text{tr}\left(\textbf{R}_\text{SRIS}{\textbf{T}}_r\right)+\sigma_t^2\tilde{\beta}_{ap,m}\text{tr}(\bar{\mathbf{\Pi}}_{mk}\textbf{R}_{ap,m})d\text{tr}\left(\textbf{R}_\text{SRIS}{\textbf{T}}_t\right)
\end{array}
\right]
    ,
  \end{array}
   \label{derivative_11}
      \vspace{-2 pt}
   \end{equation} 
   \vspace{-10 pt}
   \hrulefill
\end{figure*}
Similarly, we obtain the derivation of $\nabla _{\mathbf{u}_{\omega_k}}\text{NMSE}({\boldsymbol{\theta}},\textbf{u})$ by considering the following differentials of with respect to $\mathbf{u}_{\omega_k}$ as
\begin{equation}
    \begin{array}{ll}
\displaystyle
d\text{tr}(\textbf{T}_{\omega_k})& \displaystyle=\text{tr}\Big{(}d\big{(}A\boldsymbol{\Theta}_{\omega_k}\bar{\textbf{R}}\boldsymbol{\Theta}_{\omega_k}^H\big{)}\Big{)}\\ &\displaystyle=A\text{tr}\left(
\boldsymbol{\theta}_{\omega_k}\bar{\textbf{R}}{\boldsymbol{\Theta}}_{\omega_k}^Hd\textbf{u}_{\omega_k}+\left({\boldsymbol{\Theta}}_{\omega_k}\bar{\textbf{R}}\boldsymbol{\theta}_{\omega_k}^H\right)^Td{\textbf{u}}^*_{\omega_k}\right),
   \end{array}
   \label{derivative_pre1_u}
\end{equation}
\begin{equation}
    \begin{array}{ll}
\displaystyle
d\text{tr}(\textbf{R}_\text{SRIS}\textbf{T}_{\omega_k})& \displaystyle=\text{tr}\Big{(}d\big{(}A^2{\textbf{R}}\boldsymbol{\Theta}_{\omega_k}\bar{\textbf{R}}\boldsymbol{\Theta}_{\omega_k}^H\big{)}\Big{)}\\ &\displaystyle=A^2\text{tr}\bigg{(}\boldsymbol{\theta}_{\omega_k}\bar{\textbf{R}}{\boldsymbol{\Theta}}_{\omega_k}^H{\textbf{R}}d{\textbf{u}}_{\omega_k}+\left({\textbf{R}}{\boldsymbol{\Theta}}_{\omega_k}\bar{\textbf{R}}\boldsymbol{\theta}_{\omega_k}^H\right)^Td{\textbf{u}}^*_{\omega_k}\bigg{)}.
   \end{array}
   \label{derivative_pre_u}
\end{equation}
Accordingly, $d\text{tr}(\bar{\textbf{T}}_{\omega_k})$ and $d\text{tr}(\textbf{R}_\text{SRIS}\bar{\textbf{T}}_{\omega_k})$ can obtained based on \eqref{T_bar_matrix}, \eqref{derivative_pre1_u}-\eqref{derivative_pre_u}.
Since ${\mathbf{u}_\omega}$ is a real-valued diagonal matrix, then so must $d{\mathbf{u}_\omega}$. With the assistance of \cite{Minka2000OldAN}, we can calculate
\begin{equation}
    \begin{array}{ll}
\displaystyle
\nabla _{\textbf{u}_{\omega_k}}\text{tr}(\bar{\textbf{g}}_\text{ap,m}\textbf{T}_{\omega_k}\bar{\textbf{g}}_\text{ap,m}^H)\displaystyle=2A\mathfrak{RE}\left\{\Big{(}\boldsymbol{\theta}_{\omega_k}\bar{\textbf{R}}{\boldsymbol{\Theta}}_{\omega_k}^H\bar{\textbf{g}}_\text{ap,m}^H\bar{\textbf{g}}_\text{ap,m}\Big{)}^T\circ\textbf{I}_L\right\},
   \end{array}
   \label{derivative_pre_11_u}
\end{equation}
\begin{equation}
    \begin{array}{ll}
\displaystyle
\nabla _{\textbf{u}_{\omega_k}}\text{tr}(\textbf{R}_\text{SRIS}\textbf{T}_{\omega_k})\displaystyle=2A^2\mathfrak{RE}\left\{\Big{(}\boldsymbol{\theta}_{\omega_k}\bar{\textbf{R}}{\boldsymbol{\Theta}}_{\omega_k}^H{\textbf{R}}\Big{)}^T\circ\textbf{I}_L\right\},
   \end{array}
   \label{derivative_pre_1_u}
\end{equation}
\begin{equation}
    \begin{array}{ll}
\displaystyle
\nabla _{\textbf{u}_{\omega_k}}\text{tr}(\bar{\textbf{g}}_\text{ap,m}\bar{\textbf{T}}_{\omega_k}\bar{\textbf{g}}_\text{ap,m}^H)\displaystyle=2\mathfrak{RE}\left\{\Big{(}\boldsymbol{\theta}_{\omega_k}\bar{\textbf{A}}_{u,k}{\boldsymbol{\Theta}}_{\omega_k}^H\bar{\textbf{g}}_\text{ap,m}^H\bar{\textbf{g}}_\text{ap,m}\Big{)}^T\circ\textbf{I}_L\right\},
   \end{array}
   \label{derivative_pre21_u}
\end{equation}
\begin{equation}
    \begin{array}{ll}
\displaystyle
\nabla _{\textbf{u}_{\omega_k}}\text{tr}(\textbf{R}_\text{SRIS}\bar{\textbf{T}}_{\omega_k})\displaystyle=2A\mathfrak{RE}\left\{\Big{(}\boldsymbol{\theta}_{\omega_k}\bar{\textbf{A}}_{u,k}{\boldsymbol{\Theta}}_{\omega_k}^H{\textbf{R}}\Big{)}^T\circ\textbf{I}_L\right\}.
   \end{array}
   \label{derivative_pre211_u}
\end{equation}

Similarly, we can obtain $ \nabla _{\mathbf{u}_{\omega_k}}\text{tr}(\mathbf{\Delta}_{mk})$ in \eqref{Optimization_9} with the assistance of \eqref{Delta} and \eqref{derivative_pre_11_u}-\eqref{derivative_pre211_u}.
We can also
obtain the derivative process of $\nabla _{\mathbf{u}_{\omega_k}}\text{tr}(\mathbf{Q}_{mk})$ following \eqref{derivative_11}, and the derivative is shown in \eqref{Optimization_10}. We omit the details for brevity.

\section
{Derivation of SE term approximations}
We introduce the detailed derivation $\text{SINR}_{\text{u},k}$ in this appendix \cite{8845768,9416909}. First, the estimate $\hat{\textbf{g}}_{mk}$ and the estimation error ${\textbf{e}}_{mk}$ are uncorrelated based on the properties of MMSE estimation \cite{9875036,10621117}. If users are in the same pilot sequence set, $k'\in\mathcal{P}_k$, $\hat{\textbf{g}}_{mk'}$ is correlated with ${\textbf{g}}_{mk}$. Inspired by \cite{9665300,10484981}, we can compute the terms in $\text{SINR}_{\text{u},k}$ by the following steps.
\subsection{Compute $\bar{\textbf{b}}_k$} 
First, we can obtain the elements in $\bar{\textbf{b}}_k$ as
\begin{equation}
\begin{array}{ll}
\displaystyle\bar{b}_{mk}&\displaystyle=\mathbb{E}\left\{\hat{\textbf{g}}_{mk}^H{\textbf{g}}_{mk}\right\}=\mathbb{E}\left\{\hat{\textbf{g}}_{mk}^H\hat{\textbf{g}}_{mk}\right\}=\text{tr}\big{(}\textbf{Q}_{mk}\big{)}=b_{mk},
\end{array}\label{b_k}
   \end{equation}
   then, we can have $\textbf{b}_k=[b_{1k},b_{2k},...,b_{Mk}]^T\in\mathbb{C}^{M\times1}$, in which element $ b_{mk}=\text{tr}(\textbf{Q}_{mk}),~\forall m,~\forall k$.
\subsection{Compute $\displaystyle\sum\nolimits_{k'=1}^Kp_u\eta_{k'}\textbf{a}_k^H\mathbb{E}\left\{\bar{\boldsymbol{\Omega}}_{kk'}\bar{\boldsymbol{\Omega}}_{kk'}^H\right\}\textbf{a}_k$} 
We first decompose the sum of inter-user interference into
\begin{equation}
\begin{array}{ll}
&\displaystyle\sum\nolimits_{k'=1}^Kp_u\eta_{k'}\textbf{a}_k^H\mathbb{E}\left\{\bar{\boldsymbol{\Omega}}_{kk'}\bar{\boldsymbol{\Omega}}_{kk'}^H\right\}\textbf{a}_k
\\&\displaystyle=\displaystyle\sum\nolimits_{k'=1}^{K}p_u\eta_{k'}\mathbb{E}\left\{\Big{|}\sum\nolimits_{m=1}^M a_{mk}^*\hat{\textbf{g}}_{mk}^H{\textbf{g}}_{mk'}\Big{|}^2\right\}\\&\displaystyle
=\underbrace{\sum\nolimits_{k'\in\mathcal{P}_k}p_u\eta_{k'}\mathbb{E}\left\{\Big{|}\sum\nolimits_{m=1}^M a_{mk}^*\hat{\textbf{g}}_{mk}^H{\textbf{g}}_{mk'}\Big{|}^2\right\}}_{T_1}\\& \displaystyle+\underbrace{\sum\nolimits_{k'\notin\mathcal{P}_k}p_u\eta_{k'}\mathbb{E}\left\{\Big{|}\sum\nolimits_{m=1}^M a_{mk}^*\hat{\textbf{g}}_{mk}^H{\textbf{g}}_{mk'}\Big{|}^2\right\}}_{T_2}.
\end{array}
\label{UI}
   \end{equation}
   
\subsubsection{$\displaystyle\sum\nolimits_{k'\in\mathcal{P}_k}p_u\eta_{k'}\mathbb{E}\left\{\Big{|}\sum\nolimits_{m=1}^M a_{mk}^*\hat{\textbf{g}}_{mk}^H{\textbf{g}}_{mk'}\Big{|}^2\right\}$} The procedure shows that
\begin{equation}
\begin{array}{ll}
\displaystyle T_1
&\displaystyle=\sum\nolimits_{k'\in\mathcal{P}_k}p_u\eta_{k'}\sum\nolimits_{m=1}^Ma_{mk}^*a_{mk}^T\underbrace{\mathbb{E}\Big{\{}\hat{\textbf{g}}_{mk}^H{\textbf{g}}_{mk'}{\textbf{g}}_{mk'}^H\hat{\textbf{g}}_{mk} \Big{\}}}_{T_{11}}\\&\displaystyle+\sum\nolimits_{k'\in\mathcal{P}_k}p_u\eta_{k'}\sum\nolimits_{m=1}^M\sum\nolimits_{m'\neq m}a_{mk}^* a_{m'k}^T\underbrace{\mathbb{E}\Big{\{}\hat{\textbf{g}}_{mk}^H{\textbf{g}}_{mk'}{\textbf{g}}_{m'k'}^H\hat{\textbf{g}}_{m'k}\Big{\}}}_{T_{12}},
 \end{array}
 \label{UI_part1}
   \end{equation}

The relevant terms in \eqref{UI_part1} can be given by \eqref{part_1} and \eqref{part_2} at the top of this page.
\begin{figure*}[t!]
\begin{equation}
\begin{array}{ll}
T_{11}&\displaystyle=
\displaystyle\tau_pp_p\mathbb{E}\Bigg{\{}\sum\nolimits_{k''\in\mathcal{P}_k}
\textbf{g}_{mk''}^H\textbf{Z}_{mk}^H\textbf{g}_{mk'}\textbf{g}_{mk'}^H\textbf{Z}_{mk}\textbf{g}_{mk''}\Bigg{\}}+\mathbb{E}\Bigg{\{}(\textbf{g}_{ap,m}\bar{\boldsymbol{\Theta}}_{t}\boldsymbol{\Theta}_{t}\textbf{N}_{t}\varphi_k)^H\textbf{Z}_{mk}^H\textbf{g}_{mk'}\textbf{g}_{mk'}^H\textbf{Z}_{mk}(\textbf{g}_{ap,m}\bar{\boldsymbol{\Theta}}_{t}\boldsymbol{\Theta}_{t}\textbf{N}_{t}\varphi_k)\Bigg{\}}\\& \vspace{1.4 pt}\displaystyle+\mathbb{E}\Bigg{\{}(\textbf{g}_{ap,m}\bar{\boldsymbol{\Theta}}_{r}\boldsymbol{\Theta}_{r}\textbf{N}_{r}\varphi_k)^H\textbf{Z}_{mk}^H\textbf{g}_{mk'}\textbf{g}_{mk'}^H\textbf{Z}_{mk}(\textbf{g}_{ap,m}\bar{\boldsymbol{\Theta}}_{r}\boldsymbol{\Theta}_{r}\textbf{N}_{r}\varphi_k)\Bigg{\}}+\mathbb{E}\Bigg{\{}(\textbf{N}_{m,p}\varphi_k)^H\textbf{Z}_{mk}^H\textbf{g}_{mk'}\textbf{g}_{mk'}^H\textbf{Z}_{mk}(\textbf{N}_{m,p}\varphi_k)\Bigg{\}}\\ \vspace{1.4 pt}&\displaystyle=
\underbrace{\tau_pp_p\left(\begin{array}{ll}\text{tr}\left(\boldsymbol{\Delta}_{mk'}^d\textbf{Z}_{mk}^H\right)\text{tr}\left(\boldsymbol{\Delta}_{mk'}\textbf{Z}_{mk}\right)+\text{tr}\left(\boldsymbol{\Delta}_{mk'}^c\textbf{Z}_{mk}^H\right)\text{tr}\left(\boldsymbol{\Delta}_{mk'}^d\textbf{Z}_{mk}\right)\\\displaystyle+
2\bar{\beta}_{ap,m}\tilde{\beta}_{ap,m}\tilde{\beta}_{u,k'}^2\text{tr}\left(\textbf{T}_{\omega_{k'}}\textbf{R}_\text{SRIS}^{1/2}\right)^2\text{tr}\left(\bar{\textbf{g}}_{ap,m}^H\textbf{Z}_{mk}^H\textbf{R}_{ap,m}\textbf{Z}_{mk}\bar{\textbf{g}}_{ap,m}\right)\\
\displaystyle 
+
\tilde{\beta}_{ap,m}^2\tilde{\beta}_{u,k'}^2\text{tr}\left(\textbf{R}_{ap,m}\textbf{Z}_{mk}^H\textbf{R}_{ap,m}\textbf{Z}_{mk}
\right)\text{tr}\left(\hat{\textbf{F}}_{\omega_{k'},\omega_{k'}}\textbf{R}_\text{SRIS}\right)
\displaystyle+
\bar{\beta}_{u,k'}\tilde{\beta}_{u,k'}\text{tr}\left(\bar{\textbf{T}}_{\omega_{k'}}\boldsymbol{\Delta}_{ap,mk}\right)\text{tr}\left(\textbf{T}_{\omega_{k'}}\boldsymbol{\Delta}_{ap,mk}^H\right)\\
\displaystyle+
2\tilde{\beta}_{ap,m}^2\bar{\beta}_{u,k'}\tilde{\beta}_{u,k'}
\text{tr}\left(\textbf{R}_{ap,m}\textbf{Z}_{mk}^H\textbf{R}_{ap,m}\textbf{Z}_{mk}
\right)\text{tr}\left(\bar{\textbf{T}}_{\omega_{k'}}\textbf{R}_\text{SRIS}\right)\text{tr}\left(\textbf{T}_{\omega_{k'}}\textbf{R}_\text{SRIS}\right)\displaystyle+
\tilde{\beta}_{u,k'}\text{tr}\left(\textbf{T}_{\omega_{k'}}\boldsymbol{\Delta}_{ap,mk}^H\right)\text{tr}\left(\boldsymbol{\Delta}_{mk'}^c\textbf{Z}_{mk}\right)
\end{array}\right)}_{[\boldsymbol{\Omega}_{kk'}]_{mm}}

\displaystyle\\&+
\displaystyle\tau_pp_p\sum\nolimits_{k''\in\mathcal{P}_k}
\text{tr}\left(\boldsymbol{\Delta}_{mk''}\textbf{Z}_{mk}^H\boldsymbol{\Delta}_{mk'}\textbf{Z}_{mk}\right)+\displaystyle
\bar{\beta}_{ap,m}\tilde{\beta}_{ap,m}\text{tr}\left(\textbf{R}_{ap,m}\textbf{Z}_{mk}
\right)\text{tr}\left(\varpi_{mk}\textbf{F}_{k',k''}\right)+\displaystyle\tilde{\beta}_{ap,m}\text{tr}\left(\textbf{R}_{ap,m}\textbf{Z}_{mk}
\right)\text{tr}\left(\textbf{F}_{k',k''}^H\boldsymbol{\Delta}_{ap,mk}^H\right)
\\& \vspace{1.4 pt}\displaystyle+\sum\nolimits_{\omega_{t,r}}\sigma_{\omega}^2\left(\begin{array}{ll}
\bar{\beta}_{ap,m}\text{tr}\left(
\textbf{T}_\omega
\bar{\textbf{g}}_{ap,m}^H\textbf{Z}_{mk}^H\boldsymbol{\Delta}_{mk'}\textbf{Z}_{mk}\bar{\textbf{g}}_{ap,m}\right)+
\tilde{\beta}_{ap,m}\text{tr}\left(\textbf{R}_{ap,m}\textbf{Z}_{mk}^H\boldsymbol{\Delta}_{mk'}\textbf{Z}_{mk}\right)\text{tr}\left(\textbf{R}_\text{SRIS}
\textbf{T}_\omega
\right)
\\\displaystyle+\tilde{\beta}_{ap,m}\text{tr}\left(\textbf{R}_{ap,m}\textbf{Z}_{mk}\right)\text{tr}\left(
\boldsymbol{\Delta}_{ap,mk}^H\bar{\textbf{F}}_{k',\omega}\right)
\displaystyle+
\bar{\beta}_{ap,m}\tilde{\beta}_{ap,m}\text{tr}\left(\textbf{R}_{ap,m}\textbf{Z}_{mk}^H\right)\text{tr}\left(\bar{\textbf{F}}_{k',\omega}^H\varpi_{mk}^H\right)\end{array}\right)
\displaystyle+\sigma^2\text{tr}\left(\mathbf{\Delta}_{mk'}\textbf{Z}_{mk}\textbf{Z}_{mk}^H\right),
\end{array}
\label{part_1}
\vspace{-2 pt}
   \end{equation}
   \vspace{-10 pt}
   \hrulefill
   \end{figure*}
\begin{figure*}[t!]
 \begin{equation}
\begin{array}{ll}
T_{12}&\displaystyle=
\displaystyle\tau_pp_p\mathbb{E}\Bigg{\{}\sum\nolimits_{k''\in\mathcal{P}_k}
\textbf{g}_{mk''}^H\textbf{Z}_{mk}^H\textbf{g}_{mk'}\textbf{g}_{m'k'}^H\textbf{Z}_{m'k}\textbf{g}_{m'k''}\Bigg{\}}+\mathbb{E}\Bigg{\{}(\textbf{g}_{ap,m}\bar{\boldsymbol{\Theta}}_{t}\boldsymbol{\Theta}_{t}\textbf{N}_{t}\varphi_k)^H\textbf{Z}_{mk}^H\textbf{g}_{mk'}\textbf{g}_{m'k'}^H\textbf{Z}_{m'k}(\textbf{g}_{ap,m'}\bar{\boldsymbol{\Theta}}_{t}\boldsymbol{\Theta}_{t}\textbf{N}_{t}\varphi_k)\Bigg{\}}\\& \vspace{1.4 pt}\displaystyle+\mathbb{E}\Bigg{\{}(\textbf{g}_{ap,m}\bar{\boldsymbol{\Theta}}_{r}\boldsymbol{\Theta}_{r}\textbf{N}_{r}\varphi_k)^H\textbf{Z}_{mk}^H\textbf{g}_{mk'}\textbf{g}_{m'k'}^H\textbf{Z}_{m'k}(\textbf{g}_{ap,m'}\bar{\boldsymbol{\Theta}}_{r}\boldsymbol{\Theta}_{r}\textbf{N}_{r}\varphi_k)\Bigg{\}}+\mathbb{E}\Bigg{\{}(\textbf{N}_{m,p}\varphi_k)^H\textbf{Z}_{mk}^H\textbf{g}_{m'k'}\textbf{g}_{mk'}^H\textbf{Z}_{m'k}(\textbf{N}_{m',p}\varphi_k)\Bigg{\}}\\ \vspace{1.4 pt}&\displaystyle=
\underbrace{\tau_pp_p\left(
\begin{array}{ll}
\text{tr}\left(\boldsymbol{\Delta}_{mk'}^d\textbf{Z}_{mk}^H\right)\text{tr}\left(\boldsymbol{\Delta}_{m'k'}\textbf{Z}_{m'k}\right)+\text{tr}\left(\boldsymbol{\Delta}_{mk'}^c\textbf{Z}_{mk}^H\right)\text{tr}\left(\boldsymbol{\Delta}_{m'k'}^d\textbf{Z}_{m'k}\right)\\\displaystyle+\bar{\beta}_{u,k'}\tilde{\beta}_{u,k'}\text{tr}\left(
\bar{\textbf{T}}_{\omega_{k'}}\boldsymbol{\Delta}_{ap,mk}
\right)\text{tr}\left(
{\textbf{T}}_{\omega_{k'}}\boldsymbol{\Delta}_{ap,m'k}^H
\right)\displaystyle+\tilde{\beta}_{u,k'}\text{tr}\left(
{\textbf{T}}_{\omega_{k'}}\boldsymbol{\Delta}_{ap,mk}
\right)\text{tr}\left(\check{\boldsymbol{\Delta}}_{uk'}\boldsymbol{\Delta}_{ap,m'k}^H
\right)
\end{array}
\right)}_{[\boldsymbol{\Omega}_{kk'}]_{mm'}}
\\&
\displaystyle+
\displaystyle
\tau_pp_p\sum\nolimits_{k''\in\mathcal{P}_k}\text{tr}\left(\boldsymbol{\Delta}_{ap,mk}\textbf{K}_{m'k,k',k''}\right)\displaystyle+\sum\nolimits_{\omega=t,r}\sigma_{\omega}^2\text{tr}\left(\bar{\textbf{K}}_{m'k,k',\omega}\boldsymbol{\Delta}_{ap,mk}\right),
\end{array}
\label{part_2}
\vspace{-2 pt}
   \end{equation}
   \vspace{-10 pt}
   \hrulefill
   \end{figure*}

\subsubsection{$\displaystyle\sum\nolimits_{k'\notin\mathcal{P}_k}p_u\eta_{k'}\mathbb{E}\left\{\Big{|}\sum\nolimits_{m=1}^M a_{mk}^*\hat{\textbf{g}}_{mk}^H{\textbf{g}}_{mk'}\Big{|}^2\right\}$} The similar procedure shows that
\begin{equation}
\begin{array}{ll}
\displaystyle T_2& \displaystyle
=\sum\nolimits_{k'\notin\mathcal{P}_k}p_u\eta_{k'}\sum\nolimits_{m=1}^Ma_{mk}^*a_{mk}^T\underbrace{\mathbb{E}\Big{\{}\hat{\textbf{g}}_{mk}^H{\textbf{g}}_{mk'}{\textbf{g}}_{mk'}^H\hat{\textbf{g}}_{mk} \Big{\}}}_{T_{21}}\\&\displaystyle+\sum\nolimits_{k'\notin\mathcal{P}_k}p_u\eta_{k'}\sum\nolimits_{m=1}^M\sum\nolimits_{m'\neq m}a_{mk}^*a_{m'k}^T\underbrace{\mathbb{E}\Big{\{}\hat{\textbf{g}}_{mk}^H{\textbf{g}}_{mk'}{\textbf{g}}_{m'k'}^H\hat{\textbf{g}}_{m'k} \Big{\}}}_{T_{22}},
 \end{array}
 \label{UI_part2}
   \end{equation}
where the relevant terms in \eqref{UI_part2} can be expressed as \eqref{part_3} and \eqref{part_4} at the top of the next page. 
\begin{figure*}[t!]
\begin{equation}
\begin{array}{ll}
T_{21}&\displaystyle=
\displaystyle\tau_pp_p\mathbb{E}\Bigg{\{}\sum\nolimits_{k''\in\mathcal{P}_k}
\textbf{g}_{mk''}^H\textbf{Z}_{mk}^H\textbf{g}_{mk'}\textbf{g}_{mk'}^H\textbf{Z}_{mk}\textbf{g}_{mk''}\Bigg{\}}+\mathbb{E}\Bigg{\{}(\textbf{g}_{m}\bar{\boldsymbol{\Theta}}_{t}\boldsymbol{\Theta}_{t}\textbf{N}_{t}\varphi_k)^H\textbf{Z}_{mk}^H\textbf{g}_{mk'}\textbf{g}_{mk'}^H\textbf{Z}_{mk}(\textbf{g}_{m}\bar{\boldsymbol{\Theta}}_{t}\boldsymbol{\Theta}_{t}\textbf{N}_{t}\varphi_k)\Bigg{\}}\\ &\vspace{1.4 pt}\displaystyle+\mathbb{E}\Bigg{\{}(\textbf{g}_{m}\bar{\boldsymbol{\Theta}}_{r}\boldsymbol{\Theta}_{r}\textbf{N}_{r}\varphi_k)^H\textbf{Z}_{mk}^H\textbf{g}_{mk'}\textbf{g}_{mk'}^H\textbf{Z}_{mk}(\textbf{g}_{m}\bar{\boldsymbol{\Theta}}_{r}\boldsymbol{\Theta}_{r}\textbf{N}_{r}\varphi_k)\Bigg{\}}+\mathbb{E}\Bigg{\{}(\textbf{N}_{m,p}\varphi_k)^H\textbf{Z}_{mk}^H\textbf{g}_{mk'}\textbf{g}_{mk'}^H\textbf{Z}_{mk}(\textbf{N}_{m,p}\varphi_k)\Bigg{\}}\\ \vspace{1.4 pt}&\displaystyle=
\tau_pp_p\sum\nolimits_{k''\in\mathcal{P}_k}\left(
\text{tr}\left(\boldsymbol{\Delta}_{mk''}\textbf{Z}_{mk}^H\boldsymbol{\Delta}_{mk'}\textbf{Z}_{mk}\right)+\displaystyle
\bar{\beta}_{ap,m}\tilde{\beta}_{ap,m}\text{tr}\left(\textbf{R}_{ap,m}\textbf{Z}_{mk}
\right)\text{tr}\left(\varpi_{mk}\textbf{F}_{k',k''}\right)+\displaystyle\tilde{\beta}_{ap,m}\text{tr}\left(\textbf{R}_{ap,m}\textbf{Z}_{mk}
\right)\text{tr}\left(\textbf{F}_{k',k''}^H\boldsymbol{\Delta}_{ap,mk}^H\right)\right)
\\& \vspace{1.4 pt}\displaystyle+\sum\nolimits_{\omega_{t,r}}\sigma_{\omega}^2\left(\begin{array}{ll}
\bar{\beta}_{ap,m}\text{tr}\left(
\textbf{T}_\omega
\bar{\textbf{g}}_{ap,m}^H\textbf{Z}_{mk}^H\boldsymbol{\Delta}_{mk'}\textbf{Z}_{mk}\bar{\textbf{g}}_{ap,m}\right)+
\tilde{\beta}_{ap,m}\text{tr}\left(\textbf{R}_{ap,m}\textbf{Z}_{mk}^H\boldsymbol{\Delta}_{mk'}\textbf{Z}_{mk}\right)\text{tr}\left(\textbf{R}_\text{SRIS}
\textbf{T}_\omega
\right)
\\\displaystyle+\tilde{\beta}_{ap,m}\text{tr}\left(\textbf{R}_{ap,m}\textbf{Z}_{mk}\right)\text{tr}\left(
\boldsymbol{\Delta}_{ap,mk}^H\bar{\textbf{F}}_{k',\omega}\right)
\displaystyle+
\bar{\beta}_{ap,m}\tilde{\beta}_{ap,m}\text{tr}\left(\textbf{R}_{ap,m}\textbf{Z}_{mk}^H\right)\text{tr}\left(\bar{\textbf{F}}_{k',\omega}^H\varpi_{mk}^H\right)\end{array}\right)
\displaystyle+\sigma^2\text{tr}\left(\mathbf{\Delta}_{mk'}\textbf{Z}_{mk}\textbf{Z}_{mk}^H\right)\\&\displaystyle=[\boldsymbol{\Upsilon}_{kk'}]_{mm},
\end{array}
\label{part_3}
\vspace{-5 pt}
   \end{equation}
   \vspace{-10 pt}
   \hrulefill
   \end{figure*}
\begin{figure*}[t!]
\begin{equation}
\begin{array}{ll}
T_{22}&\displaystyle=
\displaystyle\tau_pp_p\mathbb{E}\Bigg{\{}\sum\nolimits_{k''\in\mathcal{P}_k}
\textbf{g}_{mk''}^H\textbf{Z}_{mk}^H\textbf{g}_{mk'}\textbf{g}_{m'k'}^H\textbf{Z}_{m'k}\textbf{g}_{m'k''}\Bigg{\}}+\mathbb{E}\Bigg{\{}(\textbf{g}_{m}\bar{\boldsymbol{\Theta}}_{t}\boldsymbol{\Theta}_{t}\textbf{N}_{t}\varphi_k)^H\textbf{Z}_{mk}^H\textbf{g}_{mk'}\textbf{g}_{m'k'}^H\textbf{Z}_{m'k}(\textbf{g}_{m'}\bar{\boldsymbol{\Theta}}_{t}\boldsymbol{\Theta}_{t}\textbf{N}_{t}\varphi_k)\Bigg{\}}\\& \vspace{1.4 pt}\displaystyle+\mathbb{E}\Bigg{\{}(\textbf{g}_{m}\bar{\boldsymbol{\Theta}}_{r}\boldsymbol{\Theta}_{r}\textbf{N}_{r}\varphi_k)^H\textbf{Z}_{mk}^H\textbf{g}_{mk'}\textbf{g}_{m'k'}^H\textbf{Z}_{m'k}(\textbf{g}_{m'}\bar{\boldsymbol{\Theta}}_{r}\boldsymbol{\Theta}_{r}\textbf{N}_{r}\varphi_k)\Bigg{\}}+\mathbb{E}\Bigg{\{}(\textbf{N}_{m,p}\varphi_k)^H\textbf{Z}_{mk}^H\textbf{g}_{m'k'}\textbf{g}_{mk'}^H\textbf{Z}_{m'k}(\textbf{N}_{m',p}\varphi_k)\Bigg{\}}\\ \vspace{1.4 pt}&\displaystyle=
\displaystyle\displaystyle
\tau_pp_p\sum\nolimits_{k''\in\mathcal{P}_k}\text{tr}\left(\boldsymbol{\Delta}_{ap,mk}\textbf{K}_{m'k,k',k''}\right)\displaystyle+\sum\nolimits_{\omega=t,r}\sigma_{\omega}^2\text{tr}\left(\bar{\textbf{K}}_{m'k,k',\omega}\boldsymbol{\Delta}_{ap,mk}\right)\\&\displaystyle=[\boldsymbol{\Upsilon}_{kk'}]_{mm'}.
\end{array}
\label{part_4}
\vspace{-3 pt}
   \end{equation}
   \vspace{-10 pt}
   \hrulefill
   \end{figure*}According to the above-mentioned decomposition, the sum of the inter-user interference can be given by
\begin{equation}
\begin{array}{ll}
\displaystyle\sum\nolimits_{k'=1}^Kp_u\eta_{k'}\mathbb{E}\left\{\Big{|}\sum\nolimits_{m=1}^M a_{mk}^*\hat{\textbf{g}}_{mk}^H{\textbf{g}}_{mk'}\Big{|}^2\right\}&\displaystyle=\sum\nolimits_{k'\in\mathcal{P}_k}p_u\eta_{k'}\textbf{a}_k^H\boldsymbol{\Omega}_{kk'}\textbf{a}_k\\ &\displaystyle+\sum\nolimits_{k'=1}^Kp_u\eta_{k'}\textbf{a}_k^H\boldsymbol{\Upsilon}_{kk'}\textbf{a}_k.
\end{array}
\label{UI_matrixversion}
   \end{equation}
The elements in $\boldsymbol{\Omega}_{kk'}\in\mathbb{C}^{M\times M}$ follow \eqref{part_1}-\eqref{part_2} at the top of this page. 
The elements in $\boldsymbol{\Upsilon}_{kk'} \in\mathbb{C}^{M\times M}$ follow \eqref{part_3}-\eqref{part_4} at the top of the next page with $\varpi_{mk}=\bar{\textbf{g}}_{ap,m}^H\textbf{Z}_{mk}^H\bar{\textbf{g}}_{ap,m}$ and
   \begin{equation}
    \begin{array}{ll}
\boldsymbol{\Delta}_{ap,mk}=\bar{\beta}_{ap,m}\bar{\textbf{g}}_{ap,m}^H\textbf{Z}_{mk}^H\bar{\textbf{g}}_{ap,m}+\tilde{\beta}_{ap,m}\text{tr}\left(\textbf{R}_{ap,m}\textbf{Z}_{mk}^H\right)\textbf{R}_\text{SRIS},
      \end{array}
\end{equation}
   \begin{equation}
    \begin{array}{ll}
\boldsymbol{\Delta}_{uk}=\bar{\beta}_{u,k}\bar{\textbf{g}}_{u,k}\bar{\textbf{g}}_{u,k}^H+\tilde{\beta}_{u,k}\textbf{R}_\text{SRIS}.
      \end{array}
\end{equation}
   \begin{equation}
    \begin{array}{ll}
\check{\boldsymbol{\Delta}}_{uk}={\boldsymbol{\Theta}}_{\omega_k}\left(\bar{\beta}_{u,k}\bar{\textbf{A}}_{u,k}+\tilde{\beta}_{u,k}\bar{\textbf{R}}_\text{SRIS}\right){\boldsymbol{\Theta}}_{\omega_k}^H.
      \end{array}
\end{equation}
Subsequently, we can introduce several useful formulas:
 \begin{equation}
\begin{array}{ll}
\displaystyle
\mathbb{E}\{\bar{\boldsymbol{\Theta}}_{\omega_k}\textbf{A}\bar{\boldsymbol{\Theta}}_{\omega_k}^H\textbf{B}\bar{\boldsymbol{\Theta}}_{\omega_k}\textbf{C}\bar{\boldsymbol{\Theta}}_{\omega_k}^H\}\approx\bar{\textbf{A}}{\textbf{B}}\bar{\textbf{C}}-(\bar{\textbf{A}}{\textbf{B}}\bar{\textbf{C}})\circ\textbf{I}+({\textbf{A}}\bar{\textbf{B}}{\textbf{C}})\circ\textbf{I},
\end{array}\label{PhaseError_1}
   \end{equation}
 \begin{equation}
\begin{array}{ll}
\displaystyle
\bar{\textbf{X}}=\mathbb{E}\{\bar{\boldsymbol{\Theta}}_{\omega_k}\textbf{X}\bar{\boldsymbol{\Theta}}_{\omega_k}^H\}=\phi^2\textbf{X}+(1-\phi^2)\textbf{X}\circ\textbf{I}.
\end{array}\label{PhaseError_2}
   \end{equation}
Referring to \eqref{PhaseError_1}-\eqref{PhaseError_2}, we can have 
$\text{tr}\left(\boldsymbol{\Delta}_{mk''}\textbf{Z}_{mk}^H\boldsymbol{\Delta}_{mk'}\textbf{Z}_{mk}\right)$ given as \eqref{Average1} at the top of this page with $\hat{\boldsymbol{\Delta}}_{mk}=\beta_{d,mk}\textbf{R}_{d,mk}+\tilde{\beta}_{ap,m}\text{tr}\left(\textbf{R}_\text{SRIS}\left(\bar{\beta}_{u,k}\bar{\textbf{T}}_{\omega_k}+\tilde{\beta}_{u,k}{\textbf{T}}_{\omega_k}\right)\right)\textbf{R}_{ap,m}$, $\tilde{\boldsymbol{\Delta}}_{mk}=\bar{\beta}_{ap,m}\bar{\textbf{g}}_{ap,m}{\boldsymbol{\Theta}}_{\omega_k}\bar{\boldsymbol{\Theta}}_{\omega_k}\boldsymbol{\Delta}_{u,k}\bar{\boldsymbol{\Theta}}_{\omega_k}^H{\boldsymbol{\Theta}}_{\omega_k}^H\bar{\textbf{g}}_{ap,m}^H$. Similarly, \eqref{Average2} at the top of this page delivers $\text{tr}\left(\textbf{Z}_{mk}^H\bar{\textbf{g}}_{ap,m}\textbf{T}_{\omega}^H\bar{\textbf{g}}_{ap,m}^H\textbf{Z}_{mk}{\boldsymbol{\Psi}}_{mk}
\right)$. Meanwhile, the terms $\textbf{F}_{k,k'}$, $\bar{\textbf{F}}_{k,\omega}$, $\hat{\textbf{F}}_{\omega,\omega'}$ are given by
\begin{figure*}[t!]
 \begin{equation}
    \begin{array}{ll}
\displaystyle\text{tr}\left(\boldsymbol{\Delta}_{mk''}\textbf{Z}_{mk}^H\boldsymbol{\Delta}_{mk'}\textbf{Z}_{mk}\right)&\displaystyle=\text{tr}\left(\boldsymbol{\hat{\Delta}}_{mk''}\textbf{Z}_{mk}^H\boldsymbol{\hat{\Delta}}_{mk'}\textbf{Z}_{mk}\right)+\text{tr}\left(\hat{\boldsymbol{\Delta}}_{mk''}\textbf{Z}_{mk}^H\boldsymbol{\tilde{\Delta}}_{mk'}\textbf{Z}_{mk}\right)+\text{tr}\left(\boldsymbol{\tilde{\Delta}}_{mk''}\textbf{Z}_{mk}^H\boldsymbol{\hat{\Delta}}_{mk'}\textbf{Z}_{mk}\right)\\&\displaystyle+
\bar{\beta}_{ap,m}^2\text{tr}\left(
\left(  \begin{array}{ll}
\check{\boldsymbol{\Delta}}_{uk''}\varpi_{mk}\check{\boldsymbol{\Delta}}_{uk'}
-\underbrace{\left(\check{\boldsymbol{\Delta}}_{uk''}\varpi_{mk}\check{\boldsymbol{\Delta}}_{uk'}\right)\circ\textbf{I}_L}_{\omega_{k''}=\omega_{k'}}
\\\displaystyle+
\underbrace{\left(\boldsymbol{\Theta}_{\omega_{k''}}{\boldsymbol{\Delta}}_{uk''}\boldsymbol{\Theta}_{\omega_{k''}}^H\left(\phi^2\varpi_{mk}+(1-\phi^2)\varpi_{mk}\circ\textbf{I}_L\right)\boldsymbol{\Theta}_{\omega_{k'}}{\boldsymbol{\Delta}}_{uk'}\boldsymbol{\Theta}_{\omega_{k'}}^H\right)\circ\textbf{I}_L}_{\omega_{k''}=\omega_{k'}}
\end{array}
\right)\varpi_{mk}^H
\right),
      \end{array}\label{Average1}
      \vspace{-3 pt}
\end{equation}
\vspace{-5 pt}
\hrulefill
\end{figure*}
\begin{figure*}[t!]
 \begin{equation}
    \begin{array}{ll}
\text{tr}\left(\textbf{Z}_{mk}^H\bar{\textbf{g}}_{ap,m}\textbf{T}_{\omega}^H\bar{\textbf{g}}_{ap,m}^H\textbf{Z}_{mk}{\boldsymbol{\Psi}}_{mk}
\right)&\displaystyle=\text{tr}\left(\textbf{Z}_{mk}^H\bar{\textbf{g}}_{ap,m}\boldsymbol{\Theta}_{\omega}\bar{\boldsymbol{\Theta}}_{\omega}\textbf{R}_\text{SRIS}\bar{\boldsymbol{\Theta}}_{\omega}^H{\boldsymbol{\Theta}}_{\omega}^H\bar{\textbf{g}}_{ap,m}^H\textbf{Z}_{mk}{\boldsymbol{\Psi}}_{mk}
\right)\\&\displaystyle=\tau_pp_p\sum\limits_{k'\in\mathcal{P}_k}\text{tr}\left(\textbf{Z}_{mk}^H\bar{\textbf{g}}_{ap,m}\textbf{T}_{\omega}^H\bar{\textbf{g}}_{ap,m}^H\textbf{Z}_{mk}\bar{\boldsymbol{\Delta}}_{mk'}
\right)
\\&\displaystyle+\tau_pp_p\bar{\beta}_{ap,m}\sum\limits_{k'\in\mathcal{P}_k}\text{tr}\left(\left(  \begin{array}{ll}
\textbf{T}_{\omega}\varpi_{mk}^H\check{\boldsymbol{\Delta}}_{uk'}\displaystyle-\underbrace{\left(\textbf{T}_{\omega}\varpi_{mk}^H\check{\boldsymbol{\Delta}}_{uk'}\right)\circ\textbf{I}_L}_{\omega=\omega_{k'}}
\\\displaystyle+
\underbrace{\left(\boldsymbol{\Theta}_{\omega}{\textbf{R}}_\text{SRIS}\boldsymbol{\Theta}_{\omega}^H\left(\phi^2\varpi_{mk}^H+(1-\phi^2)\varpi_{mk}^H\circ\textbf{I}_L\right)\boldsymbol{\Theta}_{\omega_{k'}}{\boldsymbol{\Delta}}_{uk'}\boldsymbol{\Theta}_{\omega_{k'}}^H\right)\circ\textbf{I}_L}_{\omega=\omega_{k'}}
\end{array}
\right)\varpi_{mk}\right)
\\&\displaystyle+
\sum\limits_{\omega'=t,r}\sigma^2_{\omega'}\bar{\beta}_{ap,m}\text{tr}\left(\left(  \begin{array}{ll}
\textbf{T}_{\omega}\varpi_{mk}^H\textbf{T}_{\omega'}
\displaystyle-\underbrace{\left(\textbf{T}_{\omega}\varpi_{mk}^H\textbf{T}_{\omega'}\right)\circ\textbf{I}_L}_{\omega=\omega'}
\\\displaystyle+
\underbrace{\left(\boldsymbol{\Theta}_{\omega}{\textbf{R}}_\text{SRIS}\boldsymbol{\Theta}_{\omega}^H\left(\phi^2\varpi_{mk}^H+(1-\phi^2)\varpi_{mk}^H\circ\textbf{I}_L\right)\boldsymbol{\Theta}_{\omega'}{\textbf{R}}_\text{SRIS}\boldsymbol{\Theta}_{\omega'}^H\right)\circ\textbf{I}_L}_{\omega=\omega'}
\end{array}
\right)\varpi_{mk}\right)
\\&\displaystyle+
\sum\limits_{\omega'=t,r}\sigma^2_{\omega'}\tilde{\beta}_{ap,m}\text{tr}\left({\textbf{R}}_\text{SRIS}{\textbf{T}}_{\omega'}\right)\text{tr}\left(\textbf{Z}_{mk}^H\bar{\textbf{g}}_{ap,m}\textbf{T}_{\omega}^H\bar{\textbf{g}}_{ap,m}^H\textbf{Z}_{mk}\textbf{R}_{ap,m}
\right)\displaystyle+\sigma^2\text{tr}\left(\textbf{Z}_{mk}^H\bar{\textbf{g}}_{ap,m}\textbf{T}_{\omega}^H\bar{\textbf{g}}_{ap,m}^H\textbf{Z}_{mk}
\right).
  \end{array}\label{Average2}
  \vspace{-6 pt}
\end{equation}
\vspace{-10 pt}
\hrulefill
\end{figure*}
\begin{equation}
    \begin{array}{ll}
\textbf{F}_{k,k'}&\displaystyle=\mathbb{E}\left\{\boldsymbol{\Theta}_{\omega_{k}}\bar{\boldsymbol{\Theta}}_{\omega_{k}}\boldsymbol{\Delta}_{uk}\bar{\boldsymbol{\Theta}}_{\omega_{k}}^H\boldsymbol{\Theta}_{\omega_{k}}^H\textbf{R}_\text{SRIS}{\boldsymbol{\Theta}}_{\omega_{k'}}\bar{\boldsymbol{\Theta}}_{\omega_{k'}}\boldsymbol{\Delta}_{uk'}\bar{\boldsymbol{\Theta}}_{\omega_{k'}}^H\boldsymbol{\Theta}_{\omega_{k'}}^H\right\}\\&\displaystyle=\check{\boldsymbol{\Delta}}_{uk}\textbf{R}_\text{SRIS}\check{\boldsymbol{\Delta}}_{uk'}\displaystyle-\underbrace{\left(\check{\boldsymbol{\Delta}}_{uk}\textbf{R}_\text{SRIS}\check{\boldsymbol{\Delta}}_{uk'}\right)\circ\textbf{I}_L}_{\omega_{k}= \omega_{k'}}\\&\displaystyle+\underbrace{A\left(\boldsymbol{\Theta}_{\omega_{k}}{\boldsymbol{\Delta}}_{uk}\boldsymbol{\Theta}_{\omega_{k}}^H\bar{\textbf{R}}\boldsymbol{\Theta}_{\omega_{k'}}{\boldsymbol{\Delta}}_{uk'}\boldsymbol{\Theta}_{\omega_{k'}}^H\right)\circ\textbf{I}_L}_{\omega_{k}= \omega_{k'}}
    \end{array}
    \label{F_matrix}
\end{equation}
\begin{equation}
    \begin{array}{ll}
\bar{\textbf{F}}_{k,\omega}&\displaystyle=\mathbb{E}\left\{\boldsymbol{\Theta}_{\omega_{k}}\bar{\boldsymbol{\Theta}}_{\omega_{k}}\boldsymbol{\Delta}_{uk}\bar{\boldsymbol{\Theta}}_{\omega_{k}}^H\boldsymbol{\Theta}_{\omega_{k}}^H\textbf{R}_\text{SRIS}{\boldsymbol{\Theta}}_{\omega}\bar{\boldsymbol{\Theta}}_{\omega}\textbf{R}_\text{SRIS}\bar{\boldsymbol{\Theta}}_{\omega}^H\boldsymbol{\Theta}_{\omega}^H\right\}\\&\displaystyle=
\check{\boldsymbol{\Delta}}_{uk}\textbf{R}_\text{SRIS}{\textbf{T}}_{\omega}\displaystyle-\underbrace{\left(\check{\boldsymbol{\Delta}}_{uk}\textbf{R}_\text{SRIS}{\textbf{T}}_{\omega}\right)\circ\textbf{I}_L}_{\omega_{k}= \omega}\\&\displaystyle+\underbrace{A\left(\boldsymbol{\Theta}_{\omega_{k}}{\boldsymbol{\Delta}}_{uk}\boldsymbol{\Theta}_{\omega_{k}}^H\bar{\textbf{R}}\boldsymbol{\Theta}_{\omega}\textbf{R}_\text{SRIS}\boldsymbol{\Theta}_{\omega}^H\right)\circ\textbf{I}_L}_{\omega_{k}= \omega},
    \end{array}
\end{equation}
\begin{equation}
    \begin{array}{ll}
\hat{\textbf{F}}_{\omega,\omega'}&\displaystyle=\mathbb{E}\left\{\boldsymbol{\Theta}_{\omega}\bar{\boldsymbol{\Theta}}_{\omega}\textbf{R}_\text{SRIS}\bar{\boldsymbol{\Theta}}_{\omega}^H\boldsymbol{\Theta}_{\omega}^H\textbf{R}_\text{SRIS}{\boldsymbol{\Theta}}_{\omega'}\bar{\boldsymbol{\Theta}}_{\omega'}\textbf{R}_\text{SRIS}\bar{\boldsymbol{\Theta}}_{\omega'}^H\boldsymbol{\Theta}_{\omega'}^H\right\}\\&\displaystyle=
{\textbf{T}}_{\omega}\textbf{R}_\text{SRIS}{\textbf{T}}_{\omega'}\displaystyle-\underbrace{\left({\textbf{T}}_{\omega}\textbf{R}_\text{SRIS}{\textbf{T}}_{\omega'}\right)\circ\textbf{I}_L}_{\omega'= \omega}\\&\displaystyle+\underbrace{A\left(\boldsymbol{\Theta}_{\omega}\textbf{R}_\text{SRIS}\boldsymbol{\Theta}_{\omega}^H\bar{\textbf{R}}\boldsymbol{\Theta}_{\omega'}\textbf{R}_\text{SRIS}\boldsymbol{\Theta}_{\omega'}^H\right)\circ\textbf{I}_L}_{\omega'= \omega},
    \end{array}
\end{equation}
Moreover, the terms $\textbf{K}_{mk,k',k''}$, $\bar{\textbf{K}}_{mk,k',\omega}$ and $\hat{\textbf{K}}_{mk,\omega',\omega}$ are given by \eqref{K_matrix}-\eqref{K_matrix_final} at the top of this page, respectively. 
\begin{figure*}[t!]
\begin{equation}
    \begin{array}{ll}
\textbf{K}_{mk,k',k''}&\displaystyle=\mathbb{E}\left\{\boldsymbol{\Theta}_{\omega_{k'}}\bar{\boldsymbol{\Theta}}_{\omega_{k'}}\boldsymbol{\Delta}_{uk'}\bar{\boldsymbol{\Theta}}_{\omega_{k'}}^H\boldsymbol{\Theta}_{\omega_{k'}}^H\boldsymbol{\Delta}_{ap,mk}^H{\boldsymbol{\Theta}}_{\omega_{k''}}\bar{\boldsymbol{\Theta}}_{\omega_{k''}}\boldsymbol{\Delta}_{uk''}\bar{\boldsymbol{\Theta}}_{\omega_{k''}}^H\boldsymbol{\Theta}_{\omega_{k''}}^H\right\}\\&\displaystyle=
\check{\boldsymbol{\Delta}}_{uk'}\boldsymbol{\Delta}_{ap,mk}^H\check{\boldsymbol{\Delta}}_{uk''}-\underbrace{\left(\check{\boldsymbol{\Delta}}_{uk'}\boldsymbol{\Delta}_{ap,mk}^H\check{\boldsymbol{\Delta}}_{uk''}-\left(\boldsymbol{\Theta}_{\omega_{k'}}\boldsymbol{\Delta}_{uk'}\boldsymbol{\Theta}_{\omega_{k'}}^H\bar{\boldsymbol{\Delta}}_{ap,mk}^H{\boldsymbol{\Theta}}_{\omega_{k''}}\boldsymbol{\Delta}_{uk''}\boldsymbol{\Theta}_{\omega_{k''}}^H\right)\right)\circ\textbf{I}_L}_{\omega_{k''}=\omega_{k'}},
    \end{array}
     \vspace{-2 pt}
     \label{K_matrix}
\end{equation}
\vspace{-5 pt}
\hrulefill
\end{figure*}
\begin{figure*}[t!]
\begin{equation}
    \begin{array}{ll}
\bar{\textbf{K}}_{mk,k',\omega}&\displaystyle=\mathbb{E}\left\{\boldsymbol{\Theta}_{\omega_{k'}}\bar{\boldsymbol{\Theta}}_{\omega_{k'}}\boldsymbol{\Delta}_{uk'}\bar{\boldsymbol{\Theta}}_{\omega_{k'}}^H\boldsymbol{\Theta}_{\omega_{k'}}^H\boldsymbol{\Delta}_{ap,mk}^H{\boldsymbol{\Theta}}_{\omega}\bar{\boldsymbol{\Theta}}_{\omega}\textbf{R}_\text{SRIS}\bar{\boldsymbol{\Theta}}_{\omega}^H\boldsymbol{\Theta}_{\omega}^H\right\}\\&\displaystyle=
\check{\boldsymbol{\Delta}}_{uk'}\boldsymbol{\Delta}_{ap,mk}^H\textbf{T}_{\omega}-\underbrace{\left(\check{\boldsymbol{\Delta}}_{uk'}\boldsymbol{\Delta}_{ap,mk}^H\textbf{T}_{\omega}-\left(\boldsymbol{\Theta}_{\omega_{k'}}\boldsymbol{\Delta}_{uk'}\boldsymbol{\Theta}_{\omega_{k'}}^H\bar{\boldsymbol{\Delta}}_{ap,mk}^H{\boldsymbol{\Theta}}_{\omega}\textbf{R}_\text{SRIS}\boldsymbol{\Theta}_{\omega}^H\right)\right)\circ\textbf{I}_L}_{\omega=\omega_{k'}}.
    \end{array}
     \vspace{-2 pt}
\end{equation}
\hrulefill
\vspace{-5 pt}
\end{figure*}
\begin{figure*}[t!]
\begin{equation} 
    \begin{array}{ll}
\hat{\textbf{K}}_{mk,\omega',\omega}&\displaystyle=\mathbb{E}\left\{\boldsymbol{\Theta}_{\omega'}\bar{\boldsymbol{\Theta}}_{\omega'}\textbf{R}_\text{SRIS}\bar{\boldsymbol{\Theta}}_{\omega'}^H\boldsymbol{\Theta}_{\omega'}^H\boldsymbol{\Delta}_{ap,mk}^H{\boldsymbol{\Theta}}_{\omega}\bar{\boldsymbol{\Theta}}_{\omega}\textbf{R}_\text{SRIS}\bar{\boldsymbol{\Theta}}_{\omega}^H\boldsymbol{\Theta}_{\omega}^H\right\}\\&\displaystyle=
\textbf{T}_{\omega'}\boldsymbol{\Delta}_{ap,mk}^H\textbf{T}_{\omega}-\underbrace{\left(\textbf{T}_{\omega'}\boldsymbol{\Delta}_{ap,mk}^H\textbf{T}_{\omega}-\left({\boldsymbol{\Theta}}_{\omega'}\textbf{R}_\text{SRIS}\boldsymbol{\Theta}_{\omega'}^H\bar{\boldsymbol{\Delta}}_{ap,mk}^H{\boldsymbol{\Theta}}_{\omega}\textbf{R}_\text{SRIS}\boldsymbol{\Theta}_{\omega}^H\right)\right)\circ\textbf{I}_L}_{\omega=\omega'}.
    \end{array}
     \vspace{-2 pt}
     \label{K_matrix_final}
\end{equation}
\hrulefill
\vspace{-10 pt}
\end{figure*}
  \subsection{Compute $\textbf{a}_{k}^H\mathbb{E}\left\{\bar{\mathbf{\Gamma}}_{\omega,k}\bar{\mathbf{\Gamma}}_{\omega,k}^H\right\}\textbf{a}_{k}$ and $\sigma^2\textbf{a}_{k}^H\bar{\mathbf{\Lambda}}_{k}\textbf{a}_{k}$ } With the help of \eqref{T_bar_matrix}-\eqref{EMI_t}, we can decompose the EMI term into \eqref{EMI_CF} at the top of this page
\begin{figure*}[t!]
\begin{equation}
\begin{array}{ll}
&\textbf{a}_{k}^H\mathbb{E}\left\{\bar{\mathbf{\Gamma}}_{\omega,k}\bar{\mathbf{\Gamma}}_{\omega,k}^H\right\}\textbf{a}_{k}\displaystyle=\mathbb{E}\Big{\{}\Big{|}\sum\nolimits_{m=1}^M a_{mk}^*\hat{\textbf{g}}_{mk}^{H}{\textbf{g}}_{ap,m}\bar{\boldsymbol{\Theta}}_{\omega}\boldsymbol{\Theta}_{\omega}\textbf{n}_{\omega}\Big{|}^2\Big{\}}\\&\displaystyle=\mathbb{E}\Big{\{}\sum\nolimits_{m=1}^M a_{mk}^*\hat{\textbf{g}}_{mk}^{H}{\textbf{g}}_{ap,m}\bar{\boldsymbol{\Theta}}_{\omega}\boldsymbol{\Theta}_{\omega}\textbf{n}_{\omega}\textbf{n}_{\omega}^H \boldsymbol{\Theta}_{\omega}^H\bar{\boldsymbol{\Theta}}_{\omega}^H({\textbf{g}}_{ap,m})^H\hat{\textbf{g}}_{mk}a_{mk}^T\Big{\}}+\mathbb{E}\Big{\{}\sum\nolimits_{m=1}^M\sum\nolimits_{n\neq m} a_{mk}^*\hat{\textbf{g}}_{mk}^{H}{\textbf{g}}_{ap,m}\bar{\boldsymbol{\Theta}}_{\omega}\boldsymbol{\Theta}_{\omega}\textbf{n}_{\omega}\textbf{n}_{\omega}^H \boldsymbol{\Theta}_{\omega}^H\bar{\boldsymbol{\Theta}}_{\omega}^H({\textbf{g}}_{ap,n})^H\hat{\textbf{g}}_{nk}a_{nk}^T\Big{\}}
\\&\displaystyle=\sum\nolimits_{m=1}^M\sigma_{\omega}^2a_{mk}^*a_{mk}^T\underbrace{\left(\begin{array}{ll}
\bar{\beta}_{ap,m}\text{tr}\left(\textbf{Z}_{mk}^H\bar{\textbf{g}}_{ap,m}\textbf{T}_{\omega}^H\bar{\textbf{g}}_{ap,m}^H\textbf{Z}_{mk}{\boldsymbol{\Psi}}_{mk}
\right)
+\tilde{\beta}_{ap,m}\text{tr}\left(\textbf{R}_\text{SRIS}\textbf{T}_{\omega}\right)\text{tr}\left(\textbf{Z}_{mk}^H{\textbf{R}}_{ap,m}\textbf{Z}_{mk}{\boldsymbol{\Psi}}_{mk}
\right)
\\\displaystyle+
\tau_pp_p\sum\nolimits_{k'\in\mathcal{P}_k}\left(\begin{array}{ll}
\bar{\beta}_{ap,m}\tilde{\beta}_{ap,m}\text{tr}\left(\textbf{R}_{ap,m}\textbf{Z}_{mk}
\right)\text{tr}\left(\varpi_{mk}\bar{\textbf{F}}_{k',\omega}^H\right)\displaystyle+
\bar{\beta}_{ap,m}\tilde{\beta}_{ap,m}\text{tr}\left(\textbf{R}_{ap,m}\textbf{Z}_{mk}^H
\right)\text{tr}\left(\bar{\textbf{F}}_{k',\omega}\varpi_{mk}^H\right)
\\\displaystyle+
\tilde{\beta}_{ap,m}^2\text{tr}\left(\textbf{R}_{ap,m}\textbf{Z}_{mk}^H
\right)\text{tr}\left(\textbf{R}_{ap,m}\textbf{Z}_{mk}
\right)\text{tr}\left(\bar{\textbf{F}}_{k',\omega}\textbf{R}_\text{SRIS}\right)
\end{array}
\right)
\\\displaystyle+
\sum\nolimits_{\omega '=t,r}\sigma_{\omega '}^2\left(\begin{array}{ll}
\bar{\beta}_{ap,m}\tilde{\beta}_{ap,m}\text{tr}\left(\textbf{R}_{ap,m}\textbf{Z}_{mk}
\right)\text{tr}\left(\varpi_{mk}\hat{\textbf{F}}_{\omega,\omega'}\right)
\displaystyle+
\bar{\beta}_{ap,m}\tilde{\beta}_{ap,m}\text{tr}\left(\textbf{R}_{ap,m}\textbf{Z}_{mk}^H
\right)\text{tr}\left(\hat{\textbf{F}}_{\omega,\omega'}^H\varpi_{mk}\right)
\\\displaystyle+
\tilde{\beta}_{ap,m}^2\text{tr}\left(\textbf{R}_{ap,m}\textbf{Z}_{mk}^H
\right)\text{tr}\left(\textbf{R}_{ap,m}\textbf{Z}_{mk}
\right)\text{tr}\left(\hat{\textbf{F}}_{\omega,\omega'}^H\textbf{R}_\text{SRIS}\right)
\end{array}
\right)
\end{array}
\right)}_{\gamma_{\omega,mmk}}
\\&\displaystyle+\sum\nolimits_{m=1}^M\sum\nolimits_{m'\neq m}\sigma_{\omega}^2a_{mk}^*a_{m'k}^T\underbrace{\left(
\begin{array}{ll}
\displaystyle
\tau_pp_p\sum\nolimits_{k'\in\mathcal{P}_k}\text{tr}\left(\boldsymbol{\Delta}_{ap,mk}\bar{\textbf{K}}_{m'k,k',\omega}\right)
+\displaystyle
\sum\nolimits_{\omega '=t,r}\sigma_{\omega '}^2\text{tr}\left(\boldsymbol{\Delta}_{ap,mk}\hat{\textbf{K}}_{m'k,\omega,\omega'}\right)
\end{array}
\right)}_{\gamma_{\omega,mm'k}}.
\end{array}
\label{EMI_CF}
   \vspace{-5 pt}
   \end{equation}
   \vspace{-10 pt}
      \hrulefill
\end{figure*}
with  $\mathbf{\Gamma}_{\omega,k}\in\mathbb{C}^{M\times M}$, where the diagonal elements $\gamma_{\omega,mmk}$ and the other elements $\gamma_{\omega,mm'k},~m\neq m'$ in $\mathbf{\Gamma}_{\omega,k}$ are given accordingly in \eqref{EMI_CF} . Finally, we can compute the noise power by
\begin{equation}
\begin{array}{ll}
\displaystyle\sigma^2\textbf{a}_{k}^H\bar{\mathbf{\Lambda}}_{k}\textbf{a}_{k}&\displaystyle=\mathbb{E}\bigg{\{}\sum\nolimits_{m=1}^Ma_{mk}^*\hat{\textbf{g}}_{mk}^{H}\textbf{w}_{m}\textbf{w}_{m}^H\hat{\textbf{g}}_{mk}a_{mk}^T\bigg{\}}
\\\vspace{3pt}&\displaystyle=\sum\nolimits_{m=1}^M\sigma^2a_{mk}^*\text{tr}(\textbf{Q}_{mk})a_{mk}^T
\\ &\displaystyle=
\sigma^2\textbf{a}_{k}^H\mathbf{\Lambda}_{k}\textbf{a}_{k},
\end{array}\label{lambda}
   \end{equation}
where $\mathbf{\Lambda}_{k}=\text{diag}(\textbf{b}_{k})\in\mathbb{C}^{M\times M}$, and this finishes the proof.

\end{appendices}

% or


%\appendices

%\section{}

%\section*{Acknowledgment}

\ifCLASSOPTIONcaptionsoff
  \newpage
\fi



\bibliographystyle{IEEEtran}
\bibliography{IEEEabrv,ref}

\end{document}